\documentclass[11pt,a4paper]{article}
\usepackage[utf8]{inputenc}
\usepackage{subfigure}
\usepackage{jheppub}
\usepackage{amsthm,amsbsy,amsfonts,mathrsfs,enumerate,float,wrapfig,amsmath}
\usepackage{subfigure}
\usepackage{tikz}
\usepackage{textcomp}
\usepackage{ytableau}
\usepackage{mathtools} 
\usepackage{extarrows}
\usepackage{mleftright}
\usepackage{float}
\usepackage{mathtools}
\usepackage{array}
\usepackage{bbold}
\usetikzlibrary{intersections}

\graphicspath{ {figures/} }
\usepackage[T1]{fontenc}     
\usepackage{lmodern}   

\usepackage{tikz}
\usetikzlibrary{arrows}
\usetikzlibrary{shapes.geometric,calc,arrows, positioning,shapes.misc,decorations.markings}
\tikzset{
	big arrow/.style={
		decoration={markings,mark=at position 1 with {\arrow[scale=2,#1]{>}}},
		postaction={decorate},
		shorten >=0.4pt},
	big arrow/.default=black}

\pgfdeclarelayer{edgelayer}
\pgfdeclarelayer{nodelayer}
\pgfsetlayers{edgelayer,nodelayer,main} 
\tikzstyle{none}=[inner sep=0pt]

\newcommand{\nn}{\nonumber}
\def\a{{\alpha}}

\def\b{{\beta}}

\def\r{{\gamma}}

\def\0{{\emptyset}}

\def\inf{{\infty}}

\def \le({\left(}
\def \r){\right)}

\def\cosh{{\text{cosh}}}

\def\>{{  \succcurlyeq} }
\newcommand{\bsmall}{\begin{small}   }
	\newcommand{\esmall}{ \end{small}   }

\def\PE{\mathrm{PE}}

\def\F{\mathbf{F}}
\def\AF{\mathbf{AF}}
\def\N{\mathcal{N}}

\def\balg{ \begin{align} }
\def\ealg{ \end{align} }
\def\bsp{\begin{split}}
\def\esp{\end{split}}
\def\btik{ \begin{tikzpicture} }
\def\etik{ \end{tikzpicture}  }

\def\graybox{{\color{gray}\blacksquare}}

\def\a{{\alpha}}

\def\b{{\beta}}

\def\r{{\gamma}}

\def\inf{{\infty}}

\def\N{{\mathcal{N}}}

\def \fQ { \frac{i Q}{2} }

\def\q{\mathfrak{q}}
\def\t{\mathfrak{t}}

\def\cosh{{\text{cosh}}}

\def\>{{  \succcurlyeq} }

\def\PE{\mathrm{PE}}

\def\N{ \mathcal{N}}

\def\T{ \mathcal{T}}
\def\Li{ \text{Li}  }
\def\log{ \text{log}}
\def\Weff{     \widetilde{\mathcal{W}}^{\eff}      }
\def\W{ \mathcal{W}  }

\def \r){\right)}
\def\eff{\text{eff}}
\def\F{\mathbf{F}}
\def\AF{\mathbf{AF}}
\def\N{\mathcal{N}}
\def\T{\mathcal{T}}

\def\eff{\text{eff}}
\def\CS{{\text{Chern-Simons }}}

\def\1{ \mathbf{1}}
\def\0{ \mathbf{0}}
\def\2{ \mathbf{2}}

\def\n{{\mathbf{n}}}
\def\1{{\mathbf{1}}}
\def\2{{\mathbf{2}}}
\def\3{{\mathbf{3}}}

\def\A{\textbf{A}}
\def\B{\textbf{B}}
\def\E{\textbf{E}}
\def\m{\textbf{m}}
\def\n{\textbf{n}}

\title{
3d $\mathcal{N}=2$ theories and plumbing graphs: adding matter, gauging, and new dualities
}

\author[a,b]{Shi Cheng}
 \author[b]{and Piotr Su\l kowski}

 \affiliation[a]{Department of Physics and Center for Field Theory and Particle Physics, Fudan University,
	220 Handan Road, 200433 Shanghai, China}
\affiliation[b]{Faculty of Physics, University of Warsaw, ul. Pasteura 5, 02-093 Warsaw, Poland}

\emailAdd{mirror2718@gmail.com}  \emailAdd{psulkows@fuw.edu.pl}

\abstract{Recently, a large class of 3d $\N=2$ gauge theories with mixed Chern-Simons levels, corresponding to plumbing 3-manifolds, has been identified. In this paper we generalize these theories by including in their content chiral multiples, and analyze their properties. We find that the content of such theories can be encoded in graphs, which generalize plumbing graphs, and various operations in these theories can be represented in terms of transformations of such graphs. The operations in question include gauging global symmetries, integrating out gauge nodes, which for theories without chiral multiplets corresponds to Kirby moves, and $ST$-transformations that involve chiral multiplets.  The dualities such as mirror triality and SQED-XYZ duality can be also represented in terms of graphs, and enable us to find many new dual theories by gauging global symmetries.  In particular, we find that gauged SQED-XYZ duality leads to other dualities, which take the same form as operations of linking and unlinking discussed in the context of knots-quivers correspondence. We also find that the superpotential can be encoded in an interesting class of triangle graphs that satisfy certain consistency conditions, we discuss decoupling and Higgsing of chiral multiplets, as well as interpretation of various phenomena in terms of brane webs.

	\rule{0pt}{0pt}
	\\
	\rule{0pt}{0pt}
	\\

}

\begin{document}

	\maketitle

	\section{Introduction}

The realm of 3d $\N = 2$ gauge theories is an interesting theoretical laboratory.  Such theories enjoy various dualities \cite{Intriligator:1996ex,Boer:1997ts,Aharony:1997aa,Intriligator:2013lca,Aharony_1997,Ooguri:1999bv,Giveon_2009,Kapustin:1999ha,Tong:2000ky,Aharony:2013dha,Benini:2011aa,Benini:2017dud,Nii:2020ikd,Okazaki:2021gkk,Okazaki:2021pnc,Nii:2014jsa,Kubo:2021ecs,Amariti:2020xqm,Bourgine:2021nyw,Hwang:2021ulb}, which can be tested by analysis of their partition functions \cite{Nieri:2018vc,Benvenuti:2016wet,Dimofte:2010tz,Benvenuti:2011ga,Hwang:2012jh,Beem:2012mb,Benini:2014aa,Pasquetti:2011fj,Zenkevich:2017ylb,Aprile:2018oau,Yoshida:2014ssa,Pan:2016fbl,Closset:2017zgf,Gaiotto:2015una,Fan:2020bov,Manabe:2021hxy,Kimura:2021ngu,Cheng:2021vtq,Cheng:2021aa}, and which can also be understood from perspective of string theory and M-theory.  In recent years,  various constructions of 3d $\N = 2$ theories have been proposed,  for example using M5-branes and compactification  \cite{Terashima:2014aa,Terashima:2011qi,Cecotti:2011iy,Gang:2015aa,Gadde:2013aa,Yamazaki:2012aa,Benini:2011cma,Gang:2018wek,Choi:2022dju}, as well as the associated higher-form symmetries \cite{Gaiotto:2014kfa,Eckhard:2019aa,Bergman:2020ifi,vanBeest:2022fss,Mekareeya:2022spm,Damia:2022rxw,Kaidi:2022cpf,Gukov:2022gei,Antinucci:2022vyk}.  One very interesting construction, inspired by the 3d-3d correspondence, is based on compactification of M5-branes on plumbing three-manifolds, which leads to 3d theories with mixed Chern-Simons levels \cite{Gadde:2013aa,Chung:2016aa,Gukov:2016gkn,Gukov:2017aa,Gukov:2015aa,Gukov:2019ab,Chung:2019khu,Chung:2018rea,Cheng:2018vpl}.  
In this construction,  mixed Chern-Simons levels are identified with linking numbers of one-cycles in a three-manifold, and the content of the 3d gauge theory is encoded in a plumbing graph of this plumbing three-manifold.  In fact,  one can draw many plumbing graphs representing the same manifold, which are related by operations called Kirby moves, and which correspond to dualities between corresponding 3d theories; for abelian theories Kirby moves correspond to integrating out gauge fields corresponding to nodes of plumbing graphs. While Chern-Simons levels were analyzed e.g. in \cite{Dorey:1999rb,Witten:2003ya,Kitao:1999aa,Cheng:2020aa,Closset:2023vos}, theories with mixed Chern-Simons levels have not been thoroughly studied and cannot be easily constructed using brane webs or geometric engineering.  Another interesting aspect is matter content; for example,  relations between WRT invariants and $\hat{Z}$ invariants for $SU(2)$ theories with adjoint chiral multiplets are discussed in \cite{Chung:2016aa,Gukov:2016gkn,Gukov:2017aa}.

In this paper we consider 3d $\mathcal{N}=2$ theories with mixed Chern-Simons levels corresponding to plumbing three-manifolds, and generalize them by adding chiral multiplets (which we also often refer to simply as \emph{matter}).  We consider abelian $U(1)^{N_c}$ theories with various chiral multiplets, in fundamental, bifundamental,  and more general representations.  Various operations and dualities in such theories can be represented as operations on graphs, which generalize plumbing graphs mentioned above -- in addition to nodes that represent gauge fields (which we denote by dots and refer to as gauge nodes), we introduce extra nodes (which we denote by squares and call matter nodes), which represent chiral multiplets.  In this work, we refer to such more general graphs (which include information about matter) as plumbing graphs with matter, or sometimes simply plumbing graphs (even though it is not clear if or how they capture information about putative corresponding plumbing manifolds).  We consider 3d theories on a three-sphere (so that their partition functions involve double-sine functions $s_b(\cdot)$), however we expect that analogous results should also hold for other backgrounds.

More precisely, we discuss a few types of operations on 3d theories of our interest, which have an interesting interpretation in terms of their plumbing graphs. One such operation is gauging of global symmetries, which introduces new gauge nodes into a graph. Another operation amounts to integrating out gauge fields,  and it removes nodes corresponding to such gauge fields from plumbing graphs; this is analogous to Kirby moves discussed in \cite{Gadde:2013aa}. Yet another operation is $ST$-transformation,  discussed first in \cite{Witten:2003ya} and generalized to matter fields in \cite{Dimofte:2011ju}.  We find that $ST$-transformations, which we also refer to as $ST$-moves, can be regarded as Kirby moves that involve matter nodes. The $ST$-transformations represent dualities between corresponding 3d theories and lead to nontrivial identities for their sphere partition functions.  In \cite{Cheng:2020aa}, it is shown that such $ST$-transformation could transform the theory denoted $U(1)+N_f\F+N_{a}\AF$,  with $N_f$ fundamental chiral multiplets (denoted $\F$) and $N_a$ anti-fundamental multiplets (denoted $\AF$),  to the theory $(U(1)+1\F)^{N_f+N_a}_{k_{ij}}$ with mixed Chern-Simons levels $k_{ij}$. In this note we generalize this operation to generic abelian theories with mixed Chern-Simons levels.  Furthermore, we also find that matter fields in arbitrary representations and charged under many $U(1)$ gauge fields can be mapped by $ST$-transformations to a fundamental matter that is charged only under one gauge field, which may substantially simplify plumbing graphs. Moreover,  all gauge nodes that are not attached to matter nodes can be removed (integrated out) from our graphs,  or in other words each graph can be transformed into a balanced graph (which has the same number of gauge nodes and matter nodes, and each gauge node is connect to one matter node and vice versa). In our analysis and identification of various dualities an important role is played by the parity anomaly, which imposes the condition that effective mixed \CS levels should be integer \cite{Aharony:1997aa}.

Note that recent progress on the global $SL(2,\mathbf{Z})$ symmetry and its generalized version is discussed e.g.  in \cite{Gaiotto:2014kfa,Choi:2022zal,Bhardwaj:2020ymp,Kaidi:2021xfk}.  In the context of 3d-3d correspondence, the $ST$-transformation is interpreted as changing the polarization of corresponding three-manifolds.  However,  we leave a discussion of what adding chiral fields means for corresponding three-manifolds for future work -- the graphs that we introduce in this paper represent simply the content and operations on gauge theories, and we do not discuss their interpretation for putative corresponding three-manifolds.


Our results have also intriguing relation to the knots-quivers correspondence \cite{Kucharski:2017ogk,Kucharski:2017poe}.  According to this correspondence, to a given knot one can assign a quiver, or in fact a family of quivers, so that various invariants of such quivers reproduce invariants of the corresponding knot.  These quivers also capture the content of 3d $\mathcal{N}=2$ theories corresponding to knots; each quiver node represents a chiral multiplet as well as an abelian gauge field, and the quiver matrix is interpreted as a matrix of mixed Chern-Simons levels, for details see \cite{Kucharski:2017ogk,Kucharski:2017poe,Panfil:2018faz,Ekholm:2018eee,Ekholm:2019lmb,Ekholm:2020lqy,Jankowski:2021flt,Cheng:2022rqr,Ekholm:2021gyu,Ekholm:2021irc}.  For a given quiver one can typically find many equivalent quivers, i.e. equivalent 3d $\mathcal{N}=2$ theories,  which encode the same invariants of the corresponding knot. These equivalent quivers can be of the same size \cite{Jankowski:2021flt},  or of larger size than some reference quiver \cite{Kucharski:2017poe,Ekholm:2018eee,Jankowski:2022qdp}. In the latter case, the larger quiver and the reference quivers, or equivalently corresponding matrices of mixed Chern-Simons levels, are related by operations referred to as linking and unlinking \cite{Ekholm:2019lmb}.

One of the main examples that we analyze in this paper is gauged SQED-XYZ duality. We find that it can be reformulated as a duality that transforms a theory with two chiral multiplets into a theory with three chiral multiplets (which we also refer to as 2-3 move),  for which mixed \CS levels of these theories change in the same way as in linking and unlinking operations.  It is amusing to find in the context of this paper operations of linking and unlinking analogous to those in knots-quivers correspondence.  In addition, we also find reformulations of these dualities, which are more general than linking and unlinking, and which we call exotic.  Moreover,  in the example of gauged SQED-XYZ duality and also more generally, we find that superpotentials can be encoded in relations for mixed \CS levels assigned to certain triangles that arise in plumbing graphs.

Let us briefly summarize the notation and the form of plumbing graphs with matter that we consider in this paper:

\vspace{10pt}
\noindent	
$$
\begin{tabular}{m{4cm} p{10cm}}
	~	\begin{tikzpicture}
		\node[fill=gray] (r) at (1,0) {};
	\end{tikzpicture}  
	& {a matter node that represents a free chiral multiplet $\F$}   \\ \vspace{0.5cm}
	\begin{tikzpicture}
		\filldraw (0,0) circle(2pt);
		\draw[thick] (0,0)node[below]{$k$}--(1,0);
		\node[fill=gray] (r) at (1,0) {};
	\end{tikzpicture} 
	& a decorated gauge node $U(1)_k+1\F$: a gauge node $U(1)$ with bare Chern-Simons level $k$, with a fundamental chiral multiplet charged under this gauge group  \\   \vspace{0.5cm}
	\begin{tikzpicture}
		\filldraw (0,0) circle(2pt);	\filldraw (1,0) circle(2pt);
		\draw[thick] (0,0)node[below]{$k_i$}--(1,0)node[below]{$k_j$};
		\draw[thick] (0,0)--(1,0)node[midway, above]{\small $+1$};
		\draw[thick] (1,0)--(2,0);
		\draw[thick] (0,0)--(-1,0.5);
		\draw[thick] (-1,0)--(0,0)--(-1,-0.5);
	\end{tikzpicture}
	& undecorated gauge node $U(1)_{k_i}$ with Chern-Simons level $k_i$ connected to another gauge node $U(1)_{k_j}$ through mixed Chern-Simons level $k_{ij} =+1$
\end{tabular}
$$



Some of the main results that we find, presented in terms of plumbing graphs with matter, take the following form:

\vspace{4pt}
\noindent

\begin{itemize}

\item Gauging the global symmetry of the mirror pair $1 \F \, \leftrightarrow \, U(1)_{\pm1/2} + 1\F$ is represented as
\begin{align}
	\begin{split}
		\begin{tikzpicture}
			\filldraw (0,0) node[below]{$k$}  circle(2pt);
			\draw[thick](0,0)--(0,1);
			\node[fill=gray] (r) at (0,1) {} ;
			\node at (1, -0.1) [below]{};
			\node at (1.5,0.2){		$\xleftrightarrow{~~~~}$
			};
			\filldraw (3,0) node[below]{$k \pm\frac{1}{2}$}circle(2pt);
			\draw[thick](3,0)--(4,0)node[midway,above]{$\pm1$}--(4,1);
			\filldraw (4,0) node[below]{$\pm \frac{1}{2}$}  circle(2pt);
			\node[fill=gray] (r) at (4,1) {} ;
		\end{tikzpicture}	
	\end{split} \nn
\end{align}
We also refer to this operation as $ST$-move for matter nodes. $ST$-moves reduce to ordinary Kirby moves after appropriate decoupling of matter nodes. 

\item
After gauging flavor symmetries, the SQED-XYZ duality can be promoted to a generic {2-3 move}, which we call the gauged SQED-XYZ duality:
\begin{align}\label{gaugedSQEDXYZ}
	\includegraphics[width=2in]{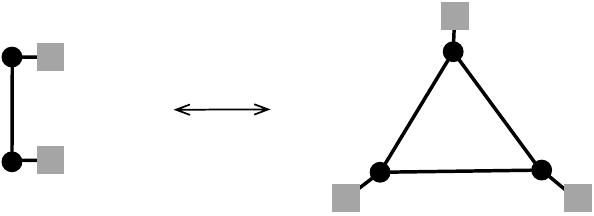} \nn
\end{align}
Mixed \CS levels for these graphs are related by unlinking, linking, or some other exotic relations \eqref{trianglesgraph}, which also encode superpotentials. 

\end{itemize}

In this work we also discuss other properties of plumbing graphs with matter, such as Higgsing and decoupling of matter fields. When discussing these operations we also incorporate Kirby moves for gauge and matter nodes. We find that these operations rely on the mass parameters and impose some constraints on allowed plumbing graphs.  We also reinterpret various properties of theories encoded in plumbing graphs in terms of brane webs.



\subsubsection*{Some open problems...}

There are various open questions that arise from this work, which we leave for future research.  First, it is of interest to generalize abelian gauge groups to higher ranks. Second,  it would be great to understand if plumbing graphs with matter correspond to some class of three-manifolds, more general than plumbing three-manifolds; a close relation between Kirby moves and $ST$-moves that we discuss indeed suggests such a possibility.  

Third, there are also other types of dualities in 3d $\mathcal{N}=2$ theories, which involve monopole operators and corresponding superpotentials.  Such a duality arises already for the basic dual pair SQED-XYZ in \eqref{SQEDXYZ}  \cite{Aharony:1997aa,Intriligator:2013lca}.  SQED, as a $U(1)_0$ theory with two fundamental chiral multiplets $Q, \tilde{Q}$ of charge $\pm1$ respectively, has chiral ring that includes monople operators $V^{\pm}$ and a meson $M=Q\tilde{Q}$. This theory is dual to the XYZ-model, with the fields mapped by $( V^+,V^-,M ) \leftrightarrow (X, Y, Z)$, so that monopole operators are chiral fields coupled by $\W=MV^+V^-$ (note that one should not view $V^\pm$ as monopoles of the magnetic theory; generically, monopole operators in electric theories and magnetic theories are different). The basic dualities in  \eqref{triviliaty}  are decoupling limits of SQED-XYZ duality; in particular,  one basic dual pair $U(1)_{-1/2} +1\F \leftrightarrow 1\F$ can be denoted in more detail by $U(1)_{-1/2} +1Q \leftrightarrow V^+$, where monopole operator $V^+$ becomes the dual free chiral multiplet $\F$. Monopole operators $\mathcal{M}^*$ can also appear in superpotentials of 3d theories and then yield new dualities,  one example being $\W = XYZ + \mathcal{M}^* + \cdots$, discussed in \cite{Benvenuti:2016wet}.  The basic idea is as follows: because of mirror symmetry,  monopole operators in the electric theory map to a product of chiral multiplets in the magnetic theory, namely $ \mathcal{M}^*~\leftrightarrow~ \prod_i \Phi_i $, and vice versa, so adding a meaningful superpotential term (which can be a cubic term $\Phi_1\Phi_2\Phi_3$) upon some operations in the magnetic theory should lead to an additional monopole term $\mathcal{M}^*$ in the superpotential of the electric theory.  Nonetheless, we do not consider such more general theories in this paper, and restrict to the simplest case of the cubic superpotential $\W=XYZ$.  Finding generalization of plumbing graphs and a graphical representation of dualities that involve monopole operators is an interesting task for future work too.

Moreover, the plumbing manifold construction of 3d theories should also be more physically understood from the perspective of 3d-3d correspondence \cite{Terashima:2011qi,Gukov:2016gkn,Gukov:2017aa,Alday:2017yxk}. The basic idea is to glue the gauge nodes along edges which correspond to the self-dual domain wall theory $T[SU(N)]$ found in \cite{Gaiotto:2008ak}. After gauging their flavor symmetries, $T[SU(N)]$ theories could be used to couple two gauge nodes $SU(N)$. For abelian theories, it should be $T[U(1)]$ theory which only contributes to a mixed \CS level. In this note, we only decorate gauge nodes by adding some matter to it, which should not change the coupling of $T[U(1)]$. Many plumbing theories that we discuss indeed exist and have been also constructed using hyperbolic 3-manifolds in \cite{Dimofte:2011ju}.

\subsubsection*{...and the plan of this paper}

This paper is organized as follows. In section \ref{secreview} we review basic properties of 3d $\mathcal{N}=2$ theories, plumbing graphs that encode a structure of an interesting class of such theories,  and 3d-3d correspondence. In section \ref{secgauging} we add chiral multiplets to 3d theories corresponding to plumbing graphs, discuss gauging of global symmetries in such theories, and introduce a large class of dualities based on $ST$-moves.  In section \ref{secsuperpotentials} we present our main example, namely gauged SQED-XYZ duality,  reinterpret it in terms of linking and unlinking operations, and discuss how superpotential is encoded in generalized plumbing graphs.  In section \ref{exchangesec} we analyze the exchange symmetry from perspective of this paper,  and in section \ref{seceffective} we discuss properties of effective superpotential.  In section \ref{secexample} we analyze other examples of 3d theories corresponding to plumbing graphs with matter. In section \ref{secbranewebs} we present a brane web interpretation of various dualities discussed earlier. In appendices we summarize properties of $q$-Pochhammer symbol and of $SL(2,\mathbf{Z})$ group.

\section{3d $\mathcal{N}=2$ theories, 3d-3d correspondence,  and plumbing graphs}  \label{secreview}  

In this section we review some basics of 3d $\mathcal{N}=2$ theories,  3d-3d correspondence -- in particular in the context of plumbing three-manifolds -- and properties of plumbing graphs.

\subsection{Chern-Simons levels and FI parameters in 3d $\mathcal{N}=2$ theories}

Chern-Simons action of an abelian theory with gauge group $U(1)^{N}$ takes form
\begin{align}
S_{CS} =k_{ij} \int A_i \wedge F_j,
\end{align}
where $F_j=d A_j$,  and $A_i$ is the gauge field for the $i$-th gauge group $U(1)$. The term $A_i \wedge F_j$ is symmetric under the exchange of $i$ and $j$, and hence $k_{ij}$ is a symmetric matrix of Chern-Simons levels. The Chern-Simons term in 3d $\N=2$ theories takes form 
$S_{CS} =k_{ij} \int V_i \cdot \Sigma_j$, where $\Sigma$ is the supersymmetric derivative of the vector multiplet $V$.

In this note, rather than using Lagrangians of 3d $\N=2$ theories, we focus on their partition functions on $S^3$ to read off information about their content.  Chern-Simons levels $k_{ij}$ and FI parameters $\xi_i$ contribute the following classical term to sphere partition functions
\begin{align}
e^{- \pi i \, k_{ij} x_i x_j  + 2 \pi i \, \xi_i x_i}   \,.
	\end{align}
According to the localization computation \cite{Hama:2011ea,Kapustin:2009kz}, for a theory on a three-sphere, the contribution of a chiral multiplet is given by a double-sine function
\begin{align}
	s_b\left( \frac{i Q}{2} -{q}_i x_i + m\right) ,
	\end{align}
where ${q}_i$ is its charge under the $i$-th gauge group $U(1)$, $m$ is its real mass parameter, and $Q=b+\frac{1}{b}$. 
In addition, the deformation parameter is defined as
$\mathfrak{q} := e^{\hbar} = e^{2 \pi i \,b^2} = e^{ 2 \pi i \, bQ} $, which is associated to the rotation symmetry along the 3d spacetime $\mathbf{R}_\q^2 \times S^1$. The definition of the double-sine function is 
\begin{align}
	s_b(x) :=   \prod_{n_1, n_2 \geqslant 0 } \frac{ n_1 b + n_2 b^{-1} + Q/2 -i\, x  }{  n_1 b + n_2 b^{-1} + Q/2 + i\, x        }  \,.	  \label{s_b}
	\end{align}


\subsection{Decoupling matter}

In our notation, the fundamental chiral multiplet $\F$ represents the matter field with ${q}=1$ and it contributes $s_b( i Q/2- x+ m)$ to the sphere partition function, while the anti-fundamental multiplet  $\AF$ has ${q}=-1$ and contributes $s_b(i Q/2 +x+ m)$. 
The decoupling of matter is equivalent to taking the limit $m \rightarrow \pm \inf$. The sphere partition functions are given by the contour integral taking values at residues, which arise from the poles of the double sine function located at
\begin{align}
i Q/2-q x+m = i \le( n_1 b +n_2 b^{-1} + Q/2 \r) \,,~~~\text{with}~n_1\,,n_2 \in \mathbf{Z} .
	\end{align}
The pole is around $x \sim  m/q$ in the large mass limit. Therefore, if $q >0$ and $m>0$, only the residues at $x\sim +\inf$ contribute, and if $q <0$ and $m>0$, only the residues at $x\sim -\inf$ contribute.
In addition, using the property that $	s_b(x) \simeq 	e^{\pm \frac{\pi i}{2} {x^2}} $ {as} $x \rightarrow \pm \inf $ and taking into account the sign of charge,  in the large mass limit we have
\begin{align}
	s_b\left( \frac{i Q}{2} -q x + m\right) ~\simeq ~ 	e^{ \frac{\pi i\, q^2 \, \text{sign[q]}\,\text{sign}[m]}{2} {x^2}}  \,,~~~~\text{if}~ m \rightarrow \pm \inf \,,
\end{align}
which shifts Chern-Simons levels.
Therefore,  if all mass parameters are positive,  the effective Chern-Simons level for the theory $U(1)_k +N_f \F+N_{a}\AF$ reads
$k^{\eff} = k+ \frac{N_{\F}}{2}- \frac{N_{\AF}}{2}. $

In particular, decoupling a single fundamental matter field $\F$ from $U(1)$ theory, the theory reduces to a pure gauge theory with Chern-Simons level shifted according to the sign of the mass parameter
\begin{align}
	U(1)_k+1\F ~\longrightarrow~ U(1)_{k + \frac{ \text{sign}(m_{})  }{2}} ,
	\label{decoup1F}
	\end{align}
so that $k^{\eff}=k + \frac{ \text{sign}(m_{})  }{2}$.  In particular, if $k=1/2$,  such a decoupling leads to $U(1)_{1}$ or $U(1)_0$.
Similarly, decoupling the anti-fundamental matter field $\AF$ leads to
\begin{align}
	U(1)_k+1\AF ~\longrightarrow~ U(1)_{	k - \frac{ \text{sign}(m_{})  }{2}} .
	\label{decoup1AF}
\end{align}

The decoupling in \eqref{decoup1F} and \eqref{decoup1AF} can also be easily illustrated using brane webs, in which the Chern-Simons level is associated to a relative angle $k^{\eff}=\tan \theta$ between branes. Decoupling matter is equivalent to sending the corresponding D5-brane to either positive infinity or negative infinity, which shifts the effective Chern-Simons levels differently \cite{Cheng:2021vtq}.


\subsection{Plumbing three-manifolds and plumbing graphs}

In this section we briefly review properties of plumbing three-manifolds, which arise in an interesting instance of 3d-3d correspondence.  One feature of plumbing manifolds is that they can be cut and glued by Dehn surgeries, and their structure and these operations can be represented by plumbing graphs, which are linking graphs of one-cycles in a given manifold.  
Plumbing graphs consist of (black) nodes and lines connecting them. Each node denotes a circle (one-cycle) in the three-manifold under consideration,  and the number attached to it denotes a self-linking number of this one-cycle. A line connecting two cycles -- together with a number attached to it -- denotes their linking number.  An example of such a graph is shown in this figure:
\begin{align}\label{nodegraph}
	\bsp
	\includegraphics[width=3.5in]{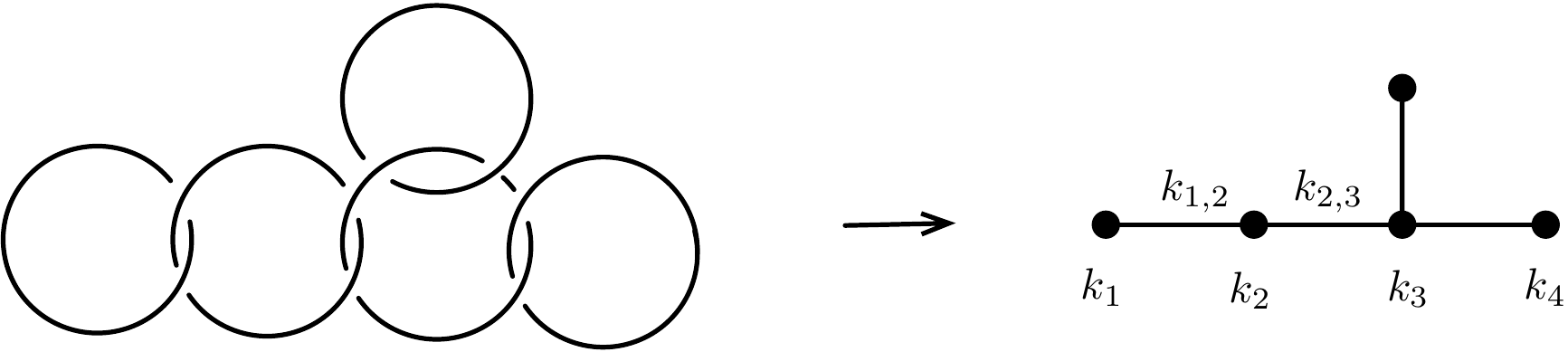}
	\esp
\end{align}


In fact,  a plumbing graph for a given three-manifold is not unique.  Equivalent graphs, which represent the same manifold, are related by operations called Kirby moves, which in the simplest cases can be summarized as follows:
\begin{align}\label{kirbyknown}
	\bsp
	&\begin{tikzpicture}
		\draw[fill] (0,0) circle(2pt)	;
		\draw[thick] (-0.5,0.5)--(0,0)node[below=0.1cm]{$k$}--(-0.5,-0.5);
		\node[thick] at (1.3,0) {$\rightleftarrows$};
		\draw[fill] (3,0) circle(2pt)	;
		\draw[thick] (2.5,0.5)--(3,0)node[below=0.1cm]{~~~~$k \pm 1$}--(2.5,-0.5);
		\draw[thick] (3,0)--(4,0) ;
		\draw[fill] (4,0) node[below=0.1cm]{~~~~$\pm1$} circle(2pt)	;
		\draw[orange,thick] (4,0) circle(6pt)	;
	\end{tikzpicture} 
	\, \qquad\qquad \\
	&\begin{tikzpicture}
		\draw[ fill=gray, rounded corners] (-1,-0.7) rectangle (-0.3,0.7);
		\draw[ fill=gray, rounded corners] (1.8,-0.7) rectangle (2.5,0.7);
		\node[thick] at (0.8,0) {$\rightleftarrows$};
		\draw[fill] (3,0) circle(2pt)	;
		\draw[thick] (2.5,0.5)--(3,0)node[below=0.1cm]{$k $}--(2.5,-0.5);
		\draw[thick] (3,0)--(4,0) ;
		\draw[fill] (4,0) node[below=0.1cm]{$0$} circle(2pt)	;
		\draw[orange,thick] (4,0) circle(6pt)	;
	\end{tikzpicture} 
	\\
	&\begin{tikzpicture}
		\draw[fill] (0,0) circle(2pt)	;
		\draw[fill] (1,0) circle(2pt)	;
		\draw[fill] (2,0) circle(2pt)	;
		\draw[thick] (-0.5,0.5)--(0,0)node[below=0.1cm]{$k_1$}--(-0.5,-0.5);
		\draw[thick]  (0,0)--(1,0)node[below=0.1cm]{$0$};
		\draw[thick]  (1,0)--(2,0)node[below=0.1cm]{$k_2$};
		\draw[thick] (2.5,0.5)--(2,0)--(2.5,-0.5)	;
		\node[thick] at (3.7,0) {$\rightleftarrows$};
		\draw[thick] (5,0.5)--(6,-0.5)  ;
		\draw[thick] (4.7,0)--(5.7+0.5,0)  ;
		\draw[thick] (5,-0.5)--(6,0.5)  ;
		\draw[fill] (5.5,0) circle(2pt)	;
		\node[thick] at (5.5,-0.8) {$k_1+k_2$} ;
		\draw[orange,thick] (1,0) circle(6pt)	;
	\end{tikzpicture} 
	\, \qquad \qquad 	 \\
	&\begin{tikzpicture}
		\draw[fill] (0,0) circle(2pt)	;
		\draw[fill] (1,0) circle(2pt)	;
		\draw[fill] (2,0) circle(2pt)	;
		\draw[thick] (-0.5,0.5)--(0,0)node[below=0.5cm]{$k_1 \pm1$}--(-0.5,-0.5);
		\draw[thick]  (0,0)--(1,0)node[below=0.5cm]{$\pm1$};
		\draw[thick]  (1,0)--(2,0)node[below=0.5cm]{$k_2\pm 1$};
		\draw[thick] (2.5,0.5)--(2,0)--(2.5,-0.5)	;
		\node[thick] at (3.5,0) {$\rightleftarrows$};
		\draw[thick] (4.5,0.5)--(5.0,0)  ;
		\draw[thick] (4.5,-0.5)--(5,0)--(6,0)  ;
		\draw[thick] (6.5,0.5)--(6,0)-- (6.5,-0.5)  ;
		\draw[fill] (5,0) circle(2pt)	;
		\draw[fill] (6,0) circle(2pt)	;
		\node[thick] at (5,-0.5) {$k_1$} ;
		\node[thick] at (6,-0.5) {$k_2$} ;
		\draw[orange,thick] (1,0) circle(6pt)	;
	\end{tikzpicture} 
	\esp
\end{align}
The operation in the first line is usually called a blow-up or a blow-down of a gauge node. The second operation means that if a gauge node has vanishing self-linking number and it is connected to only one other gauge node, then these two gauge nodes can be removed from the graph. The third operation means that if a node with vanishing self-linking is connected to two other nodes, then these three nodes can be merged into one node. The fourth operation has an analogous interpretation.  It follows that Kirby moves introduce additional nodes or remove redundant nodes from plumbing graphs.   We use orange circles to mark nodes introduced by Kirby moves. 


\subsection{3d $\mathcal{N}=2$ theories and Kirby moves}

The 3d-3d correspondence was initiated among others in \cite{Dimofte:2011ju}, where one class of 3d $\N=2$ theories was constructed by wrapping M5-branes on hyperbolic three-manifolds.  Hyperbolic three-manifolds can be decomposed into tetrahedra,  denoted in what follows by $\Delta$, which in this construction correspond to 3d free chiral multiplets; the theory of a single free chiral multiplet is denoted by $\T_\Delta$. Gluing these tetrahedra into the whole three-manifold corresponds to combining chiral multiplets, using superpotentials,  into a nontrivial 3d $\N=2$ theory. Different tetrahedra decompositions of hyperbolic three-manifolds are then supposed to give rise to dualities between corresponding $\N=2$ theories. 

In \cite{Gadde:2013aa,Gukov:2016gkn,Gukov:2017aa} the above construction was adapted to plumbing manifolds.  In this paper we consider 3d $\N=2$ theories that arise from such a construction in case of a single M5-brane, and generalize them by including chiral multiplets.
Compactifing one M5-brane on a plumbing three-manifold leads to an abelian 3d $\mathcal{N}=2$ theory.  According to the 3d-3d dictionary,  each node $i$ in a plumbing graph of such a three-manifold corresponds to a gauge group $U(1)_{k_i}$ (hence we also call the nodes in plumbing graphs as gauge nodes), and each line connecting nodes $i$ and $j$ and denoted by $k_{ij}$ corresponds to a mixed Chern-Simons level $k_{ij}$ in the resulting theory.
An adjoint matter can be also assigned to a gauge node, however it decouples in abelian theories. Therefore,  in this case we only get pure supersymmetric Chern-Simons theories $\T_\Omega$ labeled by plumbing graphs $\Omega$, whose sphere partition functions take form
\begin{align}
Z_{S_b^3}(\T_\Omega ) =\prod_{i=1}^{N_c}	\int dx_i \, e^{- \pi i \, k_{ij} x_i x_j  + 2 \pi i \, \xi_i x_i}   \,.
	\end{align}



We now show that Kirby moves that remove gauge nodes can be interpreted as integrating out corresponding vector multiplets in a 3d $\mathcal{N}=2$ theory \cite{Gadde:2013aa} (which we also refer to as integrating out gauge nodes).   
From the path integral perspective, integrating out a vector multiplet $\hat{V}$ leads to a new theory with a new action
\begin{align}
 Z =  \int \mathcal{D}{\Phi} \int \mathcal{D}{V}   \Big(\int \mathcal{D}\hat{V}   e^{S[\Phi,V,\hat{V}]}  \Big)=     \int \mathcal{D}{\Phi} \int \mathcal{D}{V}  ~ e^{\hat{S}[\Phi,V]} 
\end{align}
where $\Phi$ and $V$ denote respectively other matter fields and vector multiplets. The localization reduces path integrals to some contour integrals. More specifically,  one can integrate out a particular gauge field $U(1)_x\subset U(1)_x  \times \le( \prod_i U(1)_{x_i}  \r)$\footnote{We use the same notation to denote $U(1)_k$ and $U(1)_x$, where $k$ in $U(1)_k$ denotes the Chern-Simons level, while $x$ in $U(1)_{x}$ denotes the variable in the corresponding contour integral. } in an abelian gauge theory, if it does not interact with any chiral multiplet
\begin{align}\label{integnode}
	Z[x_i] =\int d x\,\hat{Z}[x_i,x] =  \int d x\,  e^{ -\pi i\, k x^2+ 2 \pi i (\dots) x} \hat{Z}[ x_i] \,,
\end{align}
where $ \hat{Z}[ x_i] $ is the part that  does not depend on $x$. The terms in $(...)$ can be viewed as the FI coupling for this gauge field $U(1)_x$ with other gauge fields $U(1)_{x_i}$.  

For abelian theories the contribution of a pure gauge node $U(1)_k$ is given by a gaussian integral,  which shifts Chern-Simons levels for other gauge nodes
\begin{align}\label{integform}
	\bsp
	\int dx\, e ^{-\pi i\, k x^2 +  2 \pi i\, x \sum_i p_i x_i   } =
	\begin{cases}
		&e^{- \pi i \le(-\frac{1}{k}  \r) \le(\sum_i p_i x_i\r)^2} \,,  ~\quad \text{if} ~k\neq 0  \\
		&   \delta\le( \sum_i p_i x_i \r)   \,,  \qquad ~\qquad \text{if} ~k= 0  
	\end{cases}
\esp
\end{align}
where $p_i$ denotes mixed Chern-Simons levels between $U(1)_x$ and $U(1)_{x_i}$.  Therefore, when the gauge node $U(1)_k$ is integrated out,  the mixed Chern-Simons levels between the $U(1)_{x_i}$ and  $U(1)_{x_j}$ are shifted as follows
\begin{align}\label{kirbymove}
	k_{ij} x_i x_i  \rightarrow \le( k_{ij} -\frac{p_i p_j}{k} \r)x_i x_j .
\end{align}
As mentioned earlier, we use an orange circle to denote the pure gauge node that we integrate out, or equivalently that is introduced by Kirby moves
\begin{align}
	\begin{split}
		\begin{tikzpicture}
			\draw[fill] (0,0) node[below]{$x$} circle(2pt)	;
			\draw[fill] (1,0) circle(2pt)	;
			\draw[fill] (1,0.5) circle(2pt)	;
			\draw[fill] (1,-0.5) circle(2pt)	;
			\draw[orange,thick] (0,0) circle(10pt)	;
			\draw[thick] (-1,0.5)--(0,0)--(-1,-0.5);
			\draw[thick](0,0)--(1,0.5)--(2,0.5)  ;
			\draw[thick](0,0)--(1,0)--(2,0)  ;
			\draw[thick](0,0)--(1,-0.5)--(2,-0.5)  ;
		\end{tikzpicture} 
	\end{split}
\end{align}
One can check that all Kirby moves mentioned in \eqref{kirbyknown} can be interpreted as integrating out pure gauge nodes.


Note that for a gauge node corresponding to $U(1)_0$, i.e. for $k=0$, the relation \eqref{kirbymove} does not hold, and instead a delta function shows up, 
as indicated in \eqref{integform}. This delta function introduces a constraint,  which further reduces the number of gauge nodes.  For example, integrating $U(1)_0$ corresponding to a gauge node connected to two other gauge nodes that represent $U(1)_{x_1}$ and $U(1)_{x_2}$, one gets
\begin{align}
	\delta( x_1- x_2	) =\int d x\, e^{ \pm 2 \pi i \, ( x_1 -x_2) x} ,
\end{align}
which we denote as
\begin{align}\label{deltasplit}
	\bsp
	\begin{tikzpicture}
		\draw[thick ] (-1.5,0)--(1,0)--(3.5,0) ;
		\node[below] at (0,0 ){	\small	$\mp1$};
		\node[below] at (2,0 ){	\small	$\pm1$};
		\filldraw (-1,0)   circle(2pt) node[above]{$x_1$} node[below]{$k_1$}; 
		\filldraw (1,0)   circle(2pt)node[above]{$x$} node[below]{$0$};
		\filldraw (3,0) circle(2pt) node[above]{$x_2$}node[below]{$k_2$}; 
		\node at (5,0) {$\longrightarrow$} ;
		\draw[thick] (6.5,0)--(7.5,0)  ;
		\filldraw (7,0) circle(2pt) node[above]{$x_{1\,\text{or} \,2}$} node[below]{$k_1+k_2$}; 
	\end{tikzpicture}
	\esp
\end{align}
and which reproduces the third Kirby move in \eqref{kirbyknown}.



For future reference, let us summarize a number of more general Kirby moves, stating explicitly how they affect linking numbers (equivalently mixed Chern-Simons levels):  
\begin{itemize}
	\item
The simplest case is integrating out a gauge node connected to only one other gauge node, which is also the operation in the first line of \eqref{kirbyknown}:
\begin{align}
	\begin{tikzpicture}
		\draw [thick]  (-0.5,0.5)--(0,0) node[below=0.2]{$a$}--(-0.5,-0.5); 
		\filldraw (0,0) circle(2pt);
		\filldraw (3,0) circle(2pt);
		\filldraw (2,0) circle(2pt);
		\node at (0.9,0)  {$\rightleftarrows$} ;
		\draw [thick]  (1.5,0.5)--(2,0) node[below=0.2]{~$a\pm1$}--(1.5,-0.5); 
		\draw[thick](2,0)--(3,0)node[midway, above]{\small $+1$} node[below=0.2]{$\pm1$}  ;
	\end{tikzpicture}
\end{align}
In this process linking numbers change as follows
\begin{align}
	\begin{bmatrix}
		~	a~
	\end{bmatrix} 
	~~ \rightleftarrows ~~
	\begin{bmatrix}
		a \pm 1   & 1 \\
		1 & \pm 1
	\end{bmatrix}  \,.
\end{align}
Moreover, for arbitrary Chern-Simons levels, from \eqref{kirbymove} we find the following Kirby move
\begin{align}
	\bsp
	\begin{tikzpicture}
		\filldraw (3,0) circle(2pt);
		\filldraw (2,0) circle(2pt);
		\draw [thick]  (3+0.5,0.5)--(3,0) node[below=0.1]{$k_1$}--(3+0.5,-0.5); 
		\draw[thick](3,0)--(2,0) node[below=0.1]{$k$} node[midway, above]{$a$} ;
				\node at (4.5,0)  {$\rightleftarrows$} ;
				\draw [thick]  (5+1,0.5)--(5.5,0) node[right=0.2]{\small $k_1-\frac{a^2}{k}$\qquad}--(5+1,-0.5); 
		\filldraw (5.5,0) circle(2pt);
	\end{tikzpicture}
\esp
\end{align}

	\item Integrating out a pure gauge node connected to two other gauge nodes leads to
	\begin{align}\label{example1integ}
		\bsp
		\begin{tikzpicture}[scale=0.8]
			\draw[fill] (0,0) node[left]{$\mp1$} circle(2pt)	;
			\draw[fill] (1,-1) node[right]{$k_2$} circle(2pt)	;
			\draw[fill] (1,1) node[right]{$k_1$} circle(2pt)	;
			\draw[thick](1,-1)--(0,0)--(1,1)  ;
			\node at (2.5,0) {$\rightleftarrows$};
			\draw[fill] (4,-1) node[right]{$k_2 \pm 1$} circle(2pt)	;
			\draw[fill] (4,1) node[right]{$k_1 \pm 1$} circle(2pt)	;
			\node[right]at (4,0) {$ \pm 1$} ;
			\draw[thick](4,-1)--(4,1)  ;
		\end{tikzpicture} 
		\esp
	\end{align}
	which is the fourth operation in \eqref{kirbyknown}.
	For arbitrary Chern-Simons levels we find	
	\begin{align}
		\bsp
		\begin{tikzpicture}[scale=1]
			\draw[fill] (0,0) node[left]{$k$} circle(2pt)	;
			\draw[fill] (1,-1) node[right]{$k_2$} circle(2pt)	;
			\draw[fill] (1,1) node[right]{$k_1$} circle(2pt)	;
			\draw[thick](1,-1)--(0,0)node[midway, below]{\small $b$}--(1,1)node[midway, above]{\small $a$}  ;
			\node at (2.5,0) {$\rightleftarrows$};
			\draw[fill] (4,-1) node[right]{$k_2 -\frac{b^2}{k}$} circle(2pt)	;
			\draw[fill] (4,1) node[right]{$k_1-  \frac{a^2}{k}$} circle(2pt)	;
			\node[right]at (4,0) {\small $ - \frac{ab}{k}$} ;
			\draw[thick](4,-1)--(4,1)  ;
		\end{tikzpicture} 
		\esp
	\end{align}

	\item  Integrating a pure gauge node connected to three other gauge nodes leads to
	\begin{align}
		\bsp
		\begin{tikzpicture}[scale=0.8]
			\draw[fill] (0,0) node[left]{$\mp1$} circle(2pt)	;
			\draw[fill] (1,-1) node[right]{$k_3$} circle(2pt)	;
			\draw[fill] (1,1) node[right]{$k_1$} circle(2pt)	;
			\draw[fill] (1,0) node[right]{$k_2$} circle(2pt)	;
			\draw[thick](1,-1)--(0,0)--(1,1)  ;	\draw[thick](0,0)--(1,0)  ;
			\node at (2.5,0) {~~$\rightleftarrows$};
			\draw[fill] (4,-1) node[below]{$k_3 \pm 1$} circle(2pt)	;
			\draw[fill] (5.7,0) node[right]{$k_2 \pm 1$} circle(2pt)	;
			\draw[fill] (4,1) node[above]{$k_1 \pm 1$} circle(2pt)	;
			\draw[thick](4,-1)--(4,1)node[midway, left]{$ \pm 1$} --(5.7,0)node[midway, above]{~$ \pm 1$}--(4,-1)node[midway, below]{$ \pm 1$}  ;
		\end{tikzpicture} \,,
	\esp
	\end{align}
	For arbitrary Chern-Simons levels we find
	\begin{align}
		\bsp
		\begin{tikzpicture}[scale=1]
			\draw[fill] (0,0) node[left]{$k$} circle(2pt)	;
			\draw[fill] (1,-1) node[right]{$k_3$} circle(2pt)	;
			\draw[fill] (1,1) node[right]{$k_1$} circle(2pt)	;
			\draw[fill] (1,0) node[right]{$k_2$} circle(2pt)	;
			\draw[thick](1,-1)--(0,0)node[midway, below]{\small$c$}--(1,1)node[midway, above]{\small$a$} ;	\draw[thick](0,0)--(1,0)node[midway,above=-0.1]{\small~~$b$}  ;
			\node at (2.5,0) {~~$\rightleftarrows$};
			\draw[fill] (4,-1) node[below]{$k_3 -\frac{c^2}{k}$} circle(2pt)	;
			\draw[fill] (5.7,0) node[right]{$k_2 -\frac{b^2}{k}$} circle(2pt)	;
			\draw[fill] (4,1) node[above]{$k_1 -\frac{a^2}{k} $} circle(2pt)	;
			\draw[thick](4,-1)--(4,1)node[midway, left]{$ \small -\frac{ac}{k} $} --(5.7,0)node[midway, above=-0.1]{~$  \small -\frac{ab}{k} $}--(4,-1)node[midway, below=-0.1]{$ \small -\frac{bc}{k} $}  ;
		\end{tikzpicture} \,,
	\esp
	\end{align}

	\item Kirby move involving a node connected to four other nodes leads to a tetrahedron with edges labeled by $\pm1$
	\begin{align}
		\bsp
		\begin{tikzpicture}[scale=0.8]
			\draw[fill] (0,0) node[left]{$\mp1$} circle(2pt)	;
			\draw[fill] (1,-1+0.5) node[right]{$k_3$} circle(2pt)	;
			\draw[fill] (1,1.5) node[right]{$k_1$} circle(2pt)	;
			\draw[fill] (1,0.5) node[right]{$k_2$} circle(2pt)	;
			\draw[fill] (1,-2+0.5) node[right]{$k_4$} circle(2pt)	;
			\draw[thick](0,0)--(1,-2+0.5)  ;
			\draw[thick](1,-0.5)--(0,0)--(1,1.5)  ;	\draw[thick](0,0)--(1,0.5)  ;
			\node at (2.5,0) {~~$\rightleftarrows$};
			\draw[fill] (4,-1) node[below]{$k_4 \pm 1$} circle(2pt)	;
			\draw[fill] (6,1) node[above]{$k_2 \pm 1$} circle(2pt)	;
			\draw[fill] (6,-1) node[below]{$k_3 \pm 1$} circle(2pt)	;
			\draw[fill] (4,1) node[above]{$k_1 \pm 1$} circle(2pt)	;
			\draw[thick](4,-1)--(4,1)--(6,-1)--(6,1)--(5.1,0.1)  
			(4.9,-0.1)--(4,-1);
			\draw[thick](4,-1)--(6,-1) ;
			\draw[thick](4,1)--(6,1) ;
		\end{tikzpicture} 
	\esp
	\end{align}
	\item Configurations with more gauge nodes $U(1)_x\times U(1)_{x_1} \times U(1)_{x_2}\times \cdots \times U(1)_{x_n}$, after integrating out $U(1)_x$,  result in more general polyhedra.
\end{itemize}


Finally, let us present an operation, which resembles an operation of framing in knot knot.  In the context of knots-quivers correspondence, changing framing of a knot by $f$ shifts all entries of a matrix that encodes a quiver corresponding to this knot by $f$.  In our context, there are Kirby moves which have analogous effect, i.e. they shift all mixed Chern-Simons levels by $f$.  One way to get such a result is to add a gauge node $U(1)_{-1/f}$ with \CS level $-1/f$,  and to connect it to all other gauge nodes $i$ by mixed \CS levels
\begin{align}\label{framngcs}
	k_{0\, i} =+1 \,,\quad \forall \, i = 1, 2 ,\ldots 
\end{align} 
Then the Kirby move takes form
\begin{align}
	\bsp
	\begin{tikzpicture}[scale=0.8]
		\draw[fill] (0,0) node[left]{$\frac{1}{f}~$} circle(2pt)	;
		\draw[fill] (1,-1) node[right]{$k_3$} circle(2pt)	;
		\draw[fill] (1,1) node[right]{$k_1$} circle(2pt)	;
		\draw[fill] (1,0) node[right]{$k_2$} circle(2pt)	;
		\draw[thick](1,-1)--(0,0)node[midway, below]{\tiny{+1}}--(1,1) node[midway, above]{\tiny{+1}}  ;	\draw[thick](0,0)--(1,0)node[midway, below]{\tiny{+1}} ;
		\node at (3,0) {$\longrightarrow$};
		\draw[fill] (5,-1) node[below]{$k_3 + f$} circle(2pt)	;
		\draw[fill] (6.7,0) node[right]{$k_2 + f$} circle(2pt)	;
		\draw[fill] (5,1) node[above]{$k_1 + f$} circle(2pt)	;
		\draw[thick](5,-1)--(5,1)node[midway, left]{$ f$} --(6.7,0)node[midway, above=-0.1]{$~~  f$}--(5,-1)node[midway, below=-0.1]{$~~f$}  ;
	\end{tikzpicture} 
\esp
\end{align}
and, after integrating out the framing gauge node, mixed \CS levels become
\begin{align}\label{framingkij}
	k_{ij}^{\eff}~~ \rightarrow~~ 	k_{ij}^{\eff}  + f \,, \quad \forall ~i\,,j =1, 2 ,\ldots
\end{align}

The same result follows once we introduce $f$ new gauge nodes $U(1)_{-1}$ and connect each of them to all original gauge nodes, via mixed \CS levels $+1$.  Integrating out these new framing nodes one by one also leads to \eqref{framingkij}. 

Yes another way to shift all \CS levels by $f$ is to add two gauge nodes $U(1)_f$ and  $U(1)_0$ and then to integrate them out:
\begin{align}
	\bsp
	\begin{tikzpicture}[scale=0.8]
		\draw[fill] (-1,0) node[below]{$f~$	} circle(2pt)	;
		\draw[thick] (-1,0)--(0,0);
		\draw[fill] (0,0) node[below]{$0~$} circle(2pt)	;
		\draw[fill] (1,-1) node[right]{$k_3$} circle(2pt)	;
		\draw[fill] (1,1) node[right]{$k_1$} circle(2pt)	;
		\draw[fill] (1,0) node[right]{$k_2$} circle(2pt)	;
		\draw[thick](1,-1)--(0,0)--(1,1)  ;	\draw[thick](0,0)--(1,0)  ;
		\node at (3,0) {$\longrightarrow$};
		\draw[fill] (5,-1) node[below]{$k_3 + f$} circle(2pt)	;
		\draw[fill] (6.7,0) node[right]{$k_2 + f$} circle(2pt)	;
		\draw[fill] (5,1) node[above]{$k_1 + f$} circle(2pt)	;
		\draw[thick](5,-1)--(5,1)node[midway, left]{$ f$} --(6.7,0)node[midway, above=-0.1]{$~~  f$}--(5,-1)node[midway, below=-0.1]{$~~f$}  ;
	\end{tikzpicture} 
\esp
\end{align}
In particular,  for $f=0$,  these two extra nodes simply decouple.


\subsection{Lens spaces} 

Let us discuss in more detail lens spaces $L(p,q)$, which are interesting examples of plumbing manifolds, see e.g. \cite{Jejjala:2022lrm}.  For $q=1$,  they can be defined as quotients $L(p,1)  = S^3/{\mathbf{Z}_p}$,  and can be also represented as fiber bundles with circle fibers
\[\mathcal{O}(-p) \rightarrow \mathbf{P}^1 = L(p,1) =S^2_p \times S^1 \,.\]
A plumbing graph of $L(p,1)$ lens space consists of a single node, which represents one-cycle with a self-linking number  $-p$
\begin{align}
	\bsp
	\btik[scale=0.8]
	\draw[ thick,->] (0.5,0) arc (0:360:0.5) ;
	\node at(0,-0.8) {$-p$} ;
	\etik
	\esp
\end{align}
Dehn surgeries of lens spaces give other plumbing manifolds, e.g.  $L(p,q) = S^3_{-p/q}(\text{unknot})$.

Wrapping an M5-brane on $L(p,1)$ lens space gives rise to a pure 3d $\N=2$ theory with gauge group $U(1)_p$ and a decoupled chiral multiplet in the adjoint representation \cite{Gukov:2015aa}. Wrapping an M5-brane on $L(p,q)$ engineers a theory with gauge group $U(1)_k$ with \CS level $k =-p/q$ and an adjoint chiral multiplet, which also decouples, so that its plumbing graph consists only of a single node.  Using Kirby moves,  such plumbing graph can be transformed into a linear graph
\begin{align}\bsp
	\begin{tikzpicture}
		\filldraw (-2,0) node [below] { ${p}/{q}$ }circle (2pt); 
		\node at (-1,0) {$=$};
		\coordinate (A)at (0,0);
		\coordinate (B)at (1,0);
		\coordinate (C)at (2,0);
		\coordinate (D)at (3.7,0);
		\coordinate (E)at (5,0);
		\filldraw (A) node[below]{$a_1$}  circle(2pt);
		\filldraw (B) node[below]{$a_2$}  circle(2pt);
		\filldraw (C) node[below]{$a_3$}  circle(2pt);
		\filldraw (E) node[below]{$a_n$}  circle(2pt);
		\node 	at (D) {$\cdots\cdots$};
		\draw[thick] (0,0)--(3,0)  ;
		\draw[thick] (4.5,0)--(5,0)  ;
		;
	\end{tikzpicture}\esp
\end{align}
where the linking number for each line is $+1$,  and $a_i$ are given by the continuous fraction decomposition
\begin{align}\label{lensfiltration}
	\frac{p}{q} = a_1 -\frac{1}{  
		a_2- \frac{1 }{
			{  a_3 -\frac{ 1}{\dots\,-
					\frac{1}{a_n}
				}
			}  
	}  } \,.
\end{align}
The 3d gauge theory corresponding to this graph is $U(1)_{a_1} \times U(1)_{a_2} \times \cdots \times U(1)_{a_n}$,  and integrating out gauge nodes from left to right also reproduces the relation (\ref{lensfiltration}). In particular, for $L(p,1)$ and linking numbers $\pm1$,  we get $a_1= p\pm 1\,, a_2 =\pm 2\,,a_3 =\pm 2\,, \cdots, \,, a_n=\pm1$, where $n$ can be any positive number. 

Furthermore, for $p/q$ and $a_i$ related as in (\ref{lensfiltration}),  there is also an equivalence
\begin{align}\bsp
	\begin{tikzpicture}
		\filldraw (-2,0) node [below] { $0$ }circle (2pt); 
		\node at (-1,0) {$=$};
		\coordinate (A )at (0,0);
		\coordinate (B)at (1,0);
		\coordinate (C)at (2,0);
		\coordinate (D)at (3.7,0);
		\coordinate (E)at (5,0);
		\coordinate (F)at (6,0);
		\filldraw (A) node[below]{$a_n$}  circle(2pt);
		\filldraw (B) node[below]{$a_{n-1}$}  circle(2pt);
		\filldraw (C) node[below]{$a_{n-2}$}  circle(2pt);
		\filldraw (E) node[below]{$a_1$}  circle(2pt);
		\filldraw (F) node[below]{$q/p$}  circle(2pt);
		\node 	at (D) {$\cdots\cdots$};
		\draw[thick] (0,0)--(3,0)  ;
		\draw[thick] (4.5,0)--(6,0)  ;
	\end{tikzpicture}  \esp
\end{align}	
There are also other equivalences, such as $L(p,q) = -L(p,p-q)$ and $L(p,-q) = -L(p,q)$, which can also be checked using Kirby moves.

A nice property of linear plumbing graphs is that they can be represented using $SL(2,\mathbf{Z})$ group, which in the standard notation is generated by $S$ and $T$ operators
\begin{align}
	&L(p,1)  \longrightarrow S T^p S  ,	\\
	&L(p,q)  \longrightarrow S T^{a_1} S T^{a_2}S \cdots ST^{a_n}S  .
\end{align}
In this representation each node $U(1)_{k}$ corresponds to the $T^k$ operator, and each line corresponds to the $S$ operator, which can be interpreted as $S$-duality domain wall theory that glues building blocks (gauge nodes)  \cite{Gaiotto:2008ak}.  The parameters $a_i$ are also determined by \eqref{lensfiltration}. More details on this group are presented in appendix \ref{SL2Zappendex}.


\section{Adding chiral multiplets and gauging global symmetries}   \label{secgauging}

The main goal of this paper is to generalize abelian 3d $\mathcal{N}=2$ theories corresponding to plumbing three-manifolds, which we summarized in the previous section, by including in their content chiral multiplets.  In what follows we denote chiral multiplets in fundamental and anti-fundamental representation respectively by $\F$ and $\AF$.  We represent the contents of such 3d $\mathcal{N}=2$ theories in graphs that generalize plumbing graphs, which have extra gray boxes ${\color{gray}\blacksquare}$ that denote fundamental chiral multiplets, and blue boxes ${\color{blue}\blacksquare}$ that denote anti-fundamental chiral multiplets. We refer to such nodes as matter nodes.  In addition,  numbers assigned to lines between gauge nodes and matter nodes denote charges of matter multiplets.  We stress that, at this stage, such graphs represent the contents of gauge theories, however it is unclear if they have some interpretation in terms of three-manifolds.  Nonetheless, various operations and dualities between these theories can still be encoded as operations on such graphs, as we show in detail in what follows.

As a warm up example, and to illustrate our notation, consider adding a chiral multiplet to the 3d $\mathcal{N}=2$ theory corresponding to a lens space.  We can integrate one by one (from right to left in the graph below) gauge fields corresponding to gauge nodes that do not interact with the chiral multiplet, which therefore yields the following equivalence of theories

\begin{align}
	\bsp
	\begin{tikzpicture}
		\filldraw (-2,0) node [below] { ${p}/{q}$ }circle (2pt); 
		\draw[thick](-2,0)--(-2,0.8) node{$\graybox$} ;
		\node at (-1,0) {$=$};
		\coordinate (A )at (0,0);
		\coordinate (B)at (1,0);
		\coordinate (C)at (2,0);
		\coordinate (D)at (3.7,0);
		\coordinate (E)at (5,0);
		\filldraw (A) node[below]{$a_1$}  circle(2pt);
		\filldraw (B) node[below]{$a_2$}  circle(2pt);
		\filldraw (C) node[below]{$a_3$}  circle(2pt);
		\filldraw (E) node[below]{$a_n$}  circle(2pt);
		\node 	at (D) {$\cdots\cdots$};
		\draw[thick] (0,0)--(3,0)  ;
		\draw[thick] (4.5,0)--(5,0)  ;
		\draw[thick](0,0)--(0,0.8) node{$\graybox$} ;
		;
	\end{tikzpicture}
	\esp
\end{align}

More generally,  operations on graphs that we discuss include generalization of Kirby moves to matter nodes.  Gauging of global symmetries is crucial to introduce such moves.  In abelian 3d $\N=2$ theories,  global symmetries include the topological symmetry for each gauge node and the flavor symmetry for each chiral multiplet. These two types of global symmetries are associated with FI parameters $\xi_i$ and mass parameters $m_j$ respectively.  For generic abelian theories, the global symmetries are thus
$\prod_i U(1)_{\xi_i} ~\times~ \prod_j U(1)_{m_j}$.  We discuss their gauging in what follows, in particular in the context of mirror triality, which we present first.

\subsection{Mirror triality}

There is an interesting action of $SL(2,\mathbf{Z})$ group on 3d $\mathcal{N}=2$ theories. Its generators, denoted as usual by $S$ and $T$, act non-trivially on the theory of a fundamental chiral multiplet $\T_\Delta$, which in the 3d-3d relation corresponds to a single tetrahedron $\Delta$ \cite{Dimofte:2011ju}.  These generators satisfy $(ST)^3=1$,  which has a manifestation in 3d $\mathcal{N}=2$ theories as a mirror triality, which relates the theory of a free chiral multiplet and $U(1)_{\pm 1/2}+1 \F$  \cite{Intriligator:1996ex,Kapustin:1999ha,Dimofte:2011ju} 
\begin{align}
	\label{triviliaty}
	\begin{split}
		\includegraphics[]{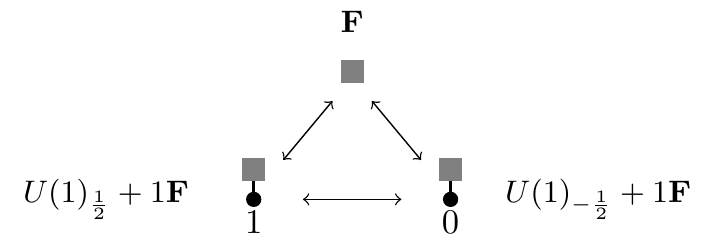}
	\end{split}
\end{align}
On the three-manifold side, the $ST$ transformation has the effect of changing the polarization of each $\Delta$. 

Let us discuss three components of the mirror triality, corresponding to the edges of the above triangle; for details see also \cite{Dimofte:2011ju}. The two theories $U(1)_{\pm 1/2}+1\F$  are dual to each other by an exchange symmetry $\q \leftrightarrow \q^{-1}$, which we discuss in more detail in section \ref{exchangesec}. The remaining two edges of the triangle correspond to two mirror pairs, which may be interpreted as corresponding respectively to $ST$ and $(ST)^2$ operations, and which relate a free chiral multiplet and a $U(1)_{\pm1/2}$ theory with a chiral multiplet:
\begin{align}
	&1 ~\F_{-1/2} ~\xlongleftrightarrow{~ST~} ~~U(1)_{1/2}+ 1~\F  \,, \\ 
	& 1 ~\F_{1/2} ~~~\xlongleftrightarrow{(ST)^2}~~ U(1)_{-1/2}+ 1~\F  \,,
	\end{align}
where the subscripts of $\F$ are the \CS levels for the background vector fields (which usually do not need to be denoted explicitly).

A theory of a fundamental chiral multiplet $\T_\Delta$, which in 3d-3d correspondence is assigned to a tetrahedron $\Delta$,  gives the contribution $s_b( i Q/2 -m)$ to sphere partition functions, where the double sine function is defined in (\ref{s_b}) and $m$ is the mass parameter.  Free chiral multiplet $\F$ with the background \CS level $k=-\frac{1}{2}$ contributes
\begin{align}
	1 ~\F_{-1/2} \rightarrow e^{-}_b\left( \frac{i Q}{2} -m  \right) := e^{ \frac{ i \pi}{ 2} \left( \frac{i Q}{2} -m   \right)^2 } s_b\left( \frac{i Q}{2} -m   \right) .
	\end{align}
Similarly,  free chiral multiplet $\F$ with a background \CS level $k=\frac{1}{2}$ contributes 
\begin{align}
	1 ~\F_{1/2} \rightarrow 
	{e}^{+}_b\left( \frac{i Q}{2} -m  \right) := e^{ -\frac{ i \pi}{ 2} \left( \frac{i Q}{2} -m   \right)^2 } s_b\left( \frac{i Q}{2} -m   \right) .
\end{align}
Partition functions of the mirror pair $1 ~\F_{-1/2} ~\xlongleftrightarrow{~ST~} ~~U(1)_{1/2}+ 1~\F$ are related by a Fourier transformation 
 \begin{align}
 	\label{gaugeF}
 	\boxed{
 		e^{-}_b \le(  \frac{iQ}{2}-z \r) =
\int dx \, e^{ - 2 \pi i \, x z   }
 e^{+}_b\le( \frac{i Q}{2} -x \r) 
   }	
 	\end{align}
where we ignore a factor $e^{ -\frac{\pi i\, Q^2 }{4} }$.  We see that the mass parameter $z$ of the free chiral multiplet is dual to the FI parameter of the theory $U(1)_{1/2} +1 \F$, which means that this mirror duality exchanges topological symmetry and flavor symmetry, $ U(1)_{F} \leftrightarrow U(1)_{T}$.

We can also understand this mirror pair reversely. Gauging the mass parameter $x$ of a matter $\F$ gives rise to a theory $U(1)_{1/2}+1\F$, and this theory is equivalent to another free matter $\F$. Therefore we have $ U(1)_{T} \leftrightarrow U(1)_{F}$. This perspective tells us that the free matter as the building block $\T_\Delta$ of 3d-3d construction is equivalent to its gauged theory, which we denote by
 $\T^{-1/2}_{\Delta} \simeq \T^{1/2}_{\Delta}/U(1)_F$,  or equivalently $1\F \simeq U(1)_{1/2}+1\F$.  Besides, the free chiral multiplet only has a flavor symmetry, and after mirroring, the matter in the dual theory $U(1)_{1/2}+1\F$ becomes massless, as we have gauged the flavor symmetry and no free parameter is left.

The second dual pair $1 \F_{1/2} \longleftrightarrow U(1)_{-1/2}+ 1\F $ is just the reverse Fourier transformation of \eqref{gaugeF}:
\begin{align}	\label{gauge2F}
	\boxed{
	e^{+}_b\le( \frac{i Q}{2} -x \r) =\int dz \, e^{  2 \pi i \, x z   } e^{-}_b \le(  \frac{iQ}{2}-z \r)  }
	\end{align}
Therefore, there is another mirror pair that we also denote by $\T^{1/2}_{\Delta} \simeq \T^{-1/2}_{\Delta}/U(1)'_F$.
The Fourier transformation is a $\mathbf{Z}_2$ group, hence applying both transformations \eqref{gaugeF} and \eqref{gauge2F} leads to the original theory, however up to an additional integral  which is a delta function that identifies $U(1)_F$ and $U(1)'_F$, namely
\begin{align}
	\T^{1/2}_{\Delta} \simeq \T^{-1/2}_{\Delta}/U(1)'_F
	\simeq \le(\T_{\Delta}^{1/2}/U(1)_F \r)/U(1)'_F  \simeq 	\T^{1/2}_{\Delta}  \,.
	\end{align} 


\subsection{
$ST$ transformations for matter nodes and gauging the mirror triality}\label{gaugetri}


As we have just discussed, there is a global symmetry left for the theory $\T_\Delta/U(1)_F$, which is the topological symmetry $U(1)_T$ for the gauge nodes.  In this section we show that gauging this global symmetry leads to a new dual pair, which can be presented graphically as follows
\begin{align}\label{blowupdownmatter}
	\begin{split}
		\begin{tikzpicture}
			\filldraw (0,0) node[below]{$k$}  circle(2pt);
			\draw[thick](0,0)--(0,1);
			\node[fill=gray] (r) at (0,1) {} ;
			\node at (1, -0.1) [below]{};
			\node at (1.5,0.2){		$~~\xrightarrow{(ST)^{\pm 1}}$};
			\filldraw (3,0) node[below]{$k \pm\frac{1}{2}$}circle(2pt);
			\draw[thick](3,0)--(4,0)node[midway,above]{$\pm1$}--(4,1);
			\filldraw (4,0) node[below]{$\pm \frac{1}{2}$}  circle(2pt);
			\node[fill=gray] (r) at (4,1) {} ;
		\end{tikzpicture}	
	\end{split}
\end{align}
We  call this operation the $ST$-move or $ST$ transformation for matter nodes -- it can be regarded as an analog of a Kirby move for gauge nodes.  The move in \eqref{blowupdownmatter} involves a matter node charged under only one gauge node, however in section \ref{genericmatter} we will discuss the $ST$ moves for chiral multiplets charged under many gauge nodes. Decoupling of matter nodes discussed in section \ref{decouplematter} also supports this interpretation and shows a close relation between $ST$ moves and Kirby moves.



\subsubsection*{Gauging one $ST$ dual pair}

To start with, we reconsider the dual pair 
$$
1 \F~~ \longleftrightarrow ~~U(1)_{1/2}+ 1\F
$$ 
which we interpret and refer to as the $ST$-move and represent graphically as
	\begin{align}\label{mirrorpairgraph1}
	\begin{split}
		\begin{tikzpicture}
			\coordinate (A) at (4,1);
			\node at (2.5,0){	$\xlongleftrightarrow{ST}$};
			\draw[thick](4,0)--(A);
			\filldraw (4,0) node[below]{$+\frac{1}{2}$}  circle(2pt);
			\node at (4,0 )[right=0.1]{$x$};
			\node[fill=gray] (r) at (4,1) {} ;
			\node at (A) [right=0.1]{$ $} ;
				\node[fill=gray] (r) at (1,0) {} ;
		\end{tikzpicture}	
	\end{split}
\end{align}
It is convenient to rewrite the Fourier transform relation between the corresponding partition functions \eqref{gaugeF} in terms of double sine functions $s_b(\cdot)$ rather than $e_b(\cdot)$, which makes \CS levels manifest
\begin{align}\label{mirrortransf11}
ST\,:\quad  	\boxed{
	s_b \le(  \frac{i Q}{2} - z \r) =  \,e^{ - \frac{\pi i}{ 2 } \le(\fQ -  z \r)^2  } \, \cdot \int dx \, 
	e^{- \frac{\pi i}{2}  x^2 - 2 \pi i \,  zx - \frac{\pi Q }{2}  x    } s_b \le(  \fQ -  x \r) } 
\end{align} 
Here $x$ is the variable for the gauge node, while $z$ is the mass parameter of the flavor symmetry $U(1)_F$ for the free chiral multiplet $\F$, and also the FI parameter of the topological symmetry $U(1)_T$ for the gauge node of the theory $U(1)_{1/2}+1\F$.  Let us gauge now the global symmetry associated with variable $z$, or in other words gauge global symmetry pair
$$
	U(1)_F ~~\xlongleftrightarrow{ST}~~ U(1)_T
$$
This gauging is equivalent to adding a contour integral on both sides of the equivalence \eqref{mirrortransf11}
\begin{align}\label{gaugingST}
	\int \, d z \, e^{ - \pi i \, k z^2 + 2 \pi i \, \xi_z z }  LHS \, = \,		\int \, d z \, e^{ - \pi i \, k z^2 + 2 \pi i \, \xi_z z }  RHS  \,,
	\end{align}
where $k$ is the \CS level, $\xi_z$ is the introduced FI parameter, and $LHS$ and  $RHS$ denote other terms on both sides of the equation. We can  turn on these parameters, because after gauging the global symmetry becomes a new gauge symmetry $U(1)$.  After the gauging we obtain a new dual pair
\begin{align}
	U(1)_k+1\F ~\longleftrightarrow~	U(1)_{k+1/2} \times \big(U(1)_{1/2}+1\F\big)
	\end{align}
which can be represented graphically as follows
 	\begin{align}
	\begin{split}
		\begin{tikzpicture}
			\coordinate (A) at (4,1);
			\filldraw (0,0) node[below]{$k$}  circle(2pt);
			\draw[thick](0,0)--(0,1);
			\node at (0,0) [left]{$z$};
			\node[fill=gray]  at (0,1) {} ;
			\node at (0, 1) [right=0.1]{$ $};
			\node at (1.5,0){	$\longleftrightarrow$};
			\filldraw (3,0) node[below]{$k +\frac{1}{2} $}node[left]{$z$}  circle(2pt);
			\draw[thick](3,0)--(4,0)--(A);
			\filldraw (4,0) node[below]{$\frac{1}{2}$}  circle(2pt);
			\node at (4,0 )[right=0.1]{$x$};
			\node at (3.5,0 )[above]{\small $+1$};
			\node[fill=gray] (r) at (A) {} ;
			\node at (A) [right=0.1]{$ $} ;
		\end{tikzpicture}	
	\end{split}
\end{align}
In this process,  the mixed \CS levels are generated\footnote{Here we use the fact that 
in the limit $b \rightarrow 0$,
$
	s_b(z) \rightarrow e^{- \frac{\pi i}{2 } z^2  } \exp \le(   \frac{ \text{Li}_2 ( e^{ 2\pi b z })  }{ \hbar  }\r)  
$.}	
 \begin{align}
	k^{\eff}_{ij}:~~	\begin{bmatrix}
		k + \frac{1}{2} 
	\end{bmatrix} 
	~~ \rightarrow ~~
	\begin{bmatrix}
		k +\frac{1}{2}    &
		~ 1 \\
		1	 & { 1 }
	\end{bmatrix} \,.	
\end{align}

We can also consider a more general case of a chiral multiplet carrying other charges
 \begin{align}\label{mirrortransf1}
 	s_b \le(  \frac{i Q}{2} -q z \r) =  e^{ - \frac{\pi i}{ 2 } \le(\fQ - q z \r)^2  } \, \cdot \int dy \, 
 	e^{- \frac{\pi i}{2}  y^2 - 2 \pi i \, q yz - \frac{\pi Q }{2} y    } s_b \le(  \fQ -  y \r)  .
 	\end{align} 
Gauging the global symmetry in this case leads to the dual pair
 	\begin{align}\label{chargeqST}
 	\begin{split}
 		\begin{tikzpicture}
 			\coordinate (A) at (4,1);
 			\filldraw (0,0) node[below]{\small	$k$}  circle(2pt);
 			\draw[thick](0,0)--(0,1)node[midway,right	]{\small$q$};
 			\node[fill=gray]  at (0,1) {} ;
 			\node at (1.5,0){	$\longrightarrow$};
 			\filldraw (3,0) node[below]{\small$k +\frac{q^2}{2} $}circle(2pt);
 			\draw[thick](3,0)--(4,0)--(A)node[midway, right]{\small$+1$};
 			\filldraw (4,0) node[below]{\small$\frac{1}{2}$}  circle(2pt);
 			\node at (3.5,0 )[above]{\small	$q$};
 			\node[fill=gray] (r) at (A) {} ;
 		\end{tikzpicture}	
 	\end{split}
 \end{align}
where the charge of the chiral multiplet is $q$, which is explicitly denoted in the graphs.  In this case the effective mixed \CS levels change as follows \footnote{We thank Lakshya Bhardwaj for a valuable discussion about this example.}
 \begin{align}
 	k^{\eff}_{ij}:~~	\begin{bmatrix}
 		k + \frac{q^2}{2} 
 	\end{bmatrix} 
~~ \rightarrow ~~
 	\begin{bmatrix}
 		 k +\frac{q^2}{2}    &
 		 ~ q \\
 		 q & { 1 }
 		\end{bmatrix} \,,
 	\end{align}
 which is obtained by taking into account the matter contribution, since $s_b\le( \fQ-q z\r)$ contributes $ \frac{q^2}{2}$ to effective \CS levels.
 We emphasize that the charge $q$ matter is turned into a $\F$, as shown in \eqref{chargeqST}.  In addition, if we ungauge the gauge node $U(1)_k$, then it becomes a flavor symmetry, and this dual pair returns to the original graph pair in \eqref{mirrorpairgraph1}.


\subsubsection*{Gauging the $(ST)^2$ dual pair}

Performing the Fourier transformation twice results in $(ST)^2$ transformation \eqref{gauge2F}
$$
1 \F ~~\xlongleftrightarrow{(ST)^2}~~ U(1)_{-1/2}+ 1\F .
$$ 
We rewrite now the relation between the corresponding partition functions as
\begin{align}\label{mirrortransf2}
(ST)^2\,:\quad	\boxed{
	s_b \le(  \frac{i Q}{2} - z \r) =  \,e^{  \frac{\pi i}{ 2 } \le(\fQ -  z \r)^2  } \, \cdot \int dx \, 
	e^{ \frac{\pi i}{2}  x^2 + 2 \pi i \, z x - \frac{\pi Q }{2}  x    } s_b \le(  \fQ - x \r) } 
\end{align} 
Similarly as above,  we can gauge the above mirror dual pair.  Turning on \CS levels and FI parameters on both sides of \eqref{mirrortransf2} and using the same formula as in \eqref{gaugingST} we get a new dual pair
\begin{align}
	U(1)_k+1\F ~\longleftrightarrow~	U(1)_{k-1/2} \times_{} \big(U(1)_{1/2}-1\F\big),
\end{align}
which can be represented graphically as
\begin{align}\label{gaugeSTST}
	\begin{split}
		\begin{tikzpicture}
			\filldraw (0,0) node[below]{$k$}  circle(2pt);
			\draw[thick](0,0)--(0,1);
			\node at (0,0.2) [right]{$z$};
			\node[fill=gray] (r) at (0,1) {} ;
			\node at (1, -0.1) [below]{};
			\node at (1.5,0){	$\longleftrightarrow$};
			\filldraw (3,0) node[below]{$k -\frac{1}{2}$}node[above]{$z$}  circle(2pt);
			\draw[thick](3,0)--(4,0)node[midway, above]{$-1$}--(4,1);
			\filldraw (4,0) node[below]{$-\frac{1}{2}$}  circle(2pt);
			\node at (4,0.2 )[right]{$y$};
			\node[fill=gray] (r) at (4,1) {} ;
		\end{tikzpicture}	
	\end{split}
\end{align}
In this process the effective \CS levels change as follows
\begin{align}
	k^{\eff}_{ij}:~~	\begin{bmatrix}
		k + \frac{1}{2} 
	\end{bmatrix} 
	~~ \longleftrightarrow ~~
	\begin{bmatrix}
		k  -\frac{1}{2}  & -1 \\
		-1 & 0
	\end{bmatrix} \,.
\end{align}

Analogously as above,  more general matter field with charge $q$ contributes $s_b( iQ/2-q z)$ to the partition function, and its $(ST)^2$ gauging leads to
\begin{align}\label{chargeqSTST}
	\begin{split}
		\begin{tikzpicture}
			\coordinate (A) at (4,1);
			\filldraw (0,0) node[below]{\small	$k$}  circle(2pt);
			\draw[thick](0,0)--(0,1)node[midway,right	]{\small$q$};
			\node[fill=gray]  at (0,1) {} ;
			\node at (1.5,0){	$\longrightarrow$};
			\filldraw (3,0) node[below]{\small$k -\frac{q^2}{2} $}circle(2pt);
			\draw[thick](3,0)--(4,0)--(A)node[midway, right]{\small$+1$};
			\filldraw (4,0) node[below]{\small$-\frac{1}{2}$}  circle(2pt);
			\node at (3.5,0 )[above]{\small	$-q$};
			\node[fill=gray] (r) at (A) {} ;
		\end{tikzpicture}	
	\end{split}
\end{align}
with mixed \CS levels changing as follows
 \begin{align}
	k^{\eff}_{ij}:~~	\begin{bmatrix}
		k - \frac{q^2}{2} 
	\end{bmatrix} 
	~~ \rightarrow ~~
	\begin{bmatrix}
		k -\frac{q^2}{2}    &
		~ -q \\
		-q & { 0 }
	\end{bmatrix} \,.
\end{align}


\subsubsection*{$ST$ and $(ST)^2$ transformations differ by Kirby moves}

In view of the relation $(ST)^3=1$,  one would expect that performing $ST$-transformation twice is equivalent to $(ST)^2$ transformation.  It turns out to be true up to a Kirby move, i.e. up to integrating out a gauge node.  We illustrate this phenomenon in some examples.

First, consider the theory $U(1)_k+1\F$, which we already discussed above.  We can perform arbitrary number of $ST$ transformations for the matter field $\F$, which leads to a chain of theories represented by the following graphs 
\begin{align}
	\label{mirrorF}
		\begin{split}
	\includegraphics[]{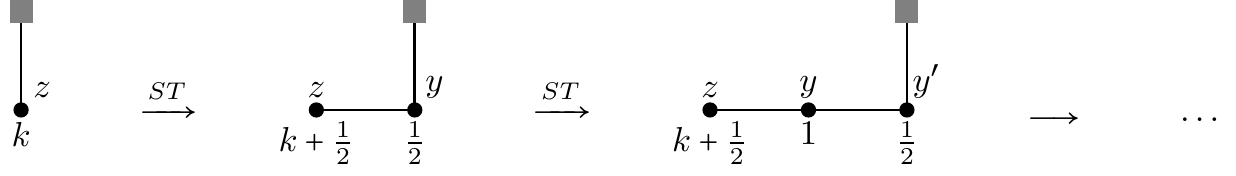}
	\end{split}
	\end{align}
for which the effective mixed \CS levels change as follows
\begin{align}
	k^{\eff}_{ij}:~~	\begin{bmatrix}
		k + \frac{1}{2} 
	\end{bmatrix} 
	~~ \rightarrow ~~
	\begin{bmatrix}
		k +\frac{1}{2}   & 1 \\
		1 & 1
	\end{bmatrix} 
~~ \rightarrow ~~
\begin{bmatrix}
	k  +\frac{1}{2}  & 1  &\\
	1 & 1 & 1\\
	&1&1
\end{bmatrix} ~~ \rightarrow ~~ \cdots 
\end{align}
The third graph in \eqref{mirrorF}, which arises after performing $ST$-move twice, should be equivalent to the graph in \eqref{gaugeSTST} that arises from the action of $(ST)^2$.  In fact, these two graphs differ by one pure gauge node, and integrating out the middle gauge node of the third graph in \eqref{mirrorF} leads to the right graph in \eqref{gaugeSTST}
 \begin{align}	
 	\begin{split}
 		\begin{tikzpicture}
 			\node at (7.5,0){	$\xlongleftrightarrow{}$};
 			\filldraw (9,0) node[below]{$k -\frac{1}{2}$}node[above]{}  circle(2pt);
 			\draw[thick](9,0)--(10,0)node[midway, above]{\small$-1$}--(10,1);
 			\filldraw (10,0) node[below]{$-\frac{1}{2}$}  circle(2pt);
 			\node[fill=gray] (r) at (10,1) {} ;
 		\filldraw (4,0) node[below]{$k +\frac{1}{2}$}node[above]{}  circle(2pt);
 		\draw[thick](4,0)--(5,0)node[midway, above]{\small$+1$}--(6,0)node[midway, above]{\small$+1$}--(6,1);
 		\filldraw (5,0) node[below]{$1$}  circle(2pt);
 			\filldraw (6,0) node[below]{$\frac{1}{2}$}  circle(2pt);
 		\node[fill=gray] (r) at (6,1) {} ;
 						\draw[thick,orange] (5,0)  circle(6pt);
 	\end{tikzpicture}	
 	\end{split}
 \end{align}
 
 Similarly, we can perform $(ST)^2$ transformations on the theory $U(1)_k+1\F$ and get a chain of theories	
 \begin{align}
 	\label{mirrorF2}
 	\begin{split}
 	\includegraphics[]{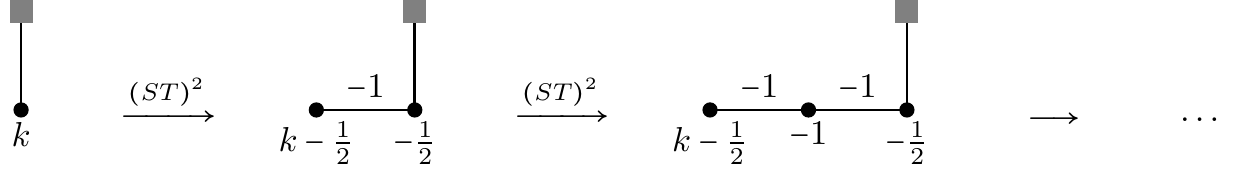}
 	\end{split}
 \end{align}
Their effective mixed \CS levels change as follows
 \begin{align}
 	k^{\eff}_{ij}:~~	\begin{bmatrix}
 		k + \frac{1}{2} 
 	\end{bmatrix} 
 	~~ \rightarrow ~~
 	\begin{bmatrix}
 		k -\frac{1}{2}   & -1 \\
 		-1 & 0
 	\end{bmatrix} 
 	~~ \rightarrow ~~
 	\begin{bmatrix}
 		k  -\frac{1}{2}  & -1  &\\
 		-1 & -1 & -1\\
 		&-1&0
 	\end{bmatrix} ~~ \rightarrow ~~ \cdots ~\,.
 \end{align}
In this case,  $ST$ and $(ST)^2 \cdot (ST)^2$ transformations also differ by a Kirby move for a gauge node
\begin{align}	
	\begin{split}
		\begin{tikzpicture}
			\node at (7.5,0){	$\xlongleftrightarrow{}$};
			\filldraw (9,0) node[below]{$k +\frac{1}{2}$}node[above]{}  circle(2pt);
			\draw[thick](9,0)--(10,0)node[midway, above]{\small$+1$}--(10,1);
			\filldraw (10,0) node[below]{$\frac{1}{2}$}  circle(2pt);
			\node[fill=gray] (r) at (10,1) {} ;
			\filldraw (4,0) node[below]{$k -\frac{1}{2}$}node[above]{}  circle(2pt);
			\draw[thick](4,0)--(5,0)node[midway, above]{\small$-1$}--(6,0)node[midway, above]{\small$-1$}--(6,1);
			\filldraw (5,0) node[below]{$-1$}  circle(2pt);
			\filldraw (6,0) node[below]{$-\frac{1}{2}$}  circle(2pt);
			\node[fill=gray] (r) at (6,1) {} ;
			\draw[thick,orange] (5,0)  circle(6pt);
		\end{tikzpicture}	
	\end{split}
\end{align}

Finally,  performing $ST$-moves three times,  we end up with a theory that seemingly differs from the original theory by three additional gauge nodes. These gauge nodes however can be integrated out, which brings us back to the original theory.


\subsubsection*{Gauged mirror triality}\label{STdiscuss}

To sum up, gauging the mirror triality \eqref{triviliaty} leads to a more generic triality
 \begin{align}\label{ststF}
\bsp
\includegraphics[width=2.5in]{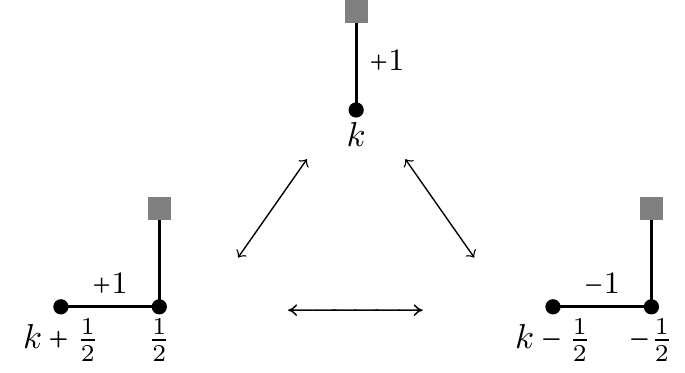}
\esp
\end{align} 
Some interesting cases of this triality arise for specific values of \CS levels. Since there is a gauge node in  \eqref{ststF} that is not attached to a matter node, one can integrate it out and get $U(1)_{k'}+1\F$ theory with a \CS level $k'$ depending on the value of $k$. Because of the parity anomaly of $k'$, only specific values of $k$ are meaningful.

Furthermore,  one can also gauge the global symmetry for a chiral multiplet with more general charge $q \neq1$. This leads to yet more general triality, which equivalently arises from combining  \eqref{chargeqST} and \eqref{chargeqSTST} 
\begin{align}\label{trichargeq}
	\bsp
	\includegraphics[width=2.5in]{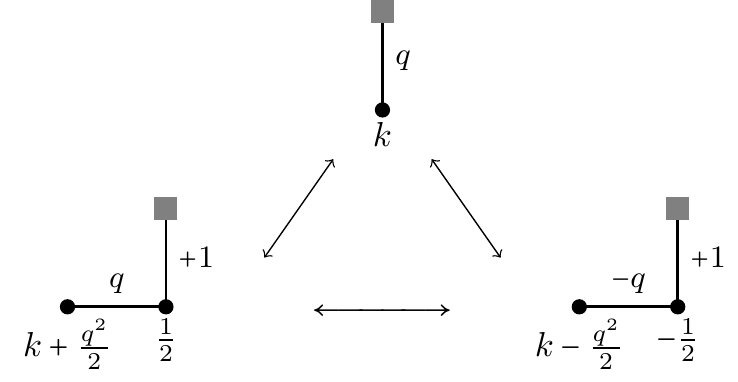}
	\esp
\end{align}

\bigskip

\noindent
\textbf{Example 1.} In this example we consider the theory $U(1)_k+1\F$ with special values of \CS levels, $k=\pm1/2\,, \pm 3/2$. In particular,  for $k=-1/2$, $ST$-move introduces an additional gauge node $U(1)_0$,  and then integrating it out in a Kirby move brings us back a free matter node
 \begin{align}
 	\begin{split}
 		\begin{tikzpicture}
 			\begin{scope}
 			\filldraw (0,0) node[below=0.1]{$-\frac{1}{2}$}  circle(2pt);
 			\draw[thick](0,0)--(0,1)node[midway,right]{\small $+1$};
 			\node at (0,0.2) [right]{ };
 			\node[fill=gray] (r) at (0,1) {} ;
 			\node at (1, -0.1) [below]{};
 			\node at (1.5,0.1){	$\xrightarrow{ST}$};
 				\draw[thick,orange] (3,0)  circle(6pt);
 			\filldraw (3,0) node[below=0.1]{$0$}node[above=0.1]{}  circle(2pt);
 			\draw[thick](3,0)--(4,0)node[midway,above]{\small $+1$}--(4,1)node[midway,right]{\small $+1$};
 			\filldraw (4,0) node[below=0.1]{$\frac{1}{2}$}  circle(2pt);
 			\node at (4.2,0 )[above]{};
 			\node[fill=gray] (r) at (4,1) {} ;
 			\end{scope}
 		\begin{scope}[xshift=4cm]
 			\node at (1, -0.1) [below]{};
 			\node at (1.5,0.1){	$\xrightarrow{\rm Kirby}$};
 			\node[fill=gray] (A) at (3,0) {}; \node[below=0.1]at (A){$\frac{1}{2}$};
 		\end{scope}
 			\end{tikzpicture}
 		\end{split}
 		\end{align}
Similarly,  for $k=1/2$, the $ST$-move reproduces another mirror pair in the mirror triality \eqref{triviliaty}. 
 
On the other hand,  applying the $ST$-move in \eqref{ststF} for $k=0$ leads to
  \begin{align}
  	\bsp
 		\begin{tikzpicture}[scale=0.8]
 			\begin{scope}
 	\filldraw (0,0) node[below=0.1]{$0	$}  circle(2pt);
 			\draw[thick](0,0)--(0,1);
 			\node at (0,0.2) [right]{ };
 			\node[fill=gray] (r) at (0,1) {} ;
 			\node at (1, -0.1) [below]{};
 			\node at (1.5,0.1){	$\xrightarrow{ST}$};
 			\draw[thick,orange] (3,0)  circle(6pt);
 			\filldraw (3,0) node[below=0.15]{$\frac{1}{2}$}node[above=0.15]{ }  circle(2pt);
 			\draw[thick](3,0)--(4,0)node[midway,above]{\small $+1$}--(4,1);
 			\filldraw (4,0) node[below=0.15]{$\frac{1}{2}$}  circle(2pt);
 			\node at (4.2,0 )[above]{ };
 			\node[fill=gray] (r) at (4,1) {} ;
 			\coordinate (A)at (5.2,0);
 			 			\coordinate (B)at (6.5,0);
 			 				\coordinate (C)at (6.5,1);
 			 					\node at (A) {$\longrightarrow$};
 			 						\filldraw (B) node[below=0.1]{$-\frac{3}{2}$}  circle(2pt);
 			 						\draw[thick] (B)--(C);	\node[fill=gray]  at (C) {} ;
 		\end{scope}
		\begin{scope}[yshift=-2.5cm]
			\filldraw (0,0) node[below=0.1]{$0	$}  circle(2pt);
			\draw[thick](0,0)--(0,1);
			\node at (0,0.2) [right]{ };
			\node[fill=gray] (r) at (0,1) {} ;
			\node at (1, -0.1) [below]{};
			\node at (1.5,0.1){	$\xrightarrow{(ST)^2}$};
			\draw[thick,orange] (3,0)  circle(6pt);
			\filldraw (3,0) node[below=0.15]{$-\frac{1}{2}$}node[above=0.15]{}  circle(2pt);
			\draw[thick](3,0)--(4,0)node[midway, above]{$-1$}--(4,1);
			\filldraw (4,0) node[below=0.15]{$-\frac{1}{2}$}  circle(2pt);
			\node at (4.2,0 )[above]{ };
			\node[fill=gray] (r) at (4,1) {} ;
			\coordinate (A)at (5.2,0);
			\coordinate (B)at (6.5,0);
			\coordinate (C)at (6.5,1);
			\node at (A) {$\longrightarrow$};
			\filldraw (B) node[below=0.1]{$\frac{3}{2}$}  circle(2pt);
			\draw[thick] (B)--(C);	\node[fill=gray]  at (C) {} ;
		\end{scope}
	\end{tikzpicture}
\esp
\end{align}
which suggests a triple duality
\begin{align}\label{equivalence032}
	\begin{split}
	\begin{tikzpicture}[scale=0.8]
		\node at (1,0.2) {$\longleftrightarrow$};
				\node at (3,0.2) {$\longleftrightarrow$};
			\draw[fill] (0,0) node[below=0.1]{$0$}  circle(2pt) coordinate (A);
		\draw[thick](0,0)--(0,1) node{$\graybox$};
			\draw[fill] (2,0) node[below=0.1]{$\frac{3}{2}$}  circle(2pt) coordinate (B);
		\draw[thick](B)--($(B)+(0,1)$) node{$\graybox$};
		\draw[fill] (4,0) node[below=0.1]{$-
			\frac{3}{2}$}  circle(2pt) coordinate (C);
		\draw[thick](C)--($(C)+(0,1)$) node{$\graybox$};
		\end{tikzpicture}
	\end{split}
	\end{align}
where the associated $k^{\eff}$ are $1/2,  2,  -1$ respectively.  However,  the theory $U(1)_0+1\F$ has parity anomaly as its effective \CS level is not integer, so we should not consider it. Therefore we are left with a dual pair $U(1)_{\pm 3/2}+1 \F$.

More generally, applying $ST$-moves in \eqref{ststF}  recursively on the matter nodes gives rise to a chain of dual theories that are in the same orbit of $ST$-moves. In particular, even for the free matter, there are orbits
 \begin{align}
 	&	\begin{tikzpicture}[scale=0.8]
 		\node[fill=gray] (r) at (1,0) {} ;
 		\node at (2,0){	$\longrightarrow$};
 		\filldraw (3,0) node[below]{$ \frac{ 1}{2}$}  circle(2pt);
 		\draw[thick](3,0)--(3,1);
 		\node[fill=gray] (r) at (3,1) {} ;
 		\node at (4,0){	$\longrightarrow$};
 		\filldraw (5,0) node[below]{$1$}  circle(2pt);
 		\filldraw (6,0) node[below]{$ \frac{ 1}{2}$}  circle(2pt);
 		\draw[thick](5,0)--(6,0);
 		\draw[thick](6,0)--(6,1);
 		\node[fill=gray] (r) at (6,1) {} ;
 		\node at (7,0){	$\longrightarrow$};
 		\filldraw (8,0) node[below]{$1$}  circle(2pt);
 		\filldraw (9,0) node[below]{$1$}  circle(2pt);
 		\filldraw (10,0) node[below]{$ \frac{ 1}{2}$}  circle(2pt);
 		\draw[thick](8,0)--(10,0);
 		\draw[thick](10,0)--(10,1);
 		\node[fill=gray] (r) at (10,1) {} ;
 		\node at (12,0){	$\longrightarrow~~\cdots$};
 	\end{tikzpicture}	\\ \nn
 	&	\begin{tikzpicture}[scale=0.8]
 		\node[fill=gray] (r) at (1,0) {} ;
 		\node at (2,0){	$\longrightarrow$};
 		\filldraw (3,0) node[below]{$ -\frac{ 1}{2}$}  circle(2pt);
 		\draw[thick](3,0)--(3,1);
 		\node[fill=gray] () at (3,1) {} ;
 		\node at (4,0){	$\longrightarrow$};
 		\filldraw (5,0) node[below]{$-	1$}  circle(2pt);
 		\filldraw (6,0) node[below]{$- \frac{ 1}{2}$}  circle(2pt);
 		\draw[thick](5,0)--(6,0) node[midway,above]{$-1$};
 		\draw[thick](6,0)--(6,1);
 		\node[fill=gray] () at (6,1) {} ;
 		\node at (7,0){	$\longrightarrow$};
 		\filldraw (8,0) node[below]{$-1$}  circle(2pt);
 		\filldraw (9,0) node[below]{$-1$}  circle(2pt);
 		\filldraw (10,0) node[below]{$- \frac{ 1}{2}$}  circle(2pt);
 		\draw[thick](8,0)--(9,0)node[midway,above]{$-1$}--(10,0)node[midway,above]{$-1$};
 		\draw[thick](10,0)--(10,1);
 		\node[fill=gray] (r) at (10,1) {} ;
 		\node at (12,0){	$\longrightarrow~~\cdots$};
 	\end{tikzpicture}	
 \end{align}
where in the first line we only perform $ST$ transformations, and in the second line we only perform $(ST)^2$ transformations. One can also randomly apply $ST$ and $(ST)^2$ transformations to get other chains; however, if we mod out the Kirby moves, these chains are equivalent and ultimately there are only two theories left $U(1)_{\pm 1/2}+1\F$.

\bigskip

\noindent
\textbf{Example 2.} 
 In this note we usually denote the anti-fundamental chiral multiplet $\AF$ by a blue box, or sometimes we still denote it by a gray box, but assign the charge of the matter on lines connecting matter nodes and gauge nodes. For the theory $U(1)_k+1\AF$, one can also gauge the global symmetry and take advantage of  identities \eqref{mirrortransf11} and \eqref{mirrortransf2}. As mentioned before, the $ST$-move of this $\AF$ turns it into $\F$. One can continue performing $ST$ transformations and gauging, which leads to a chain of theories. Let us apply $ST$-move twice to see if there is an analogous triality
 \begin{align}\label{AFtoF}
 	\begin{split}
 		\begin{tikzpicture}
 			\filldraw (0,0) node[below]{$k$}  circle(2pt);
 			\draw[thick](0,0)--(0,1);
 			\node[fill=blue] (r) at (0,1) {} ;
 			\node at (1, -0.1) [below]{};
 			\node at (1.5,0.1){	$\xlongrightarrow{ST}$};
 			\filldraw (3,0) node[below]{$k +\frac{1}{2}$} circle(2pt);
 			\draw[thick](3,0)--(4,0) node[midway, above]{$-1$}--(4,1);
 			\filldraw (4,0) node[below]{$\frac{1}{2}$}  circle(2pt);
 			\node[fill=gray] (r) at (4,1) {} ;
 				\node at (5,0.1){	$\xlongrightarrow{ST}$};
 					\filldraw (6,0) node[below]{$k +\frac{1}{2}$} circle(2pt);
 				\draw[thick](6,0)--(7,0) node[midway, above]{$-1$}--(8,0)node[midway, above]{$+1$}--(8,1);
 				\filldraw (7,0) node[below=0.15]{$1$}  circle(2pt);
 					\draw[thick,orange] (7,0)  circle(6pt);
 				\filldraw (8,0) node[below]{$\frac{1}{2}$}  circle(2pt);
 				\node[fill=gray] (r) at (8,1) {} ;
 				;
 				\node at (9,0){	$\longrightarrow$};
 			\filldraw (10,0) node[below]{$k -\frac{1}{2}$} circle(2pt);
 		\draw[thick](10,0)--(11,0) node[midway, above]{$+1$}--(11,1);
 		\filldraw (11,0) node[below]{$-\frac{1}{2}$}  circle(2pt);
 		\node[fill=gray] (r) at (11,1) {} ;
 		\end{tikzpicture}	
 	\end{split}
 \end{align}
In the third step above we integrate out the middle gauge node and end up with the fourth graph, which should be also the graph that is obtained by performing $(ST)^2$ transformation directly. Here we confirm again that $ST$ and $(ST)^2$ only differ by a Kirby move for gauge nodes.  Ultimately, by applying $ST$ transformations, we obtain the following triality for $U(1)_k+\AF$
  \begin{align}\label{ststAF}
 	\bsp
 	\includegraphics[width=2.5in]{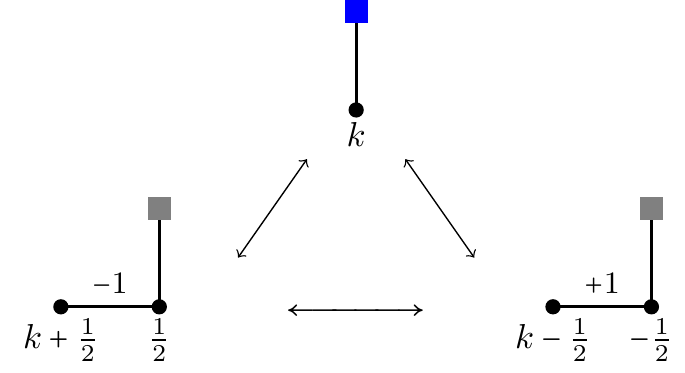}
 	\esp
 \end{align} 
which is a special case of the triality \eqref{trichargeq}  for $q=-1$.
	
We emphasize that one could also view $\AF$ as the basic matter field in this whole paper. Then the definition of $ST$ and $(ST)^2$ transformations would be switched,  however mixed \CS levels would change in the same way, and one would get the same collection of plumbing graphs.  More specifically, the $ST$-move leads to
 \begin{align}
	\begin{split}
		\begin{tikzpicture}
			\filldraw (0,0) node[below]{$k$}  circle(2pt);
			\draw[thick](0,0)--(0,1);
			\node[fill=blue] (r) at (0,1) {} ;
			\node at (1, -0.1) [below]{};
			\node at (1.5,0){	$\xlongrightarrow{ST}$};
			\filldraw (3,0) node[below]{$k +\frac{1}{2}$} circle(2pt);
			\draw[thick](3,0)--(4,0)node[midway, above]{\small$+1$}--(4,1);
			\filldraw (4,0) node[below]{$\frac{1}{2}$}  circle(2pt);
			\node[fill=blue] (r) at (4,1) {} ;
		\end{tikzpicture}	
	\end{split}
\end{align}
whose mixed bare \CS levels change as follows
 \begin{align}
 	k^{}_{ij}:~~	\begin{bmatrix}
 		k 
 	\end{bmatrix} 
 	~~ \rightarrow ~~
 	\begin{bmatrix}
 		k  +\frac{1}{2}  & 1 \\
 		1 & \frac{1}{2}
 	\end{bmatrix} \,.
 \end{align}
When acting with $ST$-move on $\AF$ many times, one also obtains a chain
 \begin{align}
 	\label{mirrorAF}
 	\begin{split}
 	\includegraphics[]{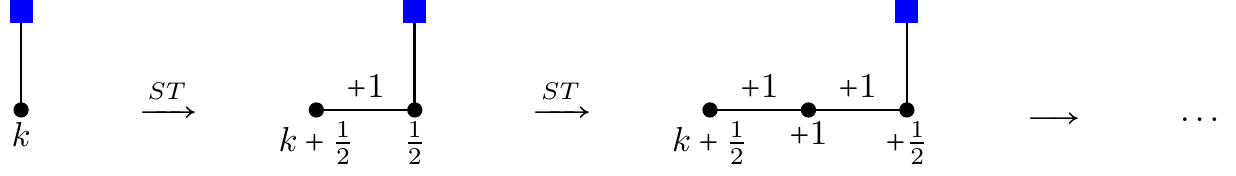}
 	\end{split}
 \end{align}
which only differs from \eqref{mirrorF} by the color of the matter node.
The associated effective mixed \CS levels change as follows
 \begin{align}
 	k^{\eff}_{ij}:~~	\begin{bmatrix}
 		k - \frac{1}{2} 
 	\end{bmatrix} 
 	~~ \rightarrow ~~
 	\begin{bmatrix}
 		k +\frac{1}{2}   & 1 \\
 		1 & 0
 	\end{bmatrix} 
 	~~ \rightarrow ~~
 	\begin{bmatrix}
 		k  + \frac{1}{2}  & 1  &\\
 		1 & 1 & 1\\
 		&1&0
 	\end{bmatrix} ~~ \rightarrow ~~ \cdots 
 \end{align}


\subsection{Chiral multiplets in other representations}\label{genericmatter}

Above, we have discussed the gauging and $ST$ transformations for chiral multiplets in the fundamental and anti-fundamental representations.  One can also gauge the flavor symmetry of matter fields in other representations, such as bifundamental, tri-fundamental, etc.  $ST$ transformations also dualize such flavor symmetries to the topological symmetry of dual theories. 

In the triality \eqref{trichargeq}, we have shown that the matter with charge $q$ under one gauge node $U(1)_k$ can be turned into $\F$.  The same statement holds for matter fields that are charged under many gauge nodes. In short, gauging the global symmetry turns matter in generic representation into a fundamental matter:
\begin{align}
	\text{matter in generic representation} ~\xlongrightarrow{\text{$ST$}}~ \text{fundamental matter}.
	\end{align}

To start with,  consider a bifundamental chiral multiplet.  The $ST$-move, performed respectively once and twice,  turns it into the following graphs
\begin{align}\label{bifundgraph}
	\bsp
\includegraphics[width=5in]{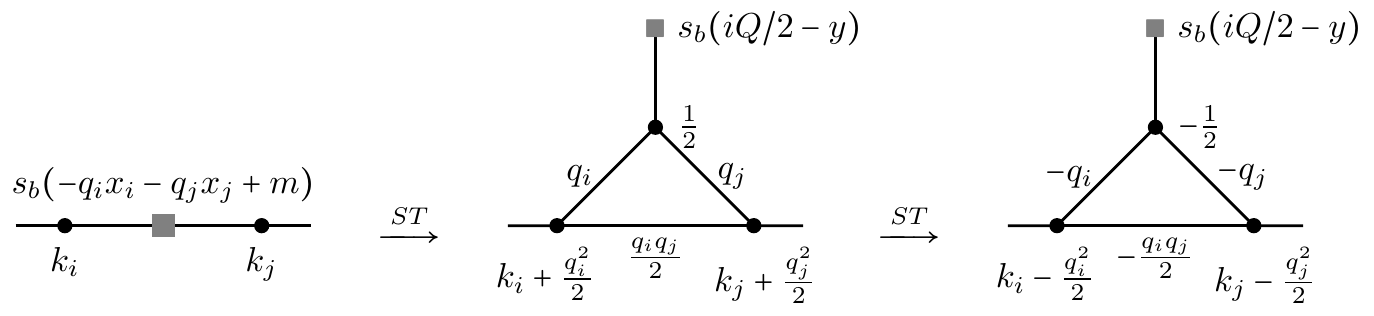}
\esp
\end{align}

We can also perform $ST$-move on chiral multiplets charged under many gauge nodes. After gauging the mass parameter becomes dynamic, which increases the number of gauge nodes by one. If a chiral multiplet is charged under several gauge nodes, after $ST$-move, the new gauge node will couple to other gauge nodes, and one can use identities \eqref{mirrortransf11} and \eqref{mirrortransf2} to compute sphere partition functions and read off mixed \CS levels.  Mixed \CS levels between two gauge nodes are the same as in the graphs in \eqref{bifundgraph}. For example, the tri-fundamental chiral multiplet transforms as follows under $ST$-move
\begin{align}
	\bsp
	\includegraphics[width=3.5in]{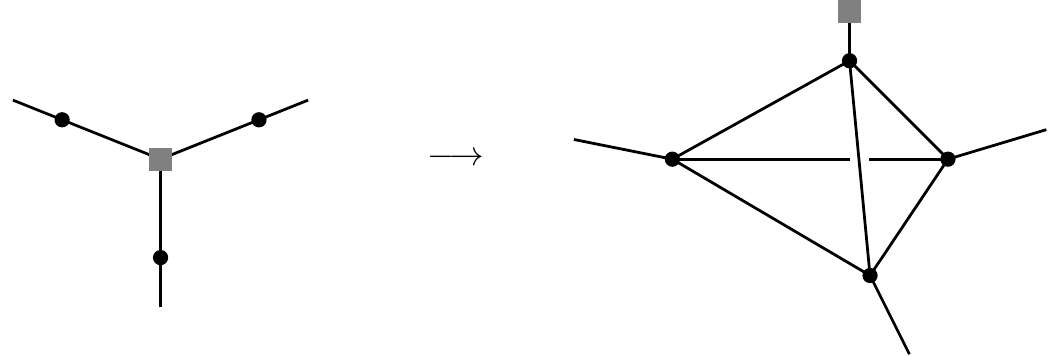}
	\esp
\end{align}
One can also decouple this tri-fundamental chiral multiplet,  as we discuss  below. This example shows that the notation of matter nodes is indeed convenient for representing $ST$ transformations.


\subsection{Decoupling and ungauging }  \label{decouplematter}

Decoupling, also referred to as integrating out a matter field,  is an operation that sends the mass parameter of a chiral multiplet to infinity, so that the corresponding matter node denoted by a gray box is deleted.  Decoupling of chiral multiplets depends on the sign of their mass parameters, and it replaces bare \CS levels of gauge nodes by effective \CS levels. Consistency of decoupling with $ST$-moves and Kirby moves imposes interesting constraints, which we discuss below.



\bigskip 

\noindent\textbf{Fundamental matter.} 
Let us start from a simple example of the gauged $ST$-dual graphs:
	\begin{align}\label{depfund}
	\begin{split}
		\begin{tikzpicture}
			\filldraw (0,0) node[below]{$k$}  circle(2pt);
			\draw[thick](0,0)--(0,1);
			\node at (0,0.2) [right]{ $z$};
			\node[fill=gray] (r) at (0,1) {} ;
			\node at (1, -0.1) [below]{};
			\node at (1.5,0){	$\xlongrightarrow{ST}$};
			\filldraw (3,0) node[below]{$k +\frac{1}{2}$}node[above]{ 	$z$}  circle(2pt);
			\draw[thick](3,0)--(4,0)node[midway,above]{\small$+1$}--(4,1);
			\filldraw (4,0) node[below]{$\frac{1}{2}$}  circle(2pt);
			\node at (4,0.2 )[right]{ $ y$};
			\node[fill=gray] (r) at (4,1) {} ;
		\end{tikzpicture}	
	\end{split}
\end{align}
Decoupling the matter field from the above graphs removes gray boxes and results in theories with effective \CS levels
 \begin{align}\label{decoupST}
 	\bsp
	\begin{tikzpicture}
	\filldraw (0,0) node[below]{$k'$}  circle(2pt);
	\node at (1,0){$\longrightarrow$};
	\filldraw (2,0) node[below]{$k +\frac{1}{2}$}   circle(2pt); 
	\draw[thick](2,0)--(3,0)node[midway, above]{\small $+1$};
	\filldraw (3,0) node[below]{$1$}  circle(2pt);
		\draw[thick,orange] (3,0)  circle(6pt);
\end{tikzpicture}
\esp
\end{align}
where $k' = k\pm 1/2$ depends on the sign of the mass parameter. We expect that the move in \eqref{decoupST} should be a Kirby move,  therefore we should choose the value $k'=k-1/2$.  It follows that the matter field $\F$ on the left graph $U(1)_k+1\F$ should have a negative mass $m<0$. This is consistent with the fact that $\F$ in the right graph $U(1)_{k+1/2}-U(1)_{1/2}+1\F$ has a positive mass. If the mass would be positive $m>0$, then we get a problematic operation on the graph:
\begin{align}
	\bsp
	\begin{tikzpicture}
		\filldraw (0,0) node[below]{$k'$}  circle(2pt);
		\node at (1,0){$\longrightarrow$};
		\filldraw (2,0) node[below]{$k +\frac{1}{2}$}   circle(2pt); 
		\draw[thick](2,0)--(3,0)node[midway, above]{\small $+1$};
		\filldraw (3,0) node[below]{$0$}  circle(2pt);
		\draw[thick,orange] (3,0)  circle(6pt);
	\end{tikzpicture}
	\esp
\end{align}
In this case the resulting gauge node $U(1)_0$ would be a problem, because according the Kirby moves in \eqref{kirbyknown} it would introduce a delta function, and hence gauge nodes connected to this $U(1)_0$ gauge node should be deleted, which obviously leads to a wrong result. Therefore,  consistency of decoupling with $ST$-move in \eqref{decoupST} implies that $\F$ must have nagative mass, whose sign is flipped by the $ST$-move
\begin{align}\label{constraintsdecoup1}
	\bsp
	& m < 0   ~~\xrightarrow{~ST~~}~~ m>0  .  \\
	\esp
\end{align}


Consider now the gauged $(ST)^2$ transformation for the second dual pair in \eqref{ststF}, whose decoupling leads to
\begin{align}
	\bsp
	\begin{tikzpicture}
		\filldraw (0,0) node[below]{$k''$}  circle(2pt);
		\node at (1,0){$\longrightarrow$};
		\filldraw (2,0) node[below]{$k -\frac{1}{2}$}   circle(2pt); 
		\draw[thick](2,0)--(3,0)node[midway, above]{\small $-1$};
		\filldraw (3,0) node[below]{$-1$}  circle(2pt);
		\draw[thick,orange] (3,0)  circle(6pt);
	\end{tikzpicture}
	\esp
\end{align}
In this case, only $m<0$ for the $(ST)^2$-transformed graph, which leads to the additional gauge node $U(1)_{-1}$, makes sense.  Integrating out this additional gauge nodes leads to $U(1)_{k+1/2} = U(1)_{k'}$ and hence $k''=k+1/2$, which corresponds to a positive mass $m>0$.
Therefore, consistency imposes the condition for $(ST)^2$ transformation
\begin{align}\label{constraintsdecoup}
	\bsp
		&	m > 0   ~~\xrightarrow{(ST)^2}~~ m<0  \,.
		\esp
	\end{align}

Note that \eqref{constraintsdecoup1} and  \eqref{constraintsdecoup} imply that the sign of the mass parameters is consistent with $(ST)^3=1$, since $m < 0  \xrightarrow{ST} m>0 \xrightarrow{(ST)^2} m<0$.   Although the matter fields in some theories are massless or their masses can be absorbed into FI parameters,  in order to impose consistency with $ST$ and $(ST)^{2}$ transformations we can introduce infinitesimal negative or positive mass parameters.  Furthermore,  if one does not  decouple the matter fields,  then the conditions in \eqref{constraintsdecoup1} and \eqref{constraintsdecoup} do not constrain the operations on graphs.

\bigskip

\noindent\textbf{Matter fields in other representations.}
The constraints in \eqref{constraintsdecoup1} and \eqref{constraintsdecoup} apply to matter fields in other representations too,  as we illustrate now for bifundamental fields.  Decoupling the matter with charges $(q_i,q_j)$ in the first graph in \eqref{bifundgraph} introduces the factor 
\begin{align}\label{bidecoup}
	&	s_b(-q_i x_i-q_j x_j +m) ~\rightarrow~ \exp \left(  \frac{- \pi i\, (q_i x_i+q_jx_j)^2 ~\text{sign}[m]}{2}  \right) \,,
\end{align}
where the mass parameter $m$ can be either positive or negative.  One can also decouple $\F$ in graphs that arise from $ST$ or $(ST)^2$ transformations in  \eqref{bifundgraph}, which should be consistent with and reduce to Kirby moves, and which imposes constraints on the sign of mass parameters before and after the transformation. It is not hard to see that the second and the third graph in \eqref{bifundgraph} should obey the same constraints as in \eqref{constraintsdecoup1} and  \eqref{constraintsdecoup}.
If the matter with charges $(q_i,q_j)$ has a positive mass $m>0$, then one is allowed to perform only $(ST)^2$ transformation that gives the third graph in \eqref{bifundgraph} for which the mass parameter is negative $m<0$, in order to ensure the decoupling of the fundamental matter $\F$ leads to an additional gauge node $U(1)_{-1}$. 
Similarly, if a matter field with charges $(q_i, q_j)$ is given a positive mass $m>0$, then only the second graph in \eqref{bifundgraph} is consistent and its decoupling leads to an additional gauge node $U(1)_{+1}$. 

This example illustrates again that $(ST)$ and $(ST)^2$ transformations flip the sign of mass parameter, also for matter charged under many gauge fields. This parameter can be only either positive or negative, respectively in the second and the third graphs in \eqref{bifundgraph}, as otherwise a wrong sign of the mass parameter during decoupling would result in a gauge node $U(1)_0$ with a vanishing \CS level, which would make the decoupled graphs inconsistent with Kirby moves. 

\bigskip

\noindent\textbf{Relations between Kirby moves and $ST$ transformations.} 
We have discussed two types of moves: Kirby moves for gauge nodes, and $ST$ transformations that can be also regarded as Kirby moves for matter nodes (even though we do not provide a corresponding geometric interpretation for plumbing manifolds). These two types of moves look quite different. The $ST$-move arises from gauging the mirror triality, while the Kirby move for gauge nodes can be interpreted as integrating out an additional gauge nodes, which are decorated by any matter node.  Nonetheless, we see from the above examples that these two types of moves are related by decoupling the matter node, i.e. sending its mass parameter to either positive or negative infinity.  If the conditions \eqref{constraintsdecoup1} and  \eqref{constraintsdecoup} are satisfied, then $ST$-moves for matter nodes reduce to Kirby moves for gauge nodes, namely 
\begin{align}\label{STrelateKirby}
	ST\text{-moves} ~~\xlongrightarrow{ m \rightarrow \pm \inf }~~\text{Kirby moves}  \,.
	\end{align}
We illustrate this statement also in the following figure, which presents a relation between theories $\mathcal{T}_1$ and $\mathcal{T}_2$,  which are related by $ST$-move, and corresponding theories $\mathcal{T}'_1$ and $\mathcal{T}'_2$ that arise in the decoupling limit:
\begin{align}\label{STKirby}
	\includegraphics[width=2.5in]{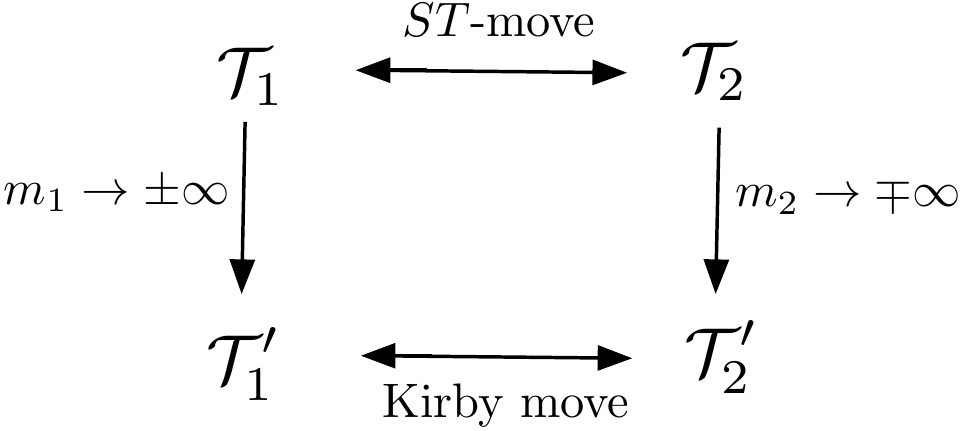} \nn
\end{align}

\bigskip

\noindent\textbf{Ungauging the gauge node.}
Ungauging of a gauge node amounts simply to deleting a gauge node that does not have matter fields attached to it, and also deleting all lines connecting to it. For example,  ungauging the undecorated gauge node in gauged mirror	triality \eqref{ststF} reduces it to the original mirror triality \eqref{triviliaty}.	
One can also ungauge a decorated gauge node in the $ST$ transformed graphs for bifundamental matter \eqref{bifundgraph}, which gives the generic  triality in \eqref{trichargeq}.  To get the gauged mirror triality \eqref{ststF}, one should set $q_i=q_j=+1$. In this case the matter fields in \eqref{bifundgraph}  should be in the bifundamental representation.




\section{Superpotentials and gauged SQED-XYZ duality}  \label{secsuperpotentials}

In this section we analyze more general abelian theories, namely those with superpotentials. Superpotentials may couple a few chiral multiplets, and we show that their presence implies some particular relations between mixed Chern-Simons levels assigned to vertices and edges of certain triangle graphs.  For theories with superpotentials, gauging and $ST$ transformations also play an important role and can be used to construct new dual theories.  
In particular, using these operations one can always transform a plumbing graph with matter into a graph, in which the numbers of gauge nodes and matter nodes are equal, and each gauge node is connected to one matter node. 
We call it a balanced condition.  A balanced graph can be denoted by $\left( U(1)+1\F\right)_{k_{ij}}^{N_f}$, where the pairs $U(1)_{k_{i}}+1\F$ are regarded as building blocks, which are connected by lines, as prescribed by mixed \CS levels $k_{ij}$.  These graphs thus look like molecules:
\begin{align}
	\bsp
	\includegraphics[width=1.5in]{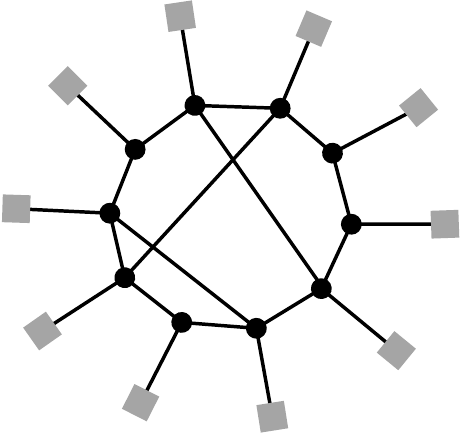}
	\esp
\end{align}

As the main result of this section, we will gauge the global symmetry for the mirror pair SQED-XYZ and identify new pairs of dual theories. Amusingly,  mixed \CS levels in some of these pairs of dual theories satisfy the same relations as in operations of linking and unlinking, identified in the context of knots-quivers correspondence \cite{Ekholm:2019lmb}.  Aside that, we also consider several other interesting examples of dualities for theories that include superpotentials.


\subsection{A basic mirror pair}

In the language of 3d $\mathcal{N}=2$ theories,  a basic example of mirror symmetry arises from $\mathcal{S}$-duality between 3d $\N=4$ theory $U(1)_0+1\F+1\AF+1 \textbf{Adj}$ (where $1\F+1\AF$, which we also denote by $(\mathrm{Q},\mathrm{\tilde{Q}})$, form a hypermultiplet, and $\textbf{Adj}$ denotes an adjoint chiral multiplet $\Phi$), and a theory of a single free hypermultiplet, which contains two chiral multiplets $1\F + 1\AF$, which we also denote by $(\mathrm{q},\mathrm{\tilde{q}})$ \cite{Intriligator:1996ex}. The former theory has a superpotential $\mathcal{W} = \mathrm{Q} \mathrm{\tilde{Q}}{\Phi}$.  We also represent this mirror pair by	 
\begin{align}\label{AdjQQqq}
	U(1)_0  + 1 (\mathrm{Q}, \mathrm{\tilde{Q}} )+ 1\Phi ~~~\longleftrightarrow~~ \le( \mathrm{q}, \mathrm{\tilde{q}}  \r) .
\end{align}
This mirror pair is well known to be a consequence of $\mathcal{S}$-duality in type IIB string theory. 

A contribution of the hypermultiplet to the sphere partition function reads
\begin{align}
\begin{split}\label{hyper}
	\text{F}_m (z) := s_b \le( z +\frac{m}{2} +  \frac{i\,Q}{4} \r)  s_b \le( -z +\frac{m}{2} + \frac{i\,Q}{4} \r)
	\end{split} \,.
	\end{align}
Note that these two chiral multiplets may have different mass parameters; absorbing the mass parameter $m$ into $z$,  this contribution $\text{F}_m (z)$ becomes $s_b \le( z +  \frac{i\,Q}{4} \r)  s_b \le( -z +{m}+ \frac{i\,Q}{4} \r)$. A manifestation of the duality \eqref{AdjQQqq} is the equality of the sphere partition functions of the mirror theories
\begin{align}\label{N4mirrortransf}
s_b(-m)	\oint d y\, e^{ - 2 \pi i\,z y} \text{F}_m(y)  = \text{F}_{-m}(z)  \,,
	\end{align}
where the term $s_b(-m)$ is a contribution of the adjoint chiral multiplet.  
For $m=0$ this relation reduces to
\begin{align}\label{move1F1AF}
	\bsp
&\oint d y\, e^{ - 2 \pi i\,z y} \, \text{F}_0(y)  = \text{F}_{0}(z)  .
 \esp
\end{align}

Applying the identity \eqref{N4mirrortransf} recursively leads to a chain of dual theories. This can also be viewed as gauging the global symmetry and then replacing the matter node, as we have discussed for the theory $U(1)_k+1\F$ in \eqref{mirrorF}.  In the present case,  for $U(1)_k+1\F+1\AF$,   the difference is that the chain is in the orbit of $\mathcal{S}$-duality
\begin{align} \label{mirrorN4exm}
	\bsp
	\begin{tikzpicture}
		\filldraw[thick] (0,0) circle(2pt);
			\filldraw[thick] (2.5,0) circle(2pt);
				\filldraw[thick] (7.5,0) circle(2pt);
				\filldraw[thick] (3.5,0) circle(2pt);
				\filldraw[thick] (6.5,0) circle(2pt);
					\filldraw[thick] (8.5,0) circle(2pt);
	\node at (-0.3,0) {$k$};
	\draw[thick](0,-0.5)node[rectangle,fill=blue] {}--(0,0.5)node[rectangle,fill=gray] {}; 
\node at (1.2,0)  {$\xlongrightarrow{\mathcal{S}}$} ;
	\node at (2+0.2,0) {$k$};
\draw[thick](2.5,0)--(3.5,0) node[right]{$0$}  ;
\draw[thick](3.5,-0.5)node[rectangle,fill=blue] {}--(3.5,0.5)node[rectangle,fill=gray] {}; 
\node[rectangle,fill=gray] at (4.3,0.5) {};
\node at (5.3,0)  {$\xlongrightarrow{\mathcal{S}}$} ;
	\node at (6+0.5,-0.3) {$k$};
\draw[thick](6.5,0)--(3.5+4,0)node[below]{$0$} --(8.5,0) node[right]{$0$}  ;
\draw[thick](3.5+5,-0.5)node[rectangle,fill=blue] {}--(3.5+5,0.5)node[rectangle,fill=gray] {}; 
\node at (10.2,0)  {$\longrightarrow ~~\cdots$} ;
	\end{tikzpicture}
\esp
	\end{align}
Here the blue node represents the anti-fundamental matter $\AF$, and the term $s_b(-m)$, which is a contribution from the free chiral multiplet,  cancels with another term $s_b(m)$ after the transformation $\mathcal{S}^2=1$.  In particular, if $k=0$, one can integrate out the pure gauge node $U(1)_0$ and get 
\begin{align}
	\begin{tikzpicture}
			\filldraw (0,0)node[right]{$0$} circle(2pt);
		\draw[thick](0,-0.5)node[rectangle,fill=blue] {}--(0,0.5)node[rectangle,fill=gray] {}; 
		\node at (1.8,0)  {$\xlongrightarrow{\mathcal{S}}$} ;
	\node[rectangle,fill=gray] at (3.5,0.5) {};		\node[rectangle,fill=gray] at (3.5,-0.5) {};
		\node[rectangle,fill=gray] at (4.3,0) {};
	\end{tikzpicture}
\end{align}
which is referred to as the 2-3 move or SQED-XYZ mirror pair.  In the orbit \eqref{mirrorN4exm},  integrating out pure nodes $U(1)_{k}$ and $U(1)_0$ and ignoring the decoupled matter node,  we are left with only two dual theories 
 \begin{align} 	\label{SQEDXYZmove}
 	\begin{tikzpicture}
 			\filldraw (0,0) circle(2pt);
 				\filldraw (4,0) circle(2pt);
 			\node at (0.3,0) {$k$};
 		\draw[thick](0,-0.5)node[rectangle,fill=blue] {}--(0,0.5)node[rectangle,fill=gray] {}; 
 		\node at (2.2,0)  {$\xlongleftrightarrow{\mathcal{S}}$} ;
 		\node at (4.5,0) {$-\frac{1}{k}$}  ;
 		\draw[thick](4,-0.5)node[rectangle,fill=blue] {}--(4,0.5)node[rectangle,fill=gray] {}; 
 		\end{tikzpicture}
 	\end{align}
 which is the dual pair $U(1)_k + 1\F +1\AF  ~\longleftrightarrow ~ U(1)_{-1/k} + 1\F +1 \AF$.
 Since there is no meaningful solution to $1/k=-k$, the theory in \eqref{SQEDXYZmove} cannot be self-dual. However, using the exchange symmetry $\q \leftrightarrow \q^{-1}$ that we will discuss in section \ref{exchangesec}, one can see that $1/k=k$ leads to solutions with $k=\pm 1$, which are special values corresponding to a mirror dual pair, see also the discussion in sections \ref{partialgauge} and \ref{secmirrorpair}. 
 
Using the identity \eqref{N4mirrortransf} recursively, we obtain sphere partition functions for the theories in the chain \eqref{mirrorN4exm} 
\begin{align}
	\bsp
	&\oint dz \, e^{- \pi i\, k z^2 + 2 \pi \,i \xi z } \text{F}_{-m}(z) \rightarrow  \\
	&	s_b(-m)\oint dz \oint dy \, e^{- \pi i\, k z^2 + 2 \pi \,i \xi z } e^{- 2 \pi i\, zy} \text{F}_{m}(y)  \rightarrow \\
	 	 &	s_b(-m) s_b(m)\oint dz \oint dy \oint dy' \, e^{- \pi i\, k z^2 + 2 \pi \,i \xi z } e^{- 2 \pi i\, zy}e^{- 2 \pi i\, y y'} \text{F}_{-m}(y') .
	 	 \esp
	\end{align}
Integrating out the variable $z$, the partition function in the second line can be reduced to
\begin{align}
	s_b(-m) \oint dy \, e^{ \frac{\pi i}{k}  (y-\xi)^2} \text{F}_{m}(y),
	\end{align}
which shows that the $\mathcal{S}$-transformation leads to $k \rightarrow - k^{-1}$.

The above relations between sphere partition functions also imply that effective \CS levels for theories in the chain  \eqref{mirrorN4exm} are 
\begin{align}
	k^{\eff}_{ij}:~~	\begin{bmatrix}
	~	k~
	\end{bmatrix} 
	~~ \rightarrow ~~
	\begin{bmatrix}
		k  & 1 \\
		1 & 0
	\end{bmatrix} 
	~~ \rightarrow ~~
	\begin{bmatrix}
		\begin{array}{ccc}
		k   & 1 &~~\\
		1 & 0 & 1\\
	~~	&1& 0	
		\end{array}
	\end{bmatrix} ~~ \rightarrow ~~ \cdots  
\end{align}

Analogously, for the plumbing theories that contain many building blocks $U(1)_{k_i}+1\F+1\AF +1\textbf{Adj}$ that are coupled by mixed \CS levels,  the $\mathcal{S}$-duality leads to
\begin{align}
	k_{ij}^{\eff} ~\xlongrightarrow{S}~
	-\le( {k_{ij}^{\eff}} \r)^{-1}  \,.
\end{align}
For such theories, since $\F$ and $\AF$ give opposite contributions to \CS levels, we get $k_{ij}^{\eff} =k_{ij} $.

We will continue the discussion of graphs in \eqref{mirrorN4exm} in section \ref{partialgauge} in more detail.


\bigskip

\noindent\textbf{SQED-XYZ mirror pair.} The mirror pair \eqref{AdjQQqq} can also be represented as the duality between $U(1)_0+1\F+1\AF$ and XYZ-model
\begin{align}\label{SQEDXYZ}
	U(1)_0 +1 \mathrm{Q} + 1 \mathrm{\tilde{Q}}   ~~~\longleftrightarrow~~ \le( \text{X}, \text{Y}, \text{Z}  \r) ,
\end{align}
where $\text{X}$, $\text{Y}$ and $\text{Z}$ are chiral multiplets.
The XYZ-model has a superpotential $\mathcal{W}=\text{X}\text{Y}\text{Z}$.
Using the formula $s_b(-x)s_b(x)=1$, one can move a term in the sphere partition functions in the equivalence \eqref{N4mirrortransf} to get the sphere partition functions for the SQED-XYZ pair
\begin{align}\label{N4mirror1}
{	\small
  \int dx\, e^{ - 2 \pi i\, z x } s_b\le( \frac{iQ}{2} -x -y \r)s_b\le( \frac{iQ}{2} +x -y \r) =s_b(y-z)s_b(y+z)	s_b \le(  \frac{i Q}{2}  -2y\r)   .}
	\end{align}
The Fourier transformation of this identity is 
\begin{align} 
  \int dz\, e^{  2 \pi i\, z x } s_b(y-z)s_b(y+z)  s_b\le(
  \frac{iQ}{2} -2 y \r)=s_b\le( \frac{iQ}{2} -x -y \r)s_b\le( \frac{iQ}{2} +x -y \r)  .
	\end{align}
In the following sections, we will use this basic dual pair and gauging the flavor symmetry to generate many other dual theories and discuss how plumbing graphs encode superpotentials.

Note that the SQED-XYZ duality can be interpreted as a process of adding a superpotential. It is also a special case of Seiberg duality,  which in general arises from a mutation of a quiver diagram for a gauge theory:
\begin{align}
	\bsp
\begin{tikzpicture}[scale=0.8]
	\node[rectangle,draw=black](v1) at (0,0){$N_a$}  ;
		\node[ ](v2) at (2,0){$N_c$}  ;
			\draw (2,0) circle(14pt)	;
			\node[rectangle,draw=black](v3) at (4,0){$N_f$}  ;
			\path[<-,thick] (v1) edge node[below]{$\mathrm{\tilde{Q}}$} (v2) ;	\draw[<-,thick] (v2)edge node[below]{$\mathrm{Q}$} (v3) ; 
	\end{tikzpicture}
\qquad\quad
\begin{tikzpicture}[scale=0.8]
	\node[rectangle,draw=black](v1) at (0,0){$N_a$}  ;
	\node[](v2) at (2,0){$N'_c$}  ;
				\draw (v2) circle(14pt)	;
	\node[rectangle,draw=black](v3) at (4,0){$N_f$}  ;
	\path[->,thick] (v1) edge node[below]{$\rm \tilde{q}$} (v2) ;	\draw[->,thick] (v2)edge node[below]{$\rm q$} (v3) ; 
\path[->,thick](v3) edge [bend right] node [above]{$\rm M$}(v1)	;
\end{tikzpicture}
\esp
	\end{align}
The theory in the left figure has no superpotential, and it is dual to a theory on the right with a superpotential $\mathcal{W} = \rm q M  \tilde{q}$.  In what follows we consider analogous processes of adding superpotentials for theories encoded in plumbing graphs with matter.


\subsection{Gauging SQED-XYZ duality}    \label{subsecgauging}

Analogously to gauging of mirror triality, we now gauge the flavor symmetry of SQED-XYZ duality, thereby introducing additional gauge nodes in diagrams that encode two sides of this duality.  We also find that the presence of superpotential implies certain non-trivial relations between mixed \CS levels assigned to a triangle in a graph representing one side of this duality.


To analyze the process of gauging, we consider the sphere partition functions of SQED and XYZ model.  As mentioned in \eqref{N4mirror1},  these partition functions are equal. For later convenience, we rewrite this equality and move one term to the left 
\begin{align}\label{partSQEDXYZ}
s_b \le( - \frac{i Q}{2}  + 2y\r)  \int dx\, e^{ - 2 \pi i\, z x } s_b\le( \frac{iQ}{2} \pm x -y \r)
	= 	s_b(y\pm z) 	,
\end{align}
where $s_b(x\pm y):=s_b(x+y)s_b(x-y)$. This relation can be also interpreted as equality of sphere partition functions for the mirror pair \eqref{AdjQQqq}. In this case, the superpotential of the theory $U(1)_0 + 1\F+1\AF + 1 \bf{Adj}$ yields non-trivial relations between its three chiral multiplets. Note that the flavor charge for the adjoint chiral multiplet is 2, and we denote flavor symmetries    associated with mass parameters $y$ and $z$ by $U(1)_y\times U(1)_z$ \footnote{Note that $U(1)_z$ is the topological symmetry on SQED side.}. 

One can draw plumbing graphs for this mirror pair, however generic Kirby moves do not lead to interesting theories. Therefore we gauge flavor symmetries, which leads to dual theories represented by the following plumbing graphs
\begin{align}\label{SQEDXYZgauge}
\bsp
\includegraphics[]{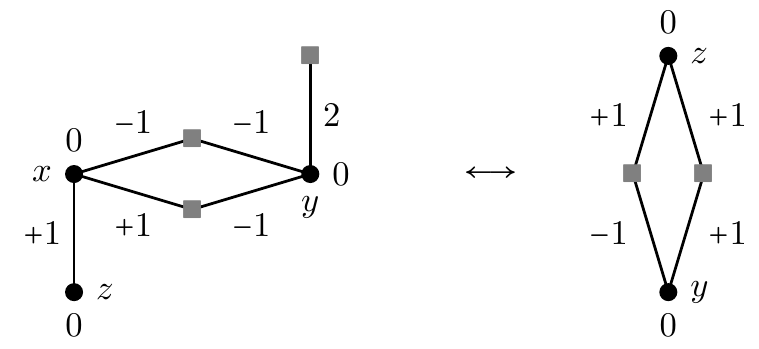}
	\esp
\end{align}
Gauging the flavor symmetry $U(1)_y\times U(1)_z$ introduces two additional gauge nodes with variables $y$ and $z$ respectively.  Note that one could move the adjoint matter to the right graph, which would represent gauging of the SQED-XYZ mirror pair; however, this would lead to the same results that we find for the above graph pair \eqref{SQEDXYZgauge}, so we only discuss this pair in what follows.  Furthermore, note that two matter nodes in the right graph have different charges $(+1,-1)$ and $(+1,+1)$, which is important, since if they would have the same or opposite charges, then mixed \CS levels could not be generated upon $ST$ transformations.

The gauging of graphs in \eqref{SQEDXYZgauge} is still not complete, as we have not turned on mixed \CS levels for new gauge nodes $U(1)_z$ and $U(1)_y$. After doing so,  we get the gauged duality represented by the following graphs
\begin{align}\label{mirrorpairplumb}
	\bsp
\includegraphics[width=3in]{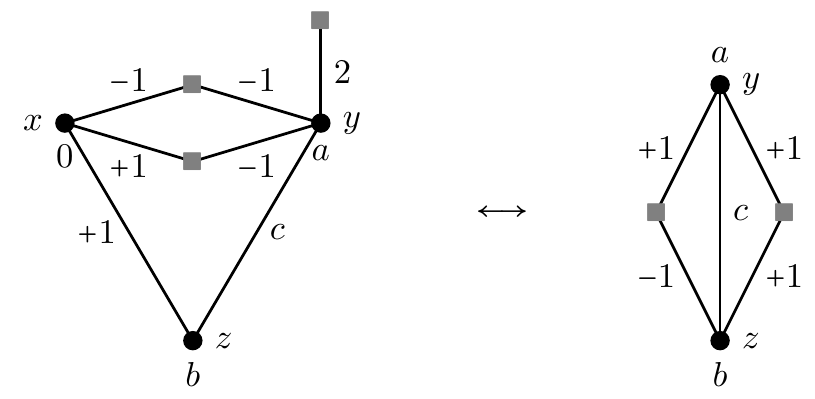}
	\esp
\end{align}
where $a$, $b$ and $c$ are \CS levels for $U(1)_y$ and $U(1)_z$ gauge groups respectively, which should be integer or half-integer. Since we are gauging global symmetries for dual theories, the introduced \CS levels and FI parameters on both sides of the duality should be the same.
However, since our purpose is to perform Kirby moves and $ST$ transformations on the graphs in \eqref{mirrorpairplumb} to see how superpotentials are encoded in these graphs, we do not discuss the values for $a$, $b$ and $c$ at this moment, as generic Kirby moves change \CS levels significantly.

The full gauging of flavor symmetries adds the following integral to sphere partition functions on both sides of the duality
\begin{align}\label{fullgauging}
	& \int \mathbf{gauge}[y,z]:=\int dy \int dz  \, 	e^{ -\pi i ( a y^2 + b z^2 + 2 c \,yz)  } e^{ 2\pi i\,( \xi_y y+ \xi_z z) } ,
	\end{align}	
so that the sphere partition functions for the gauged mirror pair are
\begin{align}\label{pairgauge}
	\bsp
\boxed{ \int \textbf{gauge}[y,z]
\int dx\, e^{ - 2 \pi i\, z x  }  		s_b \le(  -\frac{i Q}{2}  + 2y\r)
 s_b\le( \frac{iQ}{2} \pm x -y \r)    
 	=  \int \textbf{gauge}[y,z]	~ s_b(y\pm z)   } \\
\esp
\end{align}
where again $s_b(x\pm y):=s_b(x+y)s_b(x-y)$. We also refer to this relation as a 2-3 move, as the numbers of chiral multiplets on both sides are 3 and 2 respectively. After this gauging, all chiral multiplets become massless, as all free mass parameters are gauged.
 
More generally, we could also turn on non-trivial charges for $y$ and $z$,  by replacing $y \rightarrow q_1 y$ and $z \rightarrow q_2 z$, and then gauging the corresponding flavor symmetries $U(1)_y\times U(1)_z$. However, the computation shows that this does not influence the final results that we will discuss in the following sections, because after gauging, we will integrate out undecorated gauge nodes $U(1)_y$ and $U(1)_z$, which does not affect their charges.  Therefore, in the following we only discuss Kirby moves and $ST$-moves for the pair of theories in \eqref{mirrorpairplumb}.


In what follows we consider a pair of dual complicated theories represented by the graphs in \eqref{mirrorpairplumb} and transform them further, using $ST$-moves and Kirby moves, in particular transforming bifundamental matter fields into fundamental ones. Ultimately we obtain dualities between theories represented by balanced graphs (whose number of gauge nodes is the same as the number of matter nodes). We find that \eqref{mirrorpairplumb} can be transformed into operations that we call unlinking and linking, in analogy with operations on quivers in knots-quivers correspondence. In addition, we find a few other operations that we call exotic ones.  As we will see, existence of superpotentials in these theories imposes certain constraints on mixed \CS levels.



\subsubsection*{Unlinking}


To start with, we consider the graph on the right hand side in \eqref{mirrorpairplumb}. We apply $ST$-moves to two bifundamental chiral multiplets and then integrate out  pure gauge nodes denoted by orange circles:
\begin{align}\label{acbtorks}
	\bsp
\includegraphics[width=3in]{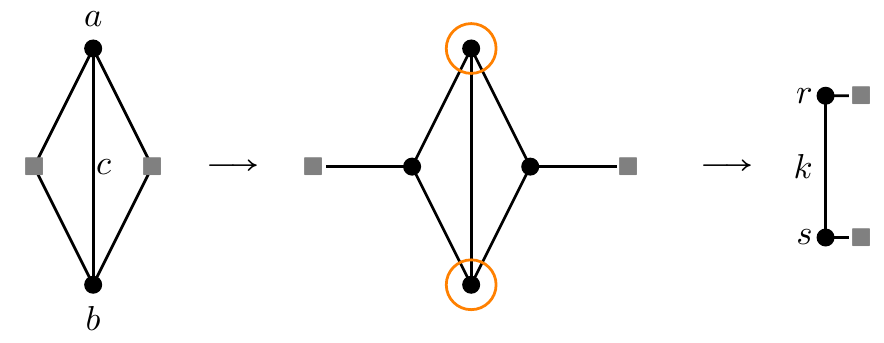}
	\esp
	\end{align}
In this way we obtain a two-node graph with effective \CS levels $r$, $s$, and $k$, which are related to $a$, $b$, and $c$ as follows:
\begin{align}
&	r =\frac{ a b -(1+c)^2 }{ 1+a +b +a b -c^2   }  \,, \quad &	s& =\frac{ a b -(1-c)^2 }{ 1+a +b +a b -c^2   }  \,, \quad &
k&=	\frac{ a -b}{ 1+a +b +a b -c^2   }  \,,  \label{abctorks-rks}
 \\
	&	a =\frac{ -1 -2 k - k^2 +rs}{ -1+k^2+r + s -r s   }  \,, \quad 
&	b &=\frac{ -1 +2 k - k^2 +rs}{ -1+k^2+r + s -r s   }  \,, \quad 
&	c&=	\frac{ r -s}{  -1+k^2+r + s -r s  }  \,. \label{abctorks}
	\end{align}

For future convenience, we also introduce the notation $\textbf{(i,j, ...)}$ to denote implementing $ST$-moves $(ST)^\textbf{i}, (ST)^\textbf{j},\ldots$ respectively on the first, second, etc.  matter node of a given balanced graph. For example, in the last graph in (\ref{acbtorks}) we do not apply further $ST$-moves on any of the two matter nodes,  so we denote this graph by $\textbf{(0,0)}$.  The assignment of the above effective \CS levels to this graph can be represented by
\begin{align}
		\textbf{ (0,0)} ~~\rightarrow~~
	k_{ij}^{\eff}\left(\btik  
	\filldraw (0,0.2) circle(1pt) ;
		\filldraw (0.4,0.2) circle(1pt) ;
		\draw[	](0.4,0.2) --(0,0.2) 	;
	\etik \right) =	\begin{bmatrix}
		r  & k  \\
		k & s \\
	\end{bmatrix} 
	\end{align}

We now transform the left hand side of \eqref{mirrorpairplumb} into a balanced graph.  Such a graph can be obtain in various ways and its mixed \CS levels are rational functions of $a,b$ and $c$, or equivalently rational functions of $k,r$ and $s$ in (\ref{abctorks-rks}).  In order to avoid partity anomaly \cite{Aharony:1997aa} we look for a balanced graph whose mixed \CS levels are integer, if $k,r$ and $s$ are integer -- in other words,  such mixed \CS levels should be polynomials in $k,r$ and $s$.  We find that such a graph can be obtain in the following process:
\begin{align}\label{unlinkCS}
	\bsp
\includegraphics[width=5in]{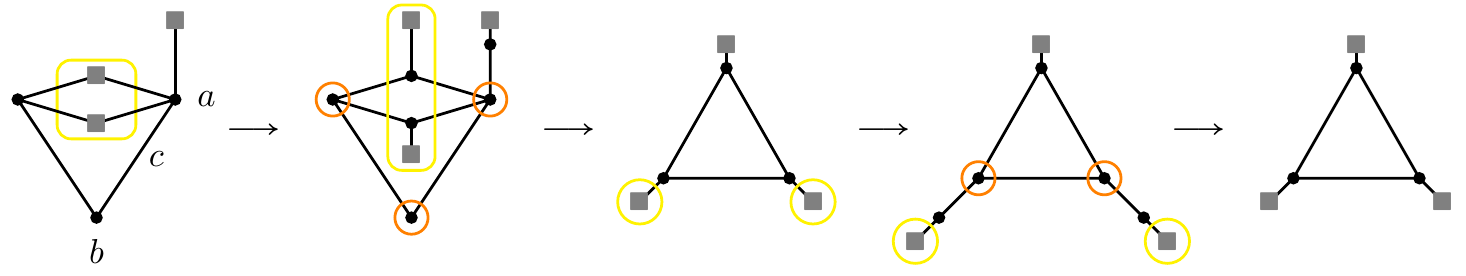}
\esp
\end{align}
where we use a yellow circle to track positions of two initially bifundamental chiral multiplets. The adjoint chiral multiplet in the first graph is turned into a fundamental chiral multiplet in the second step and then moved to the top of triangle graph in the third step. The orange circles mark the undecorated gauge nodes that we integrate out. Note that already the third graph in (\ref{unlinkCS}) is balanced, however its mixed \CS levels are fractional when expressed in terms of (integer) $k,r$, and $s$. Therefore, we apply further $ST$-moves $\bf (1,1,0)$ on this graph, in order to obtain mixed \CS levels which are polynomials in $k,r$, and $s$
\begin{align}
	\textbf{(1,1,0)} ~~\rightarrow~~
	k_{ij}^{\eff} (
	\btik[scale=0.8]  
	\filldraw (0,0) circle(1pt) ;
	\filldraw (0.2,-0.3) circle(1pt) ;
		\filldraw (-0.2,-0.3) circle(1pt) ;
	\draw[	](0,0) --(0.2,-0.3)--(-0.2,-0.3)--(0,0) 	;
	\etik
	)=	
	\begin{bmatrix}
		s  & k-1  & s+k-1\\
		k-1 & r & r+k-1\\
		s+k-1	& r+k-1&r+s +2k -1
	\end{bmatrix} 
\end{align}

To sum up, we have transformed the gauged SQED-XYZ duality \eqref{mirrorpairplumb} into the following duality, which we refer to as the 2-3 move.  In analogy with the operation in knots-quiver correspondence \cite{Ekholm:2019lmb}  we also call it unlinking, as the linking number $k$ for the two original nodes decreases by one
\begin{align}\label{unlinkingrelation}
\bsp 
\includegraphics[width=3in]{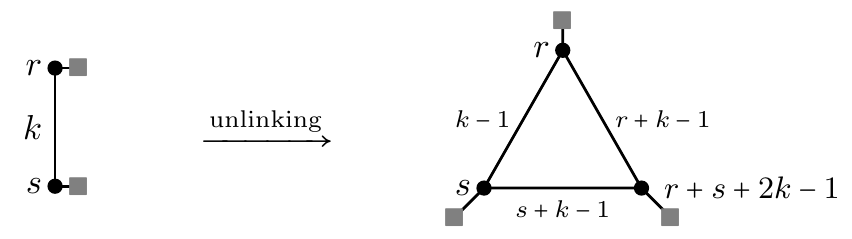}
	\esp
	\end{align}

We summarize the relation between associated mixed \CS levels as follows
\begin{align}
		k_{ij}^{\eff}\left(- \right) ~~ \xrightarrow{\text{unlinking}}~~
	k_{ij}^{\eff} (\bigtriangleup)\,.
	\end{align}
A manifestation of the superpotantial of the XYZ model are the relations between mixed \CS levels in the triangle graph in \eqref{unlinkingrelation}, i.e. the fact that all these \CS levels are determined by three parameters $k,r$ and $s$. 
Note that for $r=s=0$ and $k=1$, this dual pair reduces to the original 2-3 move for SQED-XYZ duality, whose graphs are shown in \eqref{SQEDXYZdualpari}. 

Moreover, FI parameters on both sides of the duality (\ref{unlinkingrelation}) are also related.  Denoting FI parameters for the gauge nodes $U(1)_r$ and $U(1)_s$ respectively by $x_r=e^{\xi_r}$ and $x_s=e^{\xi_s}$,  we find that the FI parameter for the new gauge node $U(1)_{r+s+2k-1}$ takes form $q^* x_r x_s$, i.e. under the duality these parameters are transformed as follows
\begin{align}
	(x_r\,, x_s) ~\longrightarrow~ (x_r\,, x_s\,, \q^{*} x_r x_s  )  
	\end{align}
where $q^*$ can be found by comparing the sphere partition functions.  Therefore,  in this sense three decorated gauge nodes of the triangle are also not independent.

Note that the integral triangle graph may not be unique for specific values $r$, $s$ and $k$, because we can perform $ST$-moves on the unlinking triangle of \eqref{unlinkingrelation}. However, these additional triangles are easy to find by scanning all graphs, as one can perform $ST$-moves at most three times on each matter node, and hence there are in total $3^{\text{rank of flavors}} = 3^3= 27$ graphs for a triangle. For example,  the \CS levels of the third graph in \eqref{unlinkCS} can  be integers for some particular values of $r$, $k$ and $s$.  Moreover,  the permutation group  of gauge nodes is a subset of the $ST$-move group $S_3 \subset 3^{\text{rank of flavors}}$. The relations between FI parameters can always be tracked by the evolution of sphere partition functions undergoing $ST$-moves.



\subsubsection*{Linking}

Another interesting form of duality, which can be obtained using some specific $ST$-moves, we call linking, also in analogy with knots-quivers correspondence  \cite{Ekholm:2019lmb}. Let us apply $ST$-moves $(\1, \1)$ on the two-node graph in \eqref{acbtorks}:
\begin{align}\label{rskSTST}
	\bsp
	\includegraphics[]{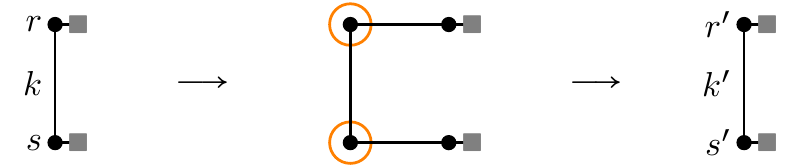}
	\esp
\end{align}
After integrating out pure gauge nodes in orange circles, we denote mixed \CS levels for the resulting two-node graph by
\begin{align}
		\textbf{ (1,1)} ~~\rightarrow~~
	k_{ij}^{\eff}\left(-  \right) =	\begin{bmatrix}
		r'  & k'  \\
		k'	 & s' \\
	\end{bmatrix}  
\end{align}
and we find the following relations 
\begin{align}
	r' = \frac{r s -k^2 -r }{s-1} \,,\quad s'=\frac{-1}{s-1} \,, \quad  k'=\frac{k}{s-1}  .
	\end{align}
Obviously, for generic $(r,s,k)$, one may not get integral $(r', s',k')$. 
 
 In this case, we find that the dual three-node graphs should undergo a different process in comparison with the unlinking case \eqref{unlinkCS}, in order to end up with a triangle with integral \CS levels:
\begin{align}\label{linkCS}
	\bsp
\includegraphics[width=5in]{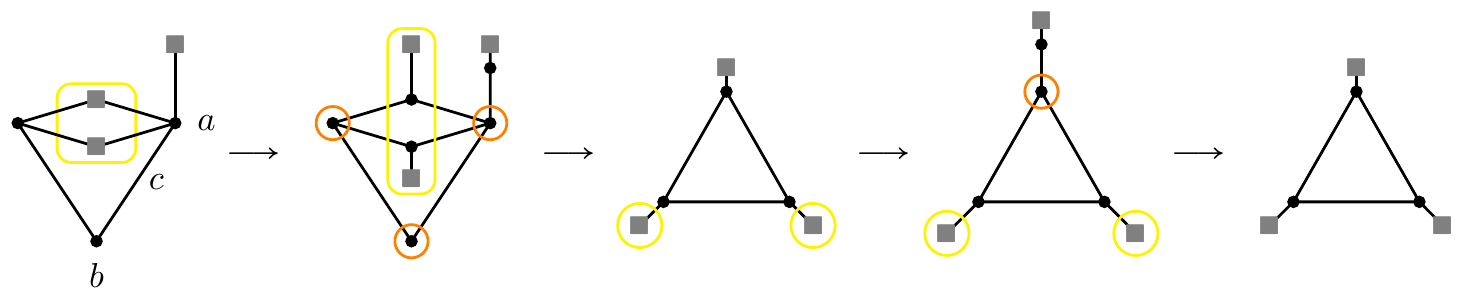}
	\esp
\end{align}
The first three graphs of \eqref{unlinkCS} are the same as that of \eqref{linkCS}, while the next two operations are different. The last triangle of \eqref{linkCS} is obtained by performing $(\0,\0, \1)$ $ST$-move on matter nodes and integrating out an undecorated gauge node. The mixed \CS levels for the last triangle are
\begin{align}\label{linkingCS}
	\textbf{ (0,0,1)} ~~\rightarrow~~
	k_{ij}^{\eff} (\bigtriangleup)=	
	\begin{bmatrix}
		r'  & k'+1  & r'+k'\\
		k'+1 & s' & s'+k'\\
		r'+k'	& s'+k'&r'+s' +2k' 
	\end{bmatrix} 
\end{align}
Assuming that $(r', s' , k')$ are integral, this triangle also has integral \CS levels too. Moreover, for specific values of  $(r', s' , k')$, there may be other triangles with integral \CS levels,  which can be easily identified by scanning all graphs in the orbit of $ST$-moves on the special triangle \eqref{linkingCS}.

To summarize, we have found another duality, which we refer to as linking:
\begin{align}\label{linkingrelation}
	\bsp 
\includegraphics[width=3in]{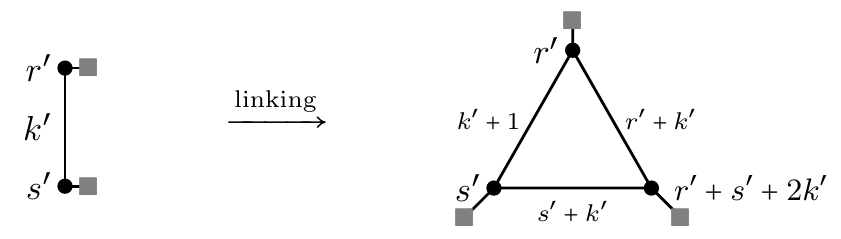}
	\esp\,.
\end{align}
As in the linking operation discussed in  \cite{Ekholm:2019lmb},  in our case the linking number $k'$ of the two-node graph also increases by one. The associated \CS levels  are related by
\begin{align}
	k_{ij}^{\eff}\left(- \right)  ~~\xrightarrow{\text{linking}}~~
	k_{ij}^{\eff} ({\bf{\bigtriangleup}}) ,
\end{align}
and the relation between FI parameters takes analogous form as in the unlinking case
\begin{align}
	(x_{r'}\,, x_{s'}) ~\longrightarrow~ (x_{r'}\,, x_{s'}\,, \q^{**} x_{r'} x_{s'}  ) ,
\end{align}
for some particular parameter $\q^{**}$.


\subsubsection*{Exotic cases}

Apart from linking and unlinking, we find also two other interesting operations that we call exotic,  which transform a two-node graph into a triangle with integral mixed \CS levels.  Let us perform $ST$-move $(\0, \1)$ only on one node of the two-node  graph:
\begin{align}\label{rskSTodd}
	\bsp
\includegraphics[]{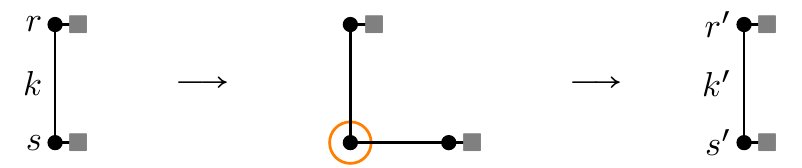}
	\esp
\end{align}
The resulting two-node graph has mixed \CS levels
\begin{align}
	\textbf{ (0,1)} ~~\rightarrow~~
	k_{ij}^{\eff}\left( - \right) =	\begin{bmatrix}
		r'  & k'  \\
		k'	 & s' \\
	\end{bmatrix} 
\end{align}
such that
\begin{align}
	r'=\frac{ rs -k^2}{s}\,,\quad k' =\frac{-k}{s}\,,\quad s'=\frac{s-1}{s} .
	\end{align}

Scanning dual triangle graphs,  we find that there are two triangle graphs with integral \CS levels that are obtained by $ST$-moves $\textbf{(2,0,1)} $ and $\textbf{(2,1,0)} $ respectively:	
\begin{align}\label{exotictransf}
	\bsp
&\textbf{(2,0,1)} \rightarrow	
	\begin{bmatrix}
		r' & k'-r' & k'-r'+1\\
	k'-r' &		r'+s'-2k'  &	r'+s'-2k'-1  \\
	k'-r'+1&			r'+s'-2k'-1 & 	r'+s'-2k'-1 \\
	\end{bmatrix}  \\
	&\textbf{(2,1,0)} \rightarrow	
		\begin{bmatrix}
	s' & k'-s'+1 & k'-s'\\
	k'-s'+1 &		r'+s'-2k'-1  &	r'+s'-2k'-1  \\
	k'-s'&			r'+s'-2k'-1 & 	r'+s'-2k' \\
\end{bmatrix} 
\esp
	\end{align}
The exotic operations that transform the rightmost graph in (\ref{rskSTodd}) into the graph captured by the first or the second matrix in (\ref{exotictransf}) we denote respectively $\textbf{E}_{r'}$ and $\textbf{E}_{s'}$.  These operations preserve only one of the original \CS levels,  respectively $r'$ or $s'$, associated with $U(1)_{r'}$ or $U(1)_{s'}$ gauge group.  Note that these two operations are related by an $ST$-move, as $\textbf{(2,0,1)}+\textbf{(0,1,2)}=\textbf{(2,1,0)}$.  Moreover, for some particular values of $(r,k,s)$ there exist other triangles with integral \CS levels, which can be obtained by further $ST$-moves on these two exotic triangles. To find such integral triangle graphs one can just scan theories in the orbit of $ST$-moves.  From the form of sphere partition functions,  we also find relations between FI parameters:
\begin{align}
	&\textbf{E}_{r'} :~~	(x_{r'}\,, x_{s'}) ~\longrightarrow~ \Big(x_{r'}\,, \q^{*} \frac{x_{s'}}{x_{r'}}\,, \q^{*} \frac{x_{s'}}{x_{r'}}\,	  \Big) \,,\\
	&\textbf{E}_{s'} :~~	(x_{r'}\,, x_{s'}) ~\longrightarrow~ \Big(x_{s'}\,, \q^{*} \frac{x_{r'}}{x_{s'}}\,, \q^{*} \frac{x_{r'}}{x_{s'}}\, \Big) \,.
\end{align}



\subsubsection*{Relations between triangle graphs}

To sum up, we have found a few theories dual to a two-node graph, which are represented by triangle graphs with integral mixed \CS levels.  We refer to these dualities respectively as unlinking, linking, and exotic, and in what follows we denote the resulting triangle graphs by $\textbf{A}$, $\textbf{B}$ and $\textbf{E}$:
\begin{align}\label{trianglesgraph}
 \includegraphics[width=6in]{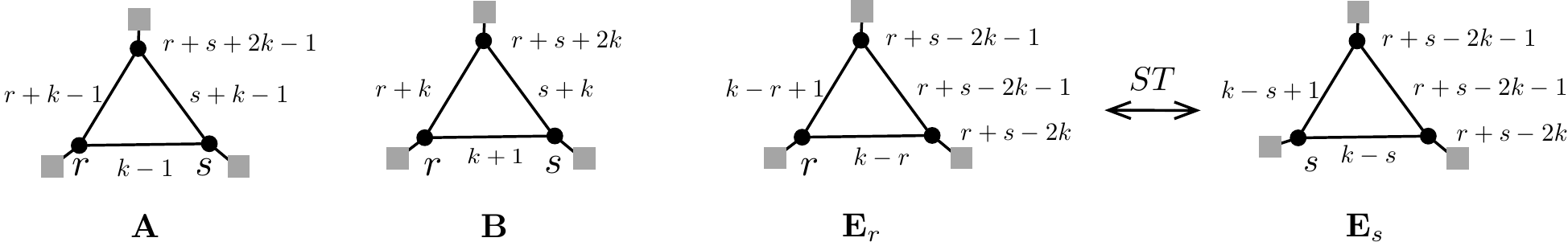}
\end{align}
As mentioned above, $\textbf{E}_r$ and $\textbf{E}_s$ are related by an $ST$-move, however other theories are not in the same orbit of $ST$-moves.  The relations between all these triangle graphs can be represented by a commutative diagram:
\begin{align}\label{trianglerelation} \includegraphics[width=1.5in]{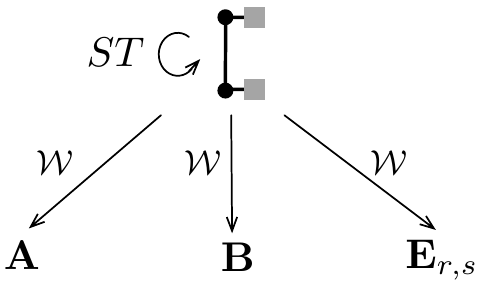}\end{align}
where $\W$ is the superpotential that is added during the gauged 2-3 move, so that the theories $\A$, $\B$ and $\E$ are be related by the process $\W \circ ST^{*}\circ \W^{-1}$.


Let us present the structure of operations in (\ref{trianglesgraph}) in more detail.  Note that for simplicity,  when we mention $ST$-moves in what follows,  we implicitly assume that Kirby moves that remove pure gauge nodes have been also performed.  First, the orbit of $ST$-moves for the two-node graph takes form
\begin{align}\label{orbitst}
	\bsp
	\includegraphics[width=2in]{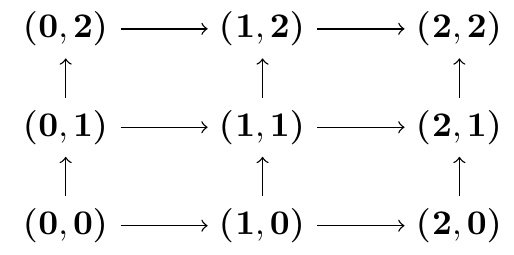}
	\esp
\end{align} 
where the $(\m,\n)$ denotes the $ST$-moves $((ST)^m,(ST)^n)$ that act on the two-node graph. Each element of this orbit gives rise to a triangle graph with integral \CS levels,  which encodes superpotential added during the 2-3 move.  Separating the above graph into layers,  we present the resulting dual triangle graphs:
\begin{align}\label{STWrelation}
	\bsp
\includegraphics[width=6in]{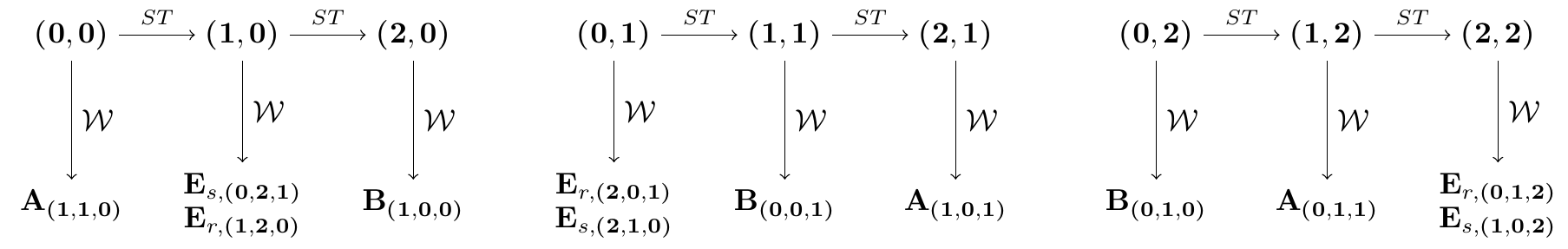}
	\esp
	\end{align} 
The indices $({\m,\n,\textbf{l}})$ of $\A$, $\B$, and $\E$ denote $ST$-moves $((ST)^m,(ST)^n, (ST)^l)$ that act on the matter nodes of three-node graphs; see \eqref{unlinkCS} and \eqref{linkCS} for examples.


\subsubsection*{Coupling external nodes}\label{fullgaugeexternal}

The gauged SQED-XYZ dual pair \eqref{mirrorpairplumb} is a local duality and it can be coupled to other nodes.  Let us discuss such a coupling to one extra external node; adding more external nodes is a trivial extension.  To begin with, adding one decorated node $U(1)_{c_i}+1\F$ extends the dual pair \eqref{mirrorpairplumb}  to \begin{align}\label{mirrorpairXYZ1}
	\bsp
	\includegraphics[width=3in]{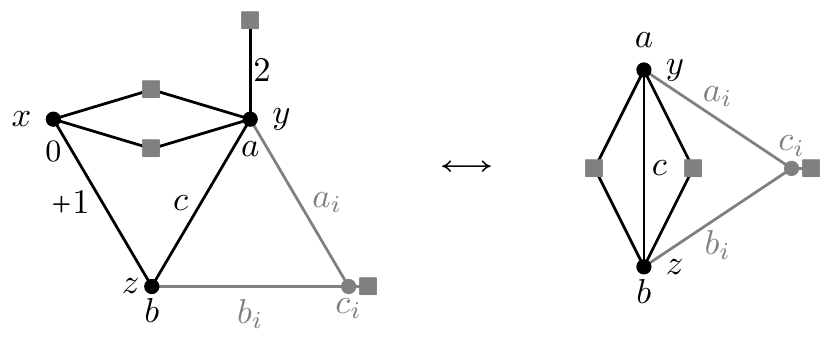}
	\esp
\end{align}
This dual pair can be transformed into the following dual pair using $ST$-moves and Kirby moves, where we use colorful lines to mark different connections:	
	\begin{align}\label{externaldual}	\includegraphics[width=3in]{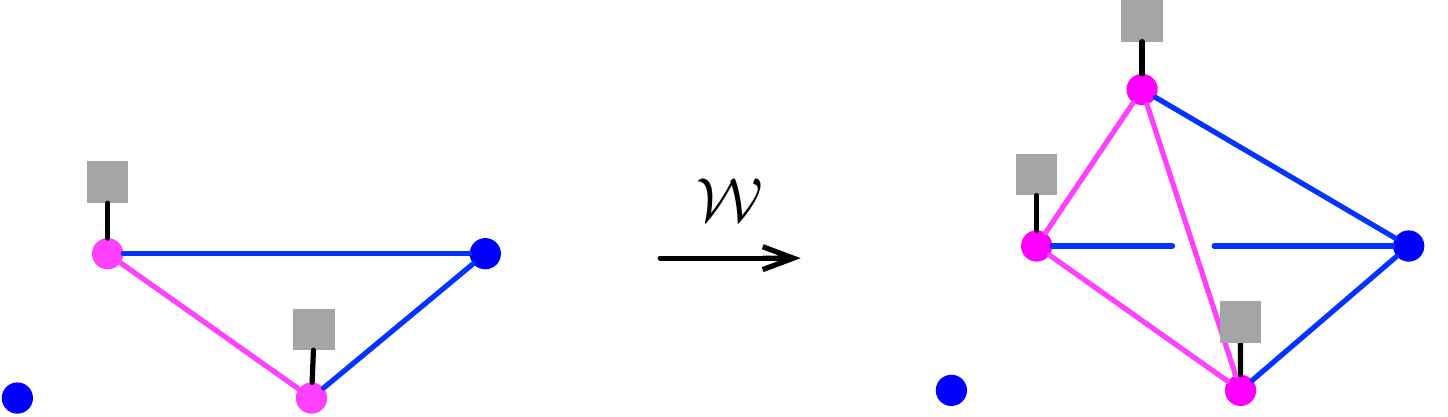} 
		\end{align}
If an external node is not coupled to the two gauge nodes (marked in red) involved in 2-3 move, then it will also not connect to the triangle graph (marked in red) encoding the superpotential.  Note that this gauged 2-3 move changes the couplings with the introduced gauge node in the dual superpotential triangle, however it does not change the couplings between the external node and the original two gauge nodes. More explicitly, for the unlinking we find
\begin{align}
	\begin{bmatrix}
		\begin{array}{c|cc}
		g_i & p_i & q_i \\ \hline
	p_i &		r  & k  \\
	q_i&		k & s \\
	\end{array}
		\end{bmatrix}  ~~\xrightarrow{\text{unlinking}}~~
	\begin{bmatrix}
		\begin{array}{c|ccc}
		g_i & p_i & q_i  & 	{ \color{blue}p_i+q_i}\\ \hline
	p_i&	r  & k-1  & 	{ r+k-1}\\
	q_i&	k-1 & s & 	{ s+k-1}\\
{	\color{blue} p_i+q_i } &		{r+k-1}	& {  s+k-1}&{ r+s +2k -1}
\end{array}
	\end{bmatrix} 
\,.
		\end{align}
Notice that there is a new line connecting the external node $U(1)_{c_i}$ and the introduced node.  The corresponding \CS level takes value $p_i+q_i$ and we denote it in blue.  Similarly,  linking leads to
 \begin{align}
 	\begin{bmatrix}
 				\begin{array}{c|cc}
 		g_i & p_i & q_i \\ \hline
 		p_i &		r  & k  \\
 		q_i&		k & s \\
 		\end{array}
 	\end{bmatrix}  ~~\xrightarrow{\text{linking}}~~
 	\begin{bmatrix}
 				\begin{array}{c|ccc}
 		g_i & p_i & q_i  & 	{ \color{blue}p_i+q_i}\\  \hline
 		p_i&	r  & k+1  & 	{ r+k}\\
 		q_i&	k+1 & s & 	{ s+k}\\
 		{	\color{blue} p_i+q_i } &		{r+k}	& {  s+k}&{ r+s +2k}
 		\end{array}
 	\end{bmatrix} \,.
 \end{align}
For exotic triangles we have 
	\begin{align}
		\begin{bmatrix}
			\begin{array}{c|cc}
				g_i & p_i & q_i \\ \hline
				p_i &		r  & k  \\
				q_i&		k & s \\
			\end{array}
		\end{bmatrix}  ~~\xrightarrow{\text{exotic-}r}~~
		\begin{bmatrix}
			\begin{array}{c|ccc}
				g_i & p_i & 		{	\color{blue} q_i-p_i } & 	{ \color{blue}q_i-p_i}\\ \hline
				p_i&	r  & k-r  & 	{ k-r+1}\\
					{	\color{blue} q_i-p_i }&	k-r & r+s-2k & 	{ r+s-2k-1}\\
				{	\color{blue} q_i-p_i } &		{k-r+1}	& {  r+s-2k-1}&{ r+s -2k -1}
			\end{array}
		\end{bmatrix} 
	\end{align}
	\begin{align}
	\begin{bmatrix}
		\begin{array}{c|cc}
			g_i & p_i & q_i \\ \hline
			p_i &		r  & k  \\
			q_i&		k & s \\
		\end{array}
	\end{bmatrix}  ~~\xrightarrow{\text{exotic-}s}~~
	\begin{bmatrix}
		\begin{array}{c|ccc}
			g_i & p_i & 		{	\color{blue} p_i-q_i } & 	{ \color{blue}p_i-q_i}\\ \hline
			p_i&	s  & k-s  & 	{ k-s+1}\\
			{	\color{blue} p_i-q_i }&	k-s & r+s-2k & 	{ r+s-2k-1}\\
			{	\color{blue} p_i-q_i } &		{k-s+1}	& {  r+s-2k-1}&{ r+s -2k -1}
		\end{array}
	\end{bmatrix} 
	\,.
\end{align}
We emphasize that $ST$-moves (and Kirby moves) performed in the analysis of linking, unlinking and exotic dual graphs,  should not affect the external node $U(1)_{c_i}$. 

Importantly,  the operations of linking and unlinking that involve extra nodes, are also consistent with general linking and unlinking operations in knots-quivers correspondence.


\subsubsection*{A case study: $(\textbf{2},\textbf{2})$ $ST$-move}

As a special case we discuss $(\textbf{2},\textbf{2})$-operation.  One reason to focus on this example is that in the following section we come accross a similar expression \eqref{LHShiggs} (which however has a different meaning).  The $(\textbf{2},\textbf{2})$ $ST$-move leads to the following graph:
	\begin{align}\label{acbspecial2}
		\bsp
		\includegraphics[width=2.in]{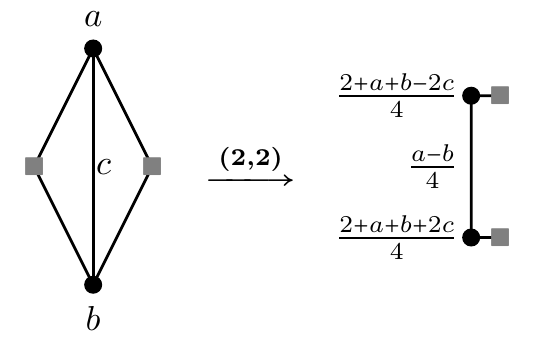}
		\esp
	\end{align}
Note that in this graph gauge nodes $U(1)_a$ and $U(1)_b$ cannot be exchanged as the charges for two bifundamental matter nodes are different, as shown in \eqref{mirrorpairplumb}.  The dual triangle graphs also have this property.  
The integral triangle graphs are obtained by $\bf(0,1,2)$ and $\bf(1,0,2)$ operations, so that only the top node in the left graph in \eqref{acbspecial2} is transformed, while the bottom one is not:
	\begin{align}\label{specialtriangle1}
	\bsp
	\includegraphics[width=3	in]{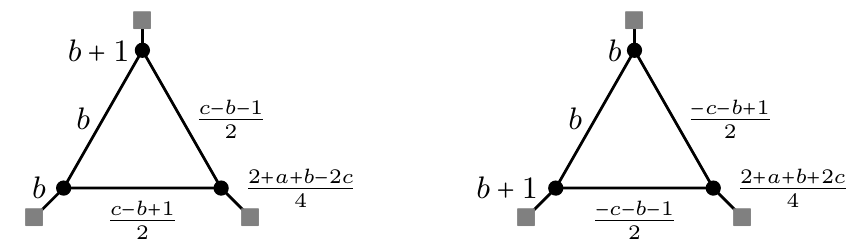}
	\esp
\end{align}
Effective mixed \CS levels for these graphs are
\begin{align}
k_{ij}^{\eff}=	\begin{bmatrix}
		\begin{array}{ccc}
b+1 & b & \frac{c-b-1}{2} \\ 
	b &		b  & \frac{c-b+1}{2} \\
	\frac{c-b-1}{2}&\frac{c-b+1}{2}		 & \frac{ 2+a+b-2c}{4} \\
	\end{array}
\end{bmatrix}
\,,\quad 
	\begin{bmatrix}
	\begin{array}{ccc}
		b & b & \frac{-c-b+1}{2} \\ 
		b &		b+1  & \frac{-c-b-1}{2} \\
		\frac{-c-b+1}{2}&\frac{-c-b-1}{2}		 & \frac{ 2+a+b+2c}{4} \\
	\end{array}
\end{bmatrix} 
	\end{align}
These two matrices only differ by the sign of $c$, and they are equal for $c=0$.  In fact,  these two triangle graphs are the exotic ones $\E_r$ and $\E_s$ that we discussed before, see \eqref{STWrelation}. To see this explicitly, we replace $(a,b,c)$ parameters by $(r,k,s)$ defined by
\begin{align} \label{exoticpara}
	a= r+s+2k -1 \,,~~ b= r+s -2k -1\,,~~ c = \pm(r-s) \,.
\end{align}	
This redefinition results in the follows \CS levels of the left graph in \eqref{acbspecial2}:
	\begin{align}\label{acbspecial2reverse}
	\bsp
	\includegraphics[width=2in]{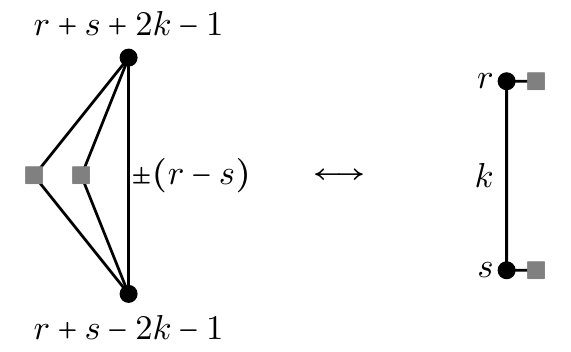}
	\esp
\end{align}
and also transforms triangles in \eqref{specialtriangle1} into exotic triangles that we discussed before.

Moreover,  in the presence of an external gauge node $U(1)_{c_i}$, the dual pair \eqref{mirrorpairXYZ1}
leads to \eqref{externaldual}, and the associated \CS levels are 
\begin{align}
	\begin{bmatrix}
	\begin{array}{c|cc}
		c_i & \frac{b_i-a_i }{2} &  \frac{-b_i-a_i }{2} \\ \hline
	 \frac{b_i-a_i }{2}&		\frac{2+a+b-2c }{4} & \frac{a-b}{4} \\
	 \frac{-b_i-a_i }{2}&		 \frac{a-b}{4} & 	\frac{2+a+b+2c }{4}  \\
	\end{array}
\end{bmatrix} 
 ~~\longleftrightarrow~~
	\begin{bmatrix}
		\begin{array}{c|ccc}
			c_i & \mp b_i & \mp b_i  & 	{	\color{blue}  \frac{\pm b_i- a_i}{2} }   \\ \hline
			\mp b_i&	b+1  & b  & \frac{ \pm c-b-1}{2 } \\
		\mp	b_i&	b & b & 	{ \frac{c-b+1}{2}}\\
			{	\color{blue}  \frac{\pm b_i-a_i}{2} } &		 \frac{\pm c-b-1}{2 } 	& {   \frac{\pm c-b+1}{2 } }&{ \frac{ 2+a+b\mp2c }{4} }
		\end{array}
	\end{bmatrix} 
\end{align}
Note that only when the parameters $a ,b, c, a_i, b_i$ are chosen properly, the dual pairs have integer \CS levels.



\subsection{Partial gauging} \label{partialgauge}

In the previous section we fully gauged the flavor symmetry group $U(1)_1\times U(1)_2$. In this section, we consider gauging of only its one subgroup $U(1)_1 \subset U(1)_1\times U(1)_2$.  In this case we also find that the presence of superpotential implies non-trivial conditions on mixed \CS levels associated to dual triangle graphs. Note that partial gauging of $U(1)_1$ can be also viewed as ungauging the $U(1)_2$ of the fully gauged $U(1)_1\times U(1)_2$.


\subsubsection*{Gauging $U(1)_z$}

In comparison with the full gauging \eqref{fullgauging}, let us consider first only gauging of the flavor symmetry for $z$:
 \begin{align}
	& \int \mathbf{gauge}(z):=\int dz  \, 	e^{ -\pi i\,   b\, z^2  + 2\pi i\, \xi_z z} \,.
\end{align}
Here the variable $y$ is viewed as a mass parameter $\langle y \rangle =m$. The associated sphere partition functions are given by adding this integral on both sides of \eqref{partSQEDXYZ}:
\begin{align}	\label{higgs1F1AF}
	\bsp
	\boxed{	s_b \le(  -\frac{i Q}{2}  + 2m\r) \int \textbf{gauge}(z)
		\int dx\, e^{ - 2 \pi i\, z x  }  	
		s_b\le( \frac{iQ}{2} \pm x -m \r)    
		=  \int \textbf{gauge}(z)	~ s_b(m\pm z)   }  \\
	\esp
\end{align}
The plumbing graphs resulting from the partial gauging are similar to the dual pair \eqref{mirrorpairplumb}, for which we only need to delete the gauge node for $y$ and lines connecting to it.  In this case the adjoint chiral multiplet becomes a free chiral multiplet and decouples, so that the plumbing graph pair can be viewed as the operation of moving a pair of $1\F$ and $1\AF$:
\begin{align}\label{higgsinggraph}
	\bsp
	\includegraphics[width=2.5in]{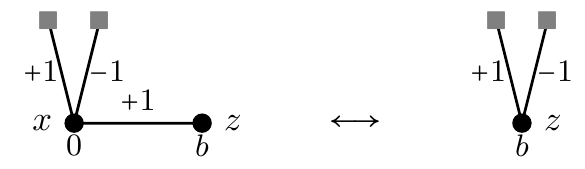}
	\esp
\end{align}
where for simplicity we do not draw the decoupled free chiral multiplet in the left graph.

The $ST$-moves of the graph on the right hand side (RHS) of \eqref{higgsinggraph} give rise to a chain (orbit) of theories with effective \CS levels:
\begin{align}\label{partialRHS}
\text{RHS:}~~ \left[	
		\begin{array}{cc}
			b & 1 \\
			1 & 0 \\
		\end{array}
\right] \,,\quad
\left[
\begin{array}{cc}
	b+1 & -1 \\
	-1 & 1 \\
\end{array} 
\right] \,,\quad
	\left[
\begin{array}{cc}
	\frac{b}{b+1} & \frac{1}{b+1} \\
	\frac{1}{b+1} & \frac{b}{b+1} \\
\end{array}
\right] \,,~~~~
\left[
\begin{array}{cc}
	\frac{b-1}{b} & -\frac{1}{b} \\
	-\frac{1}{b} & -\frac{1}{b} \\
\end{array}
\right] \,,~~~~
\left[
\begin{array}{cc}
	\frac{1}{1-b} & \frac{1}{b-1} \\
	\frac{1}{b-1} & \frac{1}{1-b} \\
\end{array}
\right]
	\end{align}
Integrating out the undecorated gauge node $U(1)_b$ of the graph on the left hand side (LHS) of \eqref{higgsinggraph} and then performing $ST$-moves give rise to a chain of theories with effective \CS levels:
\begin{align}
\text{LHS:}~~	\left[
	\begin{array}{cc}
		b+1 & b \\
		b & b \\
	\end{array}
	\right] \,,\quad
\left[
\begin{array}{cc}
	\frac{1}{1-b} & \frac{b}{b-1} \\
	\frac{b}{b-1} & \frac{1}{1-b} \\
\end{array}
\right] \,,\quad
\left[
\begin{array}{cc}
	\frac{b}{b+1} & -\frac{b}{b+1} \\
	-\frac{b}{b+1} & \frac{b}{b+1} \\
\end{array}
\right]\,,\quad
	\left[
	\begin{array}{cc}
		-\frac{1}{b} & 1 \\
		1 & 0 \\
	\end{array}
	\right] \,,\quad 
	\left[
	\begin{array}{cc}
		\frac{b-1}{b} & -1 \\
		-1 & 1 \\
	\end{array}
	\right]
	\end{align}

Since \eqref{higgsinggraph} is a dual pair, we obtain interesting dualities (which we also call 2-2 moves) by picking up integral matrices from the two chains above:
 \begin{align}\label{higgspair1}
 	\bsp
 		\text{RHS } \quad&\longleftrightarrow \quad \text{LHS}: \\	
	\left[	
 	\begin{array}{cc}
 		b & 1 \\
 		1 & 0 \\
 	\end{array}
 	\right] \,,\quad
 	\left[
 	\begin{array}{cc}
 		b+1 & -1 \\
 		-1 & 1 \\
 	\end{array} 
 	\right]  \quad &\longleftrightarrow \quad
 		\left[	
 	\begin{array}{cc}
 		b+1 & b \\
 		b & b \\
 	\end{array}
 	\right] \,.
 	\esp
 \end{align}
In principle, the LHS should be encoded in triangle graphs, whose missing node is a free chiral multiplet contributing $s_b\le( - iQ/2 +2 m\r)$, which does not couple to other gauge nodes, so that we can ignore it and the three-node graph reduces to a two-node graph (the graph for decoupled free node is shown in \eqref{freenode}). Up to this extra node,  one can check that the matrices on RHS can be obtained from those on LHS by operations of unlinking \eqref{unlinkingrelation} and linking \eqref{linkingrelation} ($\mathbf{A}$ and $\mathbf{B}$),  while matrices on LHS can be obtained from those on RHS by exotic dualities \eqref{exotictransf}. Therefore, these operations form an algebra:
\begin{align}
	~~ \E_r =\A^{-1}\,,~~\E_s =\B^{-1} \,, ~~~~\E_r = ST 	\circ \E_s  \,,~~\B = ST 	\circ \A  \,,
\end{align}
where $ST$-moves relate $\A$, $\B$ and $\E_{r,s}$, and hence they are equivalent $\A\sim \B\sim \E_{r,s}$. This feature does not appear when we fully gauge the flavor symmetry. This also implies that the superpotential does not exist for this 2-2 move in \eqref{higgsinggraph}; this is also confirmed by the equivalence between theories given by \eqref{higgsinggraph} and \eqref{SQEDXYZmove}, which do not carry superpotential.  Note that this argument may only be correct for abelian theories, as the adjoint matter does not decouple for non-abelian theories.

In addition to the above dual pairs, we can identify also other dual pairs,  if we choose other parametrizations:
\begin{align}\label{rtob1}
	r = \frac{b}{1+b} \,,~~-\frac{1}{b}\,,~~\frac{1}{1-b} \,.
	\end{align}
In this case the other three matrices in \eqref{partialRHS} give rise to further dual pairs, respectively:
\begin{align}\label{higgspair2}
	\bsp
	\text{RHS } ~~&\longleftrightarrow ~~\text{LHS}: \\	
\left[	
\begin{array}{cc}
r & 1-r \\
	1-r & r \\
\end{array}
\right] 
~~&\longleftrightarrow~~
\left[ 
\begin{array}{cc}
	r & -r \\
	-r & r \\
\end{array}
 \right]   \,,   \\ 
 \left[	
 \begin{array}{cc}
 	r+1 & r \\
 	r & r \\
 \end{array}
 \right] 
 ~~&\longleftrightarrow~~
 \left[ 
 \begin{array}{cc}
 	r+1 & -1 \\
 	-1 & 1 \\
 \end{array}
 \right]   \,, 
 ~~ \left[ 
 \begin{array}{cc}
 	r & 1 \\
 	1 & 0 \\
 \end{array}
 \right]  \,,  \\ 
 \left[	
 \begin{array}{cc}
 	r & -r \\
 	-r & r \\
 \end{array}
 \right] 
 ~~&\longleftrightarrow~~
 \left[ 
 \begin{array}{cc}
 	r & 1-r \\
 	1-r & r \\
 \end{array}
 \right]   \,.   
 \esp
	\end{align}
Ignoring the decoupled matter node,  the dual pairs in \eqref{higgspair2}, from RHS to LHS, are $\A, \E_{r,s}$ and $ \B $ respectively.  Note that the first and the third line in \eqref{higgspair2} are the same, which implies that in this case
$\B= \A^{-1}$.  On the other hand,  the second line is the same as \eqref{higgspair1}, which implies $\E_r=\A^{-1}, \E_s=\B, \E_r =ST\circ \E_s$. We emphasize that \eqref{higgspair2} includes all possible dual pairs in this partial gauging.


\subsubsection*{Gauging $U(1)_y$}

Analogously we can consider partial gauging for $U(1)_y$, keeping $U(1)_z$ as a flavor symmetry, so that  the parameter $z$ is viewed as a mass parameter, $\langle z\rangle =\tilde{m}$.  This operation is represented by the integral
\begin{align}
	& \int \mathbf{gauge}(y	):=\int dy \, 	e^{ -\pi i  a\, y^2  + 2\pi i\, \xi_y y} 
\end{align}
on both sides of \eqref{partSQEDXYZ}, so that the equivalence of sphere partition functions takes form
\begin{align}
	\bsp
	\boxed{ \int \textbf{gauge}(y)
		\int dx\, e^{ - 2 \pi i\, \tilde{m} x  }  		s_b \le(  -\frac{i Q}{2}  + 2y\r)
		s_b\le( \frac{iQ}{2} \pm x -y \r)    
		=  \int \textbf{gauge}(y)	~ s_b(y\pm \tilde{m})   } \\
	\esp
\end{align}
The corresponding plumbing graphs can be obtained by deleting the gauge node $U(1)_z$ and all lines connecting to this node:	
\begin{align}\label{higgsinggraph2}
	\bsp
	\includegraphics[width=3in]{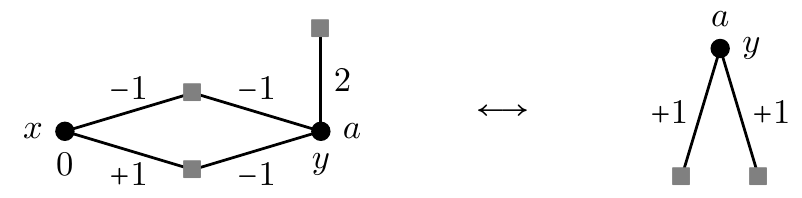}
	\esp
\end{align}
Interestingly,  in this partial gauging the adjoint matter does not decouple, and hence it is still a 2-3 move. The two-node graph is the same as the two-node graph in \eqref{higgsinggraph}, so we copy effective \CS levels for the orbit of $ST$-moves
\begin{align}\label{partical2rks}
	\bsp
&\text{RHS:} ~~	\left[	
	\begin{array}{cc}
		a+1 & 1 \\
		1 & 1 \\
	\end{array}
	\right] \,,~
	\left[
	\begin{array}{cc}
	a & -1 \\
		-1 & 0 \\
	\end{array} 
	\right] \,, ~	
	\left[
	\begin{array}{cc}
		\frac{a}{a+1} & -\frac{1}{a+1} \\
		-\frac{1}{a+1} & \frac{a}{a+1} \\
	\end{array}
	\right] \,,  ~
	\left[
	\begin{array}{cc}
		\frac{a-1}{a} & \frac{1}{a} \\
		\frac{1}{a} & -\frac{1}{a} \\
	\end{array}
	\right]\,, ~
	\left[
	\begin{array}{cc}
		\frac{1}{1-a} & \frac{1}{1-a} \\
		\frac{1}{1-a} & \frac{1}{1-a} \\
	\end{array}
	\right] \,.
	\esp
	\end{align}
The left graph (three-node graph) in (\ref{higgsinggraph2})  leads to an orbit of theories with the following effective \CS levels 
{
\begin{align}\label{partialtriangle}
	\bsp
	&\text{LHS:}\\
&
(1) : ~\left[
\begin{array}{ccc}
	a+1 & a+2 & 2 \\
	a+2 & a+4 & 2 \\
	2 & 2 & 1 \\
\end{array}
\right],\left[
\begin{array}{ccc}
	a-3 & a-2 & -2 \\
	a-2 & a & -2 \\
	-2 & -2 & 0 \\
\end{array}
\right], 
\left[
\begin{array}{ccc}
	\frac{1}{1-a} & \frac{a-2}{a-1} & -\frac{2}{a-1} \\
	\frac{a-2}{a-1} & \frac{1}{1-a} & -\frac{2}{a-1} \\
	-\frac{2}{a-1} & -\frac{2}{a-1} & -\frac{4}{a-1} \\
\end{array}
\right],
\left[
\begin{array}{ccc}
	\frac{a-4}{a} & \frac{2}{a}-1 & -\frac{4}{a} \\
	\frac{2}{a}-1 & \frac{a-1}{a} & \frac{2}{a} \\
	-\frac{4}{a} & \frac{2}{a} & -\frac{4}{a} \\
\end{array}
\right], \\
&
(2) : ~ \left[
\begin{array}{ccc}
	\frac{a}{a+1} & -\frac{a+2}{a+1} & -\frac{2}{a+1} \\
	-\frac{a+2}{a+1} & \frac{a}{a+1} & -\frac{2}{a+1} \\
	-\frac{2}{a+1} & -\frac{2}{a+1} & \frac{a-3}{a+1} \\
\end{array}
\right],
\left[
\begin{array}{ccc}
	-\frac{4}{a} & \frac{a+2}{a} & -\frac{4}{a} \\
	\frac{a+2}{a} & -\frac{1}{a} & \frac{2}{a} \\
	-\frac{4}{a} & \frac{2}{a} & \frac{a-4}{a} \\
\end{array}
\right],   \\
&
(3) : ~\left[
\begin{array}{ccc}
	-\frac{4}{a-4} & \frac{a-2}{a-4} & \frac{4}{a-4} \\
	\frac{a-2}{a-4} & \frac{1}{4-a} & -\frac{2}{a-4} \\
	\frac{4}{a-4} & -\frac{2}{a-4} & -\frac{4}{a-4} \\
\end{array}
\right],
	\left[
\begin{array}{ccc}
	-\frac{1}{a+3} & \frac{a+2}{a+3} & \frac{2}{a+3} \\
	\frac{a+2}{a+3} & -\frac{1}{a+3} & \frac{2}{a+3} \\
	\frac{2}{a+3} & \frac{2}{a+3} & \frac{a-1}{a+3} \\
\end{array}
\right],
\left[
\begin{array}{ccc}
	\frac{a}{a+4} & -\frac{a+2}{a+4} & \frac{4}{a+4} \\
	-\frac{a+2}{a+4} & \frac{a+3}{a+4} & -\frac{2}{a+4} \\
	\frac{4}{a+4} & -\frac{2}{a+4} & \frac{a}{a+4} \\
\end{array}
\right],
\left[
\begin{array}{ccc}
	\frac{a-4}{a-3} & \frac{2-a}{a-3} & \frac{2}{a-3} \\
	\frac{2-a}{a-3} & \frac{a-4}{a-3} & \frac{2}{a-3} \\
	\frac{2}{a-3} & \frac{2}{a-3} & -\frac{4}{a-3} \\
\end{array}
\right], \\
&
(4) : ~ \left[
\begin{array}{ccc}
	\frac{a+4}{4} & \frac{a+2}{4} & -1 \\
	\frac{a+2}{4} & \frac{a+4}{4} & -1 \\
	-1 & -1 & 1 \\
\end{array}
\right],
\left[
\begin{array}{ccc}
	\frac{a}{4} & \frac{a-2}{4} & 1 \\
	\frac{a-2}{4} & \frac{a}{4} & 1 \\
	1 & 1 & 0 \\
\end{array}
\right]\,.
\esp
	\end{align}
}

We can now pick up integral matrices from RHS and LHS above and form dual pairs; they are in fact identified as linking ($\mathbf{B}$) and unlinking ($\mathbf{A}$) respectively:
\begin{align}\label{partial2pair2}
	\bsp
		\text{RHS } ~~&\longleftrightarrow ~~\text{LHS}: \\	
	\left[	
	\begin{array}{cc}
		a+1 & 1 \\
		1 & 1 \\
	\end{array}
	\right] \quad 
	&\longleftrightarrow
	\quad 
		\left[
	\begin{array}{ccc}
		a+1 & a+2 & 2 \\
		a+2 & a+4 & 2 \\
		2 & 2 & 1 \\
	\end{array}
	\right] \,,   \\
	\left[
	\begin{array}{cc}
		a & -1 \\
		-1 & 0 \\
	\end{array} 
	\right] \quad
		&\longleftrightarrow
		\quad
		\left[
		\begin{array}{ccc}
			a-3 & a-2 & -2 \\
			a-2 & a & -2 \\
			-2 & -2 & 0 \\
		\end{array}
		\right] 
		\esp
\end{align}
In addition to these dual pairs, one can also introduce the following parameters to the last three matrices of the two-node graph (RHS) in \eqref{partical2rks} to make them formally integer 
\begin{align}\label{relation2} 
	a = -1 + \frac{1}{n} \,,\quad a=\frac{1}{n} \,,\quad a= 1+\frac{1}{n} \,,\quad \forall \, n \in \mathbf{Z}\,.
\end{align}
For $n=\pm 1$ we get $a= -2, 0, \pm1$, which are special since these values are allowed for the two-node graph in \eqref{higgsinggraph2}, which describes the theory $U(1)_a+1\F+1\AF$.	
We can also use another parameter $r=- n$ for these three matrices,  which yields
\begin{align} \label{rarelation2}
	r = \frac{a}{1+a} \,,~~ -\frac{1}{a}\,,~~\frac{1}{1-a}  ~~\in ~\mathbf{Z}  .
\end{align}
We rewrite the above $2\times 2$ matrices in terms of $r$ and find they are dual to $3 \times 3$ matrices in the first $(1)$ and second $(2)$ line of LHS in \eqref{partialtriangle}:
\begin{align}\label{partial2pair}
	\bsp
		\text{RHS } ~~&\longleftrightarrow ~~\text{LHS}: \\	
	\left[
	\begin{array}{cc}
		r & r-1 \\
		r-1 & r \\
	\end{array}
	\right]  ~~&\longleftrightarrow~~
	\left[
	\begin{array}{ccc}
		r & r-2 & 2 (r-1) \\
		r-2 & r & 2 (r-1) \\
		2 (r-1) & 2 (r-1) & 4 r-3 \\
	\end{array}
	\right] \,,\\
 \left[
 \begin{array}{cc}
 	r+1 & -r \\
 	-r & r \\
 \end{array}
 \right]  
 ~~&\longleftrightarrow~~
\left[
\begin{array}{ccc}
	r+1 & -2 r-1 & -2 r \\
	-2 r-1 & 4 r+1 & 4 r \\
	-2 r & 4 r & 4 r \\
\end{array}
\right]\,,~\left[
\begin{array}{ccc}
	r & 1-2 r & -2 r \\
	1-2 r & 4 r & 4 r \\
	-2 r & 4 r & 4 r+1 \\
\end{array}
\right]
\,,\\
 \left[
 \begin{array}{cc}
 	r & r \\
 	r & r \\
 \end{array}
 \right]
 ~~&\longleftrightarrow~~
\left[
\begin{array}{ccc}
	r & r+1 & 2 r \\
	r+1 & r & 2 r \\
	2 r & 2 r & 4 r \\
\end{array}
\right]\,,
\esp
	\end{align}
These dual pairs are identified respectively as $\A, \E,\B$ operations.


Furthermore,  preferred parameters for the third line $(3)$ of LHS are respectively:
\begin{align} \label{relation1}
	a= 4 +\frac{1}{n} 
	\,, 	\quad
	a= -3 +\frac{1}{n} 
	\,, 	\quad
	a= -4 +\frac{1}{n} 
	\,, 	\quad
	a= 3 +\frac{1}{n} \,, \quad \forall \, n \in \mathbf{Z} \,.
\end{align}
Rewriting the matrices in the third line in terms of these parameters does not give rise to literally integer matrices,  however they are in the same $ST$-orbit of $\A,\B,\E$ triangles (we do not show these matrices here).  
For $n=\pm 1$ in \eqref{relation1}, we get special values $a= \pm2, \pm3, \pm4,\pm5$, which are allowed for the theory $U(1)_a+1\F+1\AF$ in \eqref{higgsinggraph2}. 
There is only one overlapping value $a=-2$ for both \eqref{relation2} and \eqref{relation1}, so $a=-2$ should be a special value that leads to the maximal number of dual graphs with matrices:
\begin{align}
	\left[	
\begin{array}{cc}
	-1 & 1 \\
	1 & 1 \\
\end{array}
\right] \,,
	\left[	
\begin{array}{cc}
	-2 & -1 \\
	-1 & 0 \\
\end{array}
\right] \,,	
	\left[	
\begin{array}{cc}
	2 & 1 \\
	1 & 2 \\
\end{array}
\right] 
~\longleftrightarrow	~	\left[
	\begin{array}{ccc}
		-1 & 0 & 1 \\
		0 & -1 & 2 \\
		1 & 2 & -3 \\
	\end{array}
	\right] \,,
	\left[
	\begin{array}{ccc}
		2 & 0 & 2 \\
		0 & 2 & 2 \\
		2 & 2 & 5 \\
	\end{array}
	\right] \,,
	\left[
	\begin{array}{ccc}
		-1 & 0 & 2 \\
		0 & 2 & 2 \\
		2 & 2 & 1 \\
	\end{array}
	\right],\left[
	\begin{array}{ccc}
		-5 & -4 & -2 \\
		-4 & -2 & -2 \\
		-2 & -2 & 0 \\
	\end{array}
	\right] \,. \nn
\end{align}
Note that matrices in the fourth line $(4)$ of LHS of \eqref{partialtriangle}  cannot be integral for any $a$, so they are meaningless in this context.



\subsubsection*{External nodes}\label{partialexternalnode}

The 2-2 move and the 2-3 move that we found by partial gauging can be also coupled to external nodes.  Let us show that in the presence of external nodes, the 2-2 move arising from partial gauging is a special case of the results found by full gauging discussed in section \ref{fullgaugeexternal}.  For the 2-2 move, adding an external node $U(1)_{c_i}$ to \eqref{higgsinggraph} leads to
	\begin{align}\label{higgsinggeneric}
		\bsp
		\includegraphics[width=3.5in]{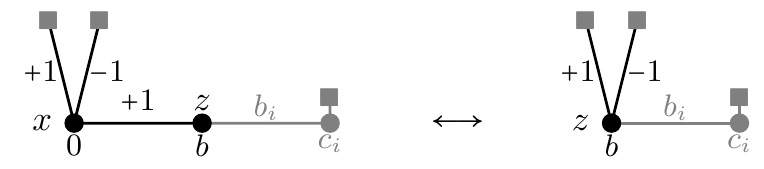}
		\esp
	\end{align}
Similarly to the full gauging, we do not need to apply a Kirby move to the external node $U(1)_{c_i}$. After scanning the orbit of theories given by $ST$-moves of the right graph in \eqref{higgsinggeneric}, we find many matrices with submatrices of the form \eqref{partialRHS}.   One subset of this orbit of theories is given by the following matrices
\begin{align}\label{RHShiggs}
	&\text{RHS:}\quad  \left[
	\begin{array}{c|cc}
		{c_i} & \pm {b_i} & 0 \\ \hline
	\pm	{b_i} & b+1 & -1 \\
		0 & -1 & 1 \\
	\end{array}
	\right] \,,~~\left[
	\begin{array}{c|cc}
		{c_i} & \pm{b_i} & 0 \\  \hline
		\pm {b_i} & b & 1 \\
		0 & 1 & 0 \\
	\end{array}
	\right]  
	\end{align} 
which are linear graphs given by $ST$-moving the $\AF$ using the triality \eqref{ststAF}. Moreover, since these four matrices are in the same orbit of $ST$-moves, it seems that $ST$-moves change the sign of $b_i$. We will discuss in the next section that this implies the possibility of Higgsing a pair of $1\F+1\AF$.

The graph on LHS of \eqref{higgsinggeneric} leads to four plumbing graphs in the orbit of $ST$-moves:
\begin{align}\label{LHShiggs}
	&\text{LHS:}\quad 
	\left[
	\begin{array}{c|cc}
		c_i & \pm b_i & \pm b_i \\ \hline
		\pm b_i & b+1 & b \\
		\pm b_i & b & b \\
	\end{array}
	\right]
\end{align} 
which also are triangle graphs. The RHS relates to LHS through unlinking and linking as well, up to an element coming from the decoupled adjoint chiral multiplet on the LHS. If this decoupled matter node is added, the LHS is encoded in $3\times 3$ matrices representing triangle graphs:
\begin{align}\label{generalhiggsingpari}
	\bsp
	 \left[
	\begin{array}{c|cc}
		{c_i} & \pm {b_i} & 0 \\ \hline
		\pm	{b_i} & b+1 & -1 \\
		0 & -1 & 1 \\
	\end{array}
	\right]  &~~ \xlongrightarrow{~\text{linking}~}~~
	\left[
	\begin{array}{c|c|cc}
		{c_i} &~0~ & \pm {b_i} &  \pm {b_i} \\ \hline
		0&1&0&0\\ \hline
		\pm	{b_i}&0 & b+1 & b \\
		 \pm {b_i}  &0& b & b \\
	\end{array}
	\right]  \,,   \\
	\left[
	\begin{array}{c|ccc}
		{c_i} & \pm {b_i} & ~0~ \\ \hline
		\pm	{b_i} & b & 1 \\
		0 & 1 & 0 \\
	\end{array}
	\right]   & ~~\xlongrightarrow{\text{unlinking}}~~
	\left[
	\begin{array}{c|c|cc}
		{c_i} &~0~ & \pm {b_i} &  \pm {b_i}  \\ \hline
		0&0&0&0\\ \hline
		\pm	{b_i}&0 & b+1 & b \\
		 \pm {b_i}  &0& b & b \\
	\end{array}
	\right]   \,.  
	\esp
\end{align}
These are therefore dual pairs between linear graphs and triangle graphs, generalizing the dual pairs \eqref{higgspair1} by coupling to an external node. One can also go from the right to the left of \eqref{generalhiggsingpari}, which would be identified with exotic operations; however, the element $1$ for the  decoupled chiral multiplet may be replaced equivalently by $0$ to satisfy the relations  for exotic triangles in \eqref{exotictransf}.

Let us discuss the generalization of the dual pair \eqref{higgspair2} to include external nodes. In this case,  both graphs in  \eqref{higgsinggeneric} can be transformed into linear or triangle graphs through $ST$- and Kirby moves:
\begin{align}\label{lineartriangle}
	\bsp
	\includegraphics[width=3in]{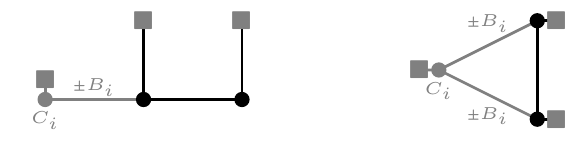}
	\esp
\end{align}
The corresponding dual pairs are represented by matrices, which in appropriate parametrization take form
\begin{align}
	\bsp
	\text{RHS} ~~&\longleftrightarrow~~ \text{LHS} \,,\\
	\left[
	\begin{array}{c|cc}
		{C_i} & B_i & -{B_i} \\ \hline
		{B_i} & r & 1-r \\
		-{B_i} & 1-r & r \\
	\end{array}
	\right]
	~~&\longleftrightarrow~~ 
	\left[
	\begin{array}{c|cc}
		C_i & B_i & -B_i \\  \hline
		B_i & r & -r \\
		-B_i & -r & r \\
	\end{array}
	\right]  \,,\\
	\left[
	\begin{array}{c|cc}
		C_i & \pm B_i & \pm B_i \\  \hline
		\pm B_i & r+1 & r \\
		\pm B_i & r & r \\
	\end{array}
	\right] 
		~~&\longleftrightarrow~~ 
		\left[
		\begin{array}{c|cc} 
			C_i & \pm B_i & 0 \\    \hline
			\pm B_i & r+1 & -1 \\
			0 & -1 & 1 \\
		\end{array}
		\right] \,,~~\left[
		\begin{array}{c|cc}
			C_i & \pm B_i & 0 \\    \hline
			\pm B_i & r & 1 \\
			0 & 1 & 0 \\
		\end{array}
		\right] \,,\\
		\left[
		\begin{array}{c|cc}
			C_i & B_i & -B_i \\  \hline
			B_i & r & -r \\
			-B_i & -r & r \\
		\end{array}
		\right]
			~~&\longleftrightarrow~~ 
			\left[
			\begin{array}{c|cc}
				C_i & B_i & -B_i \\  \hline
				B_i & r & 1-r \\
				-B_i & 1-r & r \\
			\end{array}
			\right] \,.
			\esp
	\end{align}
Note that the first and the third line are the same,  and represent a dual pair of triangle graphs. The second line represents a duality between a triangle and linear graphs.

Analogously one can discuss 2-3 move given by partial gauging the other flavor symmetry $U(1)_y$. In this case the results are a generalization of \eqref{partial2pair2} and \eqref{partial2pair}, as an interested reader can check.


\subsection{Higgsing}

Finally, we discuss Higgsing of matter fields, which happens when mass parameters are tuned to some specific values,  so that the multiplet $\F$ cancels with $\AF$
\[
\text{Higgsing:} \quad 1\F + 1\AF \rightarrow 1 
\]
In this case we can delete from the graph these two matter nodes and any lines connecting to them
\begin{align}
	\bsp
	\begin{tikzpicture}
		\draw[thick] (0,0)--(0.3,-1)node[midway,left	]{\small$+1$}--(0.6,0) node[midway,right	]{\small$-1$}  ;
		\node[rectangle,fill=gray] at (0,0) {};
		\node[rectangle,fill=gray] at (0.6,0) {};
		\node at (2,-0.5){$=~~\mathbf{1}$};
	\end{tikzpicture} \,.
	\esp
\end{align} 
The lesson one can learn from brane webs and vortex partition functions is that the Higgsing for 3d theories can be engineered by adjusting the mass parameters of these two matter fields: $m_{\F} =m_{\AF} \pm \epsilon_{-}$, where $\epsilon_- = \epsilon_1 -\epsilon_2$ and $\pm$ depends on the assignment of Omega deformation parameters $\epsilon_{1,2}$ on the spacetime of 3d theories $\mathbf{R}_{\epsilon_i}^2 \subset \mathbf{R}_{\epsilon_i}^2 \times S^1$, and then the identity $s_b( x)s_b(-x)=1$ leads to  the cancellation of these two matter fields. From the perspective of topological strings, this Higgsing is a manifestation of the geometric transition \cite{Dimofte:2010tz}.

We now illustrate that Higgsing is consistent with gauged dualities, by applying it to dual graphs and showing that they lead to the same reduced dual theories. We  consider partially gauged graphs in \eqref{higgsinggeneric}, which we copy here for convenience
\begin{align}\label{higgsinggenericcopy} 
	\bsp
	\includegraphics[width=3in]{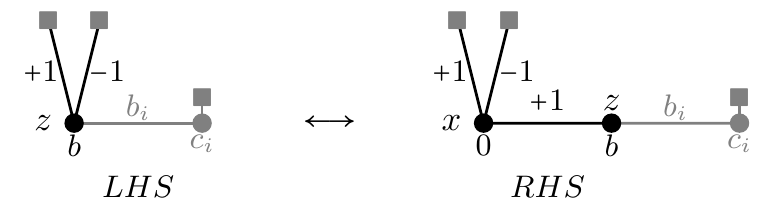}
	\esp
\end{align}
Note that $U(1)_{c_i}+1\F$ is the external node that we can add to the dual pair. The sphere partition functions for this dual pair can be easily written down by simply adding the contribution of the external node to both sides of the equivalence \eqref{higgs1F1AF}.  For $m=\frac{iQ}{2}$, Higgsing happens on the right graph (RHS), which surprisingly reduces to the external node $U(1)_{c_i}+1\F$, while the left graph (LHS) does not change. One can read off this conclusion from the sphere partition functions
\begin{align}\label{RHShiggsing}
	Z^{LHS} &=\int d z_i e^{-\pi i\, c_i z_i^2 - 2\pi i\, b_i z z_i }s_b\le(\frac{i Q}{2}-z_i \r)
	\int dz  \, 	e^{ -\pi\, i   b\, z^2 + 2 \pi i\, \xi_z z   }  ~s_b \le( \frac{i\,Q}{2} \pm z \r) \,,\\
Z^{RHS} &=	s_b \le( \frac{i\, Q}{2}\r) \int d z_i e^{-\pi i\, c_i z_i^2 - 2\pi i\, b_i z z_i }s_b\le(\frac{i Q}{2}-z_i \r)	\int dz  \, 	e^{ -\pi\, i   b\, z^2 +2 \pi i \, \xi_z z   } 	\int dx \, e^{ - 2 \pi i \, z x} s_b(\pm x)  
\nn \\
& =	s_b \le( \frac{i\, Q}{2}\r) \int d z_i e^{-\pi i\, c_i z_i^2  } s_b\le(\frac{i Q}{2}-z_i \r) \,, 
\end{align}
where integrating out $x$ gives a delta function $\delta(z)$ which imposes a constraint $z=0$. After this Higgsing, the two matter nodes in RHS graph are deleted and then we get two pure gauge nodes $U(1)_0 \times U(1)_b$. Using the second rule of Kirby moves for gauge nodes in \eqref{kirbyknown}, one can reduce the RHS graph to $U(1)_{c_i} +1\F$. Note that this reduction only happens when we Higgs these two matter nodes. If one does not turn on this special value $m=i Q/2$ by hand, then the two matter nodes cannot be deleted and the RHS graph does not reduce.  Since $Z^{LHS} =Z^{RHS}$, we are left with
\begin{align}	
	\int dz  \, 	e^{ -\pi\, i   b\, z^2 + 2 \pi i \xi_z z   }  ~s_b \le( \frac{i\,Q}{2} \pm z \r)  \times \text{(external node)}~ \sim ~s_b \le( \frac{i\, Q}{2}\r) \times \text{(external node)}	  \,.
\end{align}
The reduced dual pair arising from Higgsing matter nodes on the RHS graph is represented by the following graphs
\begin{align}\label{higgs22move}
	\bsp
	\includegraphics[width=2in]{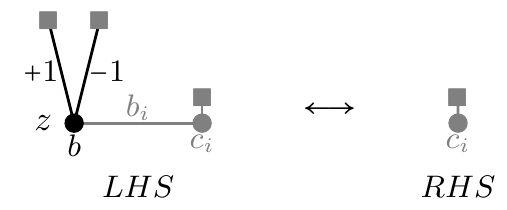}
	\esp
\end{align}
and it looks like a 3-1 move. Note that after Higgsing, the mass parameters for the $\F$ and $\AF$ on the LHS graph vanish, see \eqref{RHShiggsing}, but $\F$ does not cancel with $\AF$, so the LHS graph does not reduce. However, although the $\F$ and $\AF$ on the RHS graph have a small mass parameter $ iQ/2$, they cancel with each other, thereby reducing the graph. Because of the reduced dual pair \eqref{higgs22move}, although the sub-graph $U(1)_b+1\F+1\AF$ on LHS survives, one can just delete this part. From this perspective, when $m= i Q/2$, the LHS graph also gets reduced. In this process, the gauge node $U(1)_b$ is removed with these two matter fields $1\F+1\AF$. This phenomenon is similar to what happens when  Higging a bifundamental hypermultiplet in 5d brane webs, in which the Higgsing is interpreted as lifting a D5-brane  to infinity and hence the rank of the gauge group also reduces by one, see e.g. \cite{Dimofte:2010tz,Cheng:2021aa}.

Moreover,  we can also set $m=0$ to directly Higgs the $\F$ and $\AF$ on the left graph (LHS) of the dual pair \eqref{higgsinggenericcopy}. However,  in this case we find an undecorated gauge node $U(1)_z$ left, which needs to be integrated out to produce the same reduced dual pair \eqref{higgs22move}. The left and the right graph reduce in the following way:
\begin{align}\label{higgsingmzero} 
	\bsp
	\includegraphics[width=4in]{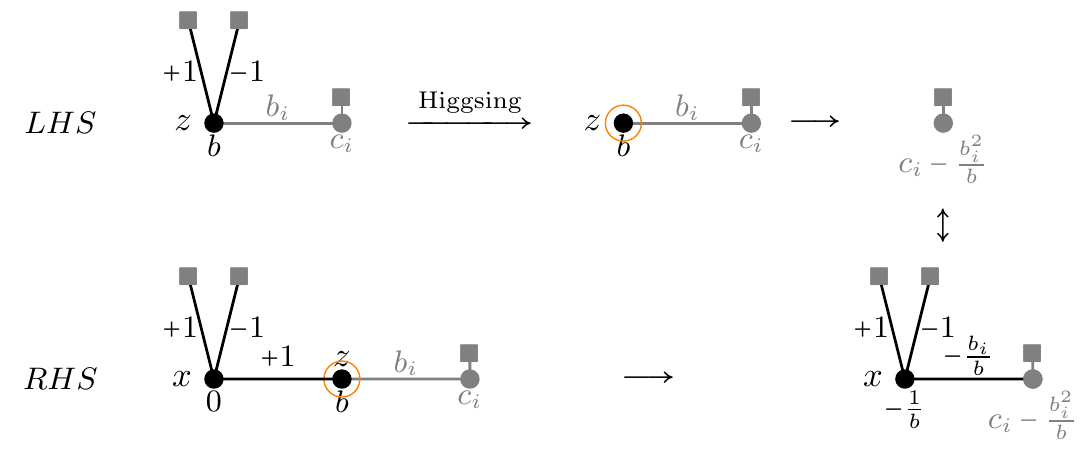}
	\esp
\end{align}
In the last step of the above operations, one ends up with the same dual pair in \eqref{higgs22move}, up to the shift of \CS levels.

We emphasize that mixed \CS levels can also indicate the Higgsing.  A hint is encoded in \eqref{RHShiggs} and \eqref{LHShiggs}, where the signs of $\pm b_i$ are flipped by $ST$-moves, which implies that lines with $\pm b_i$ could be cut and then the external node $U(1)_{c_i}$ decouples from plumbing graphs. In other words, the part $U(1)_b+1\F+1\AF$  does not really couple to the external node.   In addition, since generic Kirby moves and Higgsing can be applied commutatively on graphs in \eqref{higgsinggenericcopy} to get some linear or triangle graphs \eqref{lineartriangle} as we have discussed in section \ref{partialexternalnode}, lines $\pm B_i$ in these graphs can also be cut. 
Then the dual pair would be between a triangle and a one-node graph, which is also a 3-1 move, instead of \eqref{higgs22move}.

Apart from Higgsing, one can also consider decoupling of matter nodes from dual pairs. This should simply lead to the triality \eqref{ststF} that we discussed before. We leave this discussion as an exercise for interested readers.







\section{Exchange symmetry}\label{exchangesec}

In refined topological string theory and refined \CS theory \cite{Aganagic:2012hs,Cheng:2021aa} there is an exchange symmetry ${\q} \leftrightarrow \t^{-1}$, which in the unrefined limit reduces to $\q\leftrightarrow \q^{-1}$. The 3d theory is realized on a surface defect of 5d spacetime $\mathbf{R}^2 \times S^1 \subset \mathbf{R}_\q^2 \times \mathbf{R}^2_{\t^{-1}} \times S^1$. This exchange symmetry is equivalent to exchanging the surface defect from $\mathbf{R}^2_\q$ to $\mathbf{R}^2_{\t^{-1}}$. In other words, it exchanges $\bar{\q}$-brane and $\t$-brane. In the knot theory context,  this exchange symmetry is called mirror symmetry and it transposes Young diagrams that label colored knot polynomials.  Here we note that this exchange symmetry suggests a duality for abelian 3d theories,  which amounts to flipping the signs of bare \CS levels. We confirm this by considering vortex partition functions, or more completely holomorphic blocks of 3d theories \cite{Beem:2012mb,Yoshida:2014ssa,Dimofte:2017tpi,Cheng:2021vtq}.

For instance, the theory $U(1)_k+N_f \F +N_a \AF$ has the vortex partition function on a vacuum in Higgs branch:
\begin{align}
	Z^{\text{vortex}}( z, {\a},{\b}	) = \frac{  \prod_{j=1}^{N_f} ( \b_j;\q)_\inf }{ \prod_{i=1}^{N_a}  (\a_i ;\q)_\inf } \cdot \sum_{n=0}^{\inf} \le(  -\sqrt{\q}\r)^{k^{\eff} n^2 } z^n \cdot \frac{ \prod_{i=1}^{N_a}  (\a_i ;\q)_n }{  \prod_{j=1}^{N_f} ( \b_j;\q)_n }	 \,,
\end{align}
where contributions of $\AF$ and $\F$ are $(\a_i; \q)_n/(\a_i; \q)_\inf$ and $ (\b_j;\q)_\inf/(\b_j;\q)_n$ respectively, and $\a$ and $\b$ are associated with mass parameters $\a, \b \simeq e^{m}$.
Using the identity
$$
( \a ;\q)_n = ( \a^{-1};\q^{-1})_n \le(- \sqrt{\q^{-1}}\r)^{-n^2} \le( \sqrt{\q^{-1}} \a \r)^n \,,$$
one can derive the vortex partition function of a dual theory given by the exchange symmetry:
\begin{align}\label{exchangeU1}
	Z^{\text{vortex}}(z, \a,\b) ~\xrightarrow{ \q \,\rightarrow\, \q^{-1}}~
	Z^{\text{vortex}}( \tilde{z}, \a^{-1},\b^{-1})  \,,
\end{align}
where the vortex partition function and the effective \CS level for the dual theory are\footnote{Note that we have a normalization factor $\eta$, which can be ignored using  \eqref{normal}. For the definition of $q$-Pochharmmer symbols and theta functions, see the appendix \ref{pochappendix}.}
\begin{align}\label{thetaappear}
	\bsp
	&Z^{\text{vortex}}(\tilde{z}, \a^{-1},\b^{-1})= \eta \cdot 	\frac{  \prod_{j=1}^{N_f} ( \b_j^{-1};\q)_\inf }{ \prod_{i=1}^{N_a}  (\a_i^{-1} ;\q)_\inf } \cdot \sum_{n=0}^{\inf} \le(  -\sqrt{\q}\r)^{ \widetilde{k}^{\eff} n^2 } \ {\tilde{z}}^n \cdot \frac{ \prod_{i=1}^{N_a}  (\a_i^{-1} ;\q)_n }{  \prod_{j=1}^{N_f} ( \b_j^{-1};\q)_n }	 \,, \\
	&	 \widetilde{k}^{\eff} = -k^{\eff}-N_a +N_f =-\le(k +\frac{N_f-N_a}{2} \r) - N_a+N_f= -k +\frac{N_f-N_a}{2} \,,\\
	& \widetilde{z} =	z \cdot \frac{ \prod\limits_{i=1}^{N_a}\sqrt{\q} \a_i  }{ \prod\limits_{j=1}^{N_f} \sqrt{\q} \b_j  }  \,,\quad  \eta =\frac{ \prod\limits_{i=1}^{N_a} \theta( - \q^{1/2} \a_i )  }{ \prod\limits_{i=1}^{N_f} \theta( - \q^{1/2} \b_i )  } \,.
	\esp
\end{align}	
The term $\eta$ contains theta functions, which arise from exchanging boundary conditions for all chiral multiplets \cite{Dimofte:2017tpi,Yoshida:2014ssa,Gadde_2014,Cheng:2021vtq}. Note that 3d theories should couple to 2d $(0,2)$ theories on their boundaries to cancel anomalies. These theta functions are contributions of 2d multiplets. Adding some theta functions implies changing the matter content of the 2d theory on the boundary. Here we see that the exchange symmetry is sensitive to the boundary conditions, so considering holomorphic blocks would be more appropriate, however the physical argument would be the same.
In short, the equivalence in \eqref{exchangeU1} suggests a dual pair
\begin{align}\label{dualpariu1k}
	\bsp
	U(1)_k + N_f \F+N_a\AF &~~\longleftrightarrow~~	U(1)_{-k} + N_f \F+N_a\AF \\
	( k\,, m_i) &~~\longleftrightarrow~~	( -k\,, -m_i)
	\esp
\end{align}
In particular,  for self-dual theories \CS level should vanish, $k=0$.

For balanced plumbing theories $(U(1)+1\F)_{k_{ij}}^{N_f}$, namely the theories that contain only decorated gauge nodes, the exchange symmetry also  suggests a dual pair.  This can be deduced from the form of the vortex partition function on a Higgs vacuum:
\begin{align}
	Z^{\text{vortex}}( {z,\b}) =\prod_{j=1}^{N_f} ( \b_j;\q)_\inf \cdot \sum_{n_1\,,\cdots\,, n_{N_f} =0 }^{\inf} ~ \le( -\sqrt{\q}\r)^{\sum\limits_{i,j=1}^{N_f} k^{\eff}_{ij} n_i n_j } \cdot \prod_{i=1}^{N_f} \frac{ z_i^{n_i} }{ (\b_i;\q)_{n_i} } \,,
\end{align}
where $z_i$ is the FI parameter for each gauge node. In \cite{Cheng:2020aa}, it is shown that theories of this type become massless if one absorbs mass parameters into FI parameters, and then their contributions reduce from $(\b_j;\q)^{-1}_n$ to $(\q;\q)^{-1}_n$. The exchange symmetry also leads to a dual pair by applying $\q \leftrightarrow \q^{-1}$ on this vortex partition function:
\begin{align}\label{dualpairSTthry}
	\bsp
	(U(1)+1\F)_{k_{ij}}^{N_f}  ~~&\longleftrightarrow~~	(U(1)+1\F)_{-k_{ij}}^{N_f}  \\
	( \,k_{ij} \,, ~z_i)~~\,&\longleftrightarrow~~ (-k_{ij}  \,,~ \tilde{z}_i ) 
	\esp
\end{align}
The effective \CS levels are $k^{\eff}_{ij}= k_{ij} + \frac{1}{2} \delta_{ij}$. 
The effective \CS levels for the dual pair are related by
\begin{align}\label{mixedCSexchange}
	~k_{ij}^{\eff}~~&\longleftrightarrow~~ -k_{ij}^{\eff} +\delta_{ij} \,.
\end{align}
This duality can be confirmed by performing generic Kirby moves on plumbing theories, as we consider in examples in section \ref{secexample}. 
In particular, when $k_{ij}=0$, the theory is self-dual and trivially coupled, as it consists of only some free theories without interactions $(U(1)_0+1\F )^{ \otimes N_f}$.

This exchange symmetry has non-trivial implications. For example, integrating out the undecorated gauge node $U(1)_{k\pm1/2}$ of the gauged triality \eqref{blowupdownmatter} leads to $U(1)_{\frac{ \mp 3+2k }{2\pm 4 k }} +\F$, which should be equivalent to $U(1)_{-k}+1\F$, as suggested by the exchange symmetry. Then the equivalence ${\frac{ \mp 3+2k }{2\pm 4 k }} =-k$ leads to solutions:
\begin{align}\label{specialk1F}
	k = \pm \frac{1}{2} \,, ~\pm \frac{3}{2} .
\end{align}
In this way, these values are selected by exchange symmetry. Perhaps, they are special because the theories with these values can be related by $\mathcal{S}$-duality of type IIB string theory from the perspective of brane webs, which we discuss in  section \ref{Sduality}. Because of the factor of $\eta$ that involves theta functions, the exchange symmetry and $\mathcal{S}$-duality do not leave vortex partition functions invariant. This again confirms that vortex partition functions are not complete and it is more appropriate to consider holomorphic blocks, which include contributions from the 2d boundary theory.


\section{Effective superpotential}   \label{seceffective}

In this section we briefly discuss the structure of Coulomb branches of moduli spaces of plumbing theories, which are given by effective superpotentials.

The effective superpotential is the leading term of the vortex partition function $e^{\frac{\Weff}{\hbar}+\ldots }$.  It can be also obtained by taking the classical limit $\q = e^{\hbar} \rightarrow 1$ of a holomorphic block.	 For balanced theories $\left( U(1)+1\F\right)_{k_{ij}}^{N_f}$, the effective superpotential
takes form\footnote{If we replace $Y_i= \log\, y_i$, then twisted effective effective superpotential is written as
	$	\widetilde{\W}^{\eff}(\mathbf{y}, \mathbf{\xi})= \sum_{i=1}^{N_f} \le(  \Li_2(y_i) + \xi^{\eff}_i \log \,y_i  \r) + \sum_{i,j=1}^{N_f} \frac{ k_{ij}^{\eff}}{2} \,\log \,y_i \,\log \,y_j   $.	}
\begin{align}
	\widetilde{\W}^{\eff}(\mathbf{Y}, \mathbf{\xi})= \sum_{i=1}^{N_f} \le(  \Li_2(e^{Y_i +m_i}) + \xi^{\eff}_i Y_i  \r) + \sum_{i,j=1}^{N_f} \frac{ k_{ij}^{\eff}}{2} \, Y_i Y_j \,,	
\end{align}
where $Y_i$ is the variable for the $i$-th gauge node $U(1)_{k_i}$,  the term ${\frac{1}{2} k_{ij } Y_iY_j}$ comes from the \CS coupling between $i$-th and $j$-th gauge node, and $m_i$ is the real mass parameter for the $i$-th chiral multiplet.  Each chiral multiplet contributes a dilogarithm term ${\Li_2\le(e^{Y_i+m_i}\r)}$. In \cite{Nekrasov:2002qd,Shadchin_2007} the vacuum equations are found to take form 
\begin{align}
	\exp\left( \frac{d\,\Weff}{d\, Y_i} \right) =1\,,\quad \text{for}~\forall ~i = 1 \,, \cdots\,,N_f ,
	\end{align}
or more explicitly:
\begin{align}\label{vacuaeqn}
\sum_{j=1}^{N_f} k_{ij}^{\eff} \,Y_j + \xi_i^{\eff} =\log \left( 1- e^{Y_i +m_i}  \right)
\,,\quad \text{for}~\forall ~i = 1 \,, \cdots\,,N_f \,.
\end{align}
The terms on the right hand side come from  chiral multiplets as $\frac{d \, \Li_2(e^{Y_i+m_i})}{d\,Y_i}  =-\log\left( 1- e^{Y_i+m_i}\right)$, so we can ignore these terms when $m_i$ are very large, and then the moduli space of vacua becomes the Coulomb branch discussed in \cite{Dorey:1999rb}, in which the vacua are given by the scalar potential.

In \cite{Cheng:2021vtq}, the  $ST$-moves (transformations) of holomorphic blocks have been discussed. $ST$-moves could be represented as some operations on effective superpotentials. By looking at the gauged mirror triality \eqref{ststF}, one can see how
$ST$-moves act on effective superpotentials.  $ST$-moves only transform terms representing matter fields:
\begin{align}\label{STformatter}
(ST)^{\pm}:~~\Li_2\left(e^{Y}\right)  ~\rightarrow~ \Li_2\left(e^{Z}\right) \pm \frac{Z^2}{2} \pm Y Z  \,,
\end{align}
where the sign $\pm$ corresponds respectively to $ST$-move and $(ST)^{-1}$-move (or equivalently $(ST)^2$-move).  The same conclusion follows from analysis of sphere partition functions,  upon taking advantage of identities for double-sine functions from section \ref{gaugetri}.

If the FI parameter $\xi$ for $Y$ does not vanish, then one should shift $Z\rightarrow Z \mp \xi$ to get
\begin{align}
(ST)^{\pm}:~~	\Li_2\left(e^{Y+m}\right) +\xi Y ~ \rightarrow ~\Li_2\left(e^{Z \mp \xi}\right) \pm \frac{Z^2}{2} \pm Y Z  +\left(  -\xi \pm m\right) Z
	\pm \frac{\xi^2}{2}  \,.
\end{align}
Hence the FI parameter $\xi$ becomes the mass parameter and also contributes to the FI parameter for $Z$. The last term can be ignored as it is not dynamical. Note that the sign of FI parameter $\xi$ is  flipped by  $ST$-moves, which is consistent with decoupling of matter nodes that we discussed in section \ref{decouplematter}. The flip is interesting as it can be illustrated using 3d brane webs \cite{Cheng:2021vtq}.
Moreover, the property
\begin{align}
	 \Li_2\left(e^{X} \right)+ \Li_2\left(e^{-X} \right)= -\frac{X^2}{2} -\frac{\pi^2}{6} 
	 \end{align}
 exchanges $\F$ and $\AF$, and hence one obtains the ST-moves \eqref{ststAF} for anti-fundamental matter. Using the relation $	\Li_2\le(e^{X} \r)  =	
~\pm \frac{X^2}{2}$ for $X  \rightarrow \pm \inf $, one can also see how to decouple heavy chiral multiplets in effective superpotentials.

We emphasize that dual theories should have the same propotential that encodes BPS invariants. 3d propotential can be obtained by integrating over all gauge nodes:
\begin{align}\label{OVformula}
\exp\le(	\frac{\W^{3d} ( \mathbf{\xi})}{\hbar}  \r)=\int  \prod_i d Y_i \, \exp{\left(\frac{\Weff} {\hbar } \right) } \,,
\end{align}
which is also equivalent to the prepotential of open topological string partition functions. This prepotential encodes Ooguri-Vafa invariants $N^{OV}_{\mathcal{C}}$ of open topological string theories that engineer 3d theories. Since the parameters for gauge groups are $Y_i$, all undecorated gauge nodes can be integrated out,  and only decorated gauge nodes are left, as it is hard to integrate out the term $\Li_2(e^{Y})$ from matter fields. It is obvious from \eqref{OVformula} that theories related by Kirby moves,  which only differ by some undecorated gauge nodes, have the same prepotential. Hence theories related by Kirby moves have the same contracted Ooguri-Vafa invariants, which are given by summing over all spin indices $N^{OV}_{\mathcal{C}} = \sum_{j,r \in \mathbf{Z}/2} (-1)^{2j +2 r} N_{\mathcal{C}}^{(j,r)}$.  For more details see \cite{Aganagic:2000gs,Cheng:2020aa}. Moreover,  in terms of the Ooguri-Vafa prepotential $\W^{3d}$, the $ST$-moves on matter fields \eqref{STformatter} can be viewed as introducing Lagrangian multipliers.


\section{Examples: $U(1)_k+N_f\F$ theories}  \label{secexample}

In this section we illustrate our considerations in more involved examples of $U(1)_k+N_f\F$ theories. The theory with $N_f=1$ has been discussed in section \ref{gaugetri}, so in this section we analyze in more detail theories with $N_f=2,3$.
	

\subsection{$U(1)_k + 2\F$ }

A plumbing graph of this theory consists of two matter nodes connected to a gauge node. Performing $ST$-moves on matter nodes leads to a chain of theories:
\begin{align}\label{U1_2Fplumb}
	\begin{split}
	\includegraphics[]{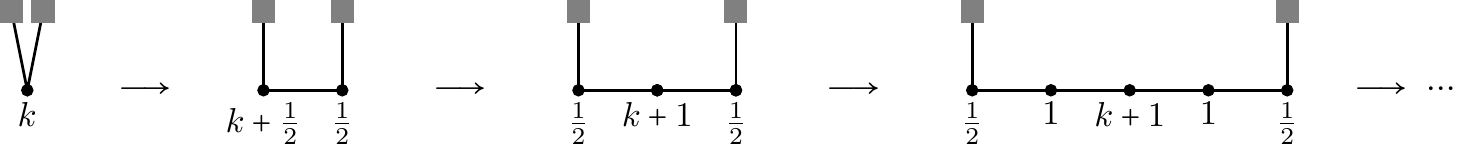}
	\end{split}
\end{align}
Although this chain contains infinitely many graphs, integrating out undecorated gauge nodes leads to several two-node graphs with different mixed \CS levels:
\begin{align}\label{2Fmatrix}
k_{ij}^{\eff} ~ =~
\begin{bmatrix}
	k+1 &~ 1 \\
	~1~& ~1 
\end{bmatrix} \,,\quad
\begin{bmatrix}
	k & -1 \\
	-1& 0 
\end{bmatrix} \,,\quad
	\begin{bmatrix}
		\frac{k}{1+k} &	-\frac{1}{1+k} \\
				-\frac{1}{1+k} & 	\frac{k}{1+k}  
	\end{bmatrix} \,,
\quad 
	\begin{bmatrix}
	\frac{k-1}{k} &~	\frac{1}{k} \\
	~	\frac{1}{k} & 	-\frac{1}{k}  
\end{bmatrix} \,, \quad 
	\begin{bmatrix}
	\frac{1}{1-k} &	\frac{1}{1-k} \\
	\frac{1}{1-k} & 	\frac{1}{1-k}  
\end{bmatrix} \,.
	\end{align}
These theories are related by $ST$-moves and hence dual to each other.  To make the last three matrices integral, $k$ needs to satisfy the following conditions respectively:
\begin{align}
	\bsp
	k = \frac{n+1}{n} \,, ~\frac{1}{n}\,, ~\frac{n-1}{n} \,, ~~\text{where}~n\in \mathbf{Z} \setminus  \{ 0\}  \,.
	\esp
	\end{align}
For the original theory $U(1)_k+2\F$, the condition is simply $k \in \mathbf{Z}$. Imposing these two conditions, the common solutions are $k= 0, \pm 1, \pm 2$. More explicitly, for $k \in \mathbf{Z}$ the first two plumbing graphs in \eqref{2Fmatrix} have obviously integral \CS levels, however other matrices may not be integral. However, for $k= \pm 1\,, \pm2$ there is a maximal numbers of equivalent graphs labeled by integral matrices:
\begin{align}
	\bsp	
& k=1 \,,\quad ~~k^{\eff}=
\begin{bmatrix}
~0 &~ ~	1 \\
~1& -1 ~
\end{bmatrix} \,,~~
\begin{bmatrix}
	~2 &~ 1~ \\
	~1~& ~1 ~
\end{bmatrix} 
\,,~~
\begin{bmatrix}
	~1 &-1~ \\
	-1~& ~0	 ~
\end{bmatrix} \,,\\
& k=-1 \,,\quad k^{\eff}=
\begin{bmatrix}
	2 & -1 \\
	-1& 1
\end{bmatrix}	
\,,~~
\begin{bmatrix}
	~0~ &~ 1~ \\
	~1~& ~1 ~
\end{bmatrix} \,,~~
\begin{bmatrix}
	-1&-1~ \\
	-1~& ~0~
\end{bmatrix} \,,
\\
	& k=2 \,,\quad~~ k^{\eff}=
\begin{bmatrix}
	-1 & -1 \\
	-1& -1 
\end{bmatrix} \,,~~
\begin{bmatrix}
	~3 &~ 1~ \\
	~1~& ~1 ~
\end{bmatrix} 
\,,~~
\begin{bmatrix}
	~2 &- 1~ \\
	-1~& ~0 ~
\end{bmatrix} \,,
\\  
 &k=-2 \,,\quad 		k^{\eff}=
\begin{bmatrix}
	~2 ~& ~1~ \\
	~1~& ~2~ 
\end{bmatrix} 
\,,~~
\begin{bmatrix}
	~1~ &~ 1~ \\
	~1~& -1 
\end{bmatrix} 
\,,~~
\begin{bmatrix}
	-2 &- 1~ \\
	-1~& ~0 ~
\end{bmatrix} \,.	
\esp
	\end{align}
For other values of $k$ only the first two matrices are integral and thus meaningful:
\begin{align}
		& k=0 \,, \pm3\,,\pm4\,,\cdots\,, 	\quad
		k^{\eff}=
			\begin{bmatrix}
			k+1 &~ 1~ \\
			~1~& ~1~ 
		\end{bmatrix} \,,
		 \quad 
	\begin{bmatrix}
		k & -1 \\
		-1& 0 
	\end{bmatrix}
	\end{align}
and corresponding plumbing graphs are 
\begin{align}\label{U1_2Fpm}
\begin{split}
	\includegraphics[]{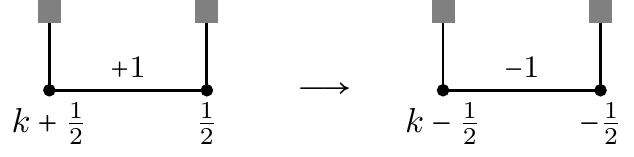}
\end{split}
\end{align}
For $k = -2\,, 0 \,, 2$, plumbing graphs are symmetric: 
\begin{align}
	k^{\eff}=
	\begin{bmatrix}
		2 &~ 1~ \\
		~1~& ~2~ 
	\end{bmatrix} \,,
	\quad 
	\begin{bmatrix}
		1~ &1~ \\
		1~& 1~ 
	\end{bmatrix}\,  ~\text{and}
~
	\begin{bmatrix}
	0 &-1 \\
	-1& 0 
\end{bmatrix} \,,
\quad
	\begin{bmatrix}
	-1 &-1 \\
	-1& -1 
\end{bmatrix}
	\,.
\end{align}	
We will reconsider these special values $k=0,\pm1, \pm 2$ from the perspective of brane webs in section \ref{secbranewebs}. 

It is easy to check that matrices of \CS levels in this example satisfy relations \eqref{dualpariu1k} and \eqref{dualpairSTthry} arising from the exchange symmetry $\q \leftrightarrow \q^{-1}$. Moreover, since graphs for these theories have two nodes, we can couple them to one additional decorated gauge node to get many dual triangle graphs satisfying unlinking, linking and exotic relations \eqref{trianglesgraph}. It would be nice to generate in this way infinitely many equivalent graphs.


\subsection{$U(1)_k + 3\F$}

Theories with $N_f=3$ have different properties than for $N_f=2$.  We now provide their classification based on \CS level $k$ and some properties of plumbing graphs. The $ST$-moves of this theory lead to the orbit:
\begin{align}\label{3fgraph}
	\begin{split}
		\includegraphics[]{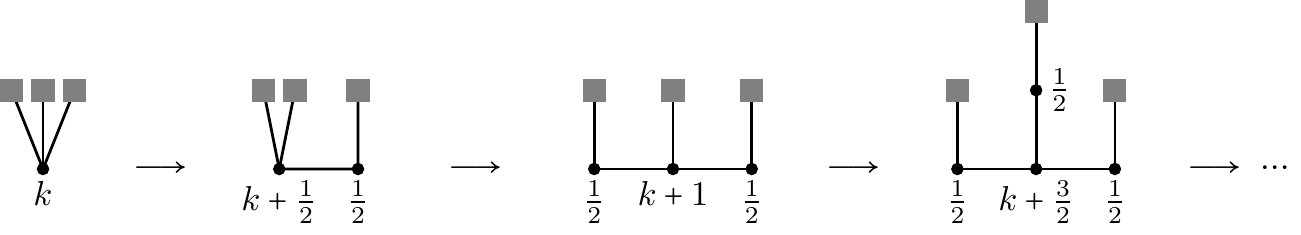}
		\end{split}
	\end{align}
Note that the third diagram in this chain was found and discussed in \cite{Cheng:2020aa} using brane webs. This graph has \CS levels	
\begin{align}\label{branewen3F}
k^{\eff}=
	\begin{bmatrix}
		k+\frac{3}{2}    & 1  &1\\
		1 & 1 &0\\
		1& 0 &1
	\end{bmatrix}  \,.
\end{align}
The fourth and all following graphs contain undecorated gauge nodes that are not attached with matter.  Integrating them out results in balanced triangle graphs, which we now classify as symmetric, partially symmetric, and bifundamental. 

\subsubsection*{Symmetric cases}

Scanning all balanced triangle graphs in \eqref{3fgraph} we find that there are two  symmetric matrices of \CS levels:	
\begin{align}
	\begin{bmatrix}
	\frac{2}{3-2k}   & 	\frac{2}{3-2k}  &	\frac{2}{3-2k}   \\
	\frac{2}{3-2k}   & 	\frac{2}{3-2k}  &	\frac{2}{3-2k}   \\
	\frac{2}{3-2k}   & 	\frac{2}{3-2k}  &	\frac{2}{3-2k}   
\end{bmatrix} \,,	\quad
\begin{bmatrix}
\frac{2k+1}{2k+3}   & -\frac{2}{2k+3}  &-\frac{2}{2k+3}  \\
-\frac{2}{2k+3}  & ~\frac{2k+1}{2k+3}    &-\frac{2}{2k+3}  \\
-\frac{2}{2k+3}  & -\frac{2}{2k+3}   &
~\frac{2k+1}{2k+3}   
\end{bmatrix}  
\end{align}
Effective \CS levels for the first matrix are integer if
\begin{align}
k=\frac{3}{2}+ \frac{1}{n} \,,\quad n \in \mathbf{Z} \setminus  \{ 0\}   \,.\label{cond1}
\end{align}
Let us pick up some special values:
\[k=\frac{1}{2}, 1, 2,\frac{5}{2}, \cdots \]
Their corresponding matrices are
\begin{align}
		\begin{bmatrix}
		~1   & ~1  & ~1~\\
		~1   & ~1  & ~1~\\
		~1   & ~1  & ~1~
	\end{bmatrix}  \,,\quad
	\begin{bmatrix}
	~2   & ~2  & ~2~\\
	~2   & ~2  & ~2~\\
	~2   & ~2  & ~2~
\end{bmatrix} \,,\quad	
	\begin{bmatrix}
	-2   & -2  & -2~\\
	-2   & -2  & -2~\\
	-2   & -2  & -2~
\end{bmatrix} \,,\quad
	\begin{bmatrix}
		-1   & -1  & -1\\
		-1   & -1  & -1\\
		-1   & -1  & -1
	\end{bmatrix}  \,,\quad \cdots \,.
	\end{align}
The vaules $k=1/2, 5/2$ are quite special, since they also ensure that the original theory $U(1)_k+3\F$ is free of parity anomaly. For these values there is also a brane web description.

For the second matrix, the integrality condition takes form
\begin{align}
k= -\frac{3}{2}  + \frac{1}{n'}  \,\quad n' \in \mathbf{Z} \setminus  \{ 0\}   \,.	\label{cond2}
\end{align}
We choose some special values:
\[k=-\frac{5}{2},-2,-1, \frac{1}{2}, \cdots\] 
for which the corresponding matrices are
\begin{align}
	\begin{bmatrix}
		~	2~   & 1~  & 1~\\
		~	1~ & 2~ & 1~\\
		~	1~ & 1~ & 2~
	\end{bmatrix}  \,,\quad
	\begin{bmatrix}
	~	3~   & 2~  & 2~\\
	~	2~ & 3~ & 2~\\
	~	2~ & 2~ & 3~
\end{bmatrix}  \,,\quad
	\begin{bmatrix}
-1~   & -2~  & -2~\\
-	2~ & -1~ & -2~\\
	-	2~ & -2~ & -1~
\end{bmatrix}  \,,\quad
	\begin{bmatrix}
	~0~   & -1~  & -1~\\
	-	1~ & ~0~ & -1~\\
	-	1~ & -1~ & ~0~
\end{bmatrix}  \,,\quad \cdots \,.
	\end{align}
For $k= -5/2 ,-1/2$, we have $k^{\eff}= -1,1$, so the original theory $U(1)_k+3\F$ is also free of anomaly and there is a brane web description. Note that there is no common value satisfying both conditions \eqref{cond1} and \eqref{cond2}. Interestingly, these special values $k=- 5/2, - 1/2$ lead also to symmetric plumbing graphs. More explicitly, the fourth graph in \eqref{3fgraph} can be turned into a balanced triangle  after integrating out the undecorated gauge node.  For $k= -5/2$ we get a triangle
\begin{align}\label{triangle3F52}
	\bsp
\includegraphics[width=2.5in]{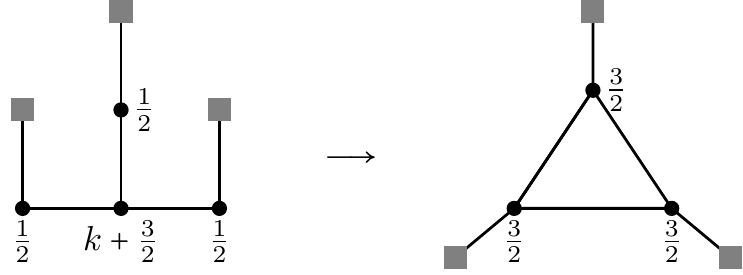}
\esp
\end{align}
with effective \CS levels 
\begin{align}
	k^{\eff}=
\begin{bmatrix}
	~	2~   & 1~  & 1~\\
	~	1~ & 2~ & 1~\\
	~	1~ & 1~ & 2~
	\end{bmatrix}  \,.
\end{align}
For $k =-1/2$ the corresponding triangle is
\begin{align}
	\bsp
\includegraphics[width=2.5in]{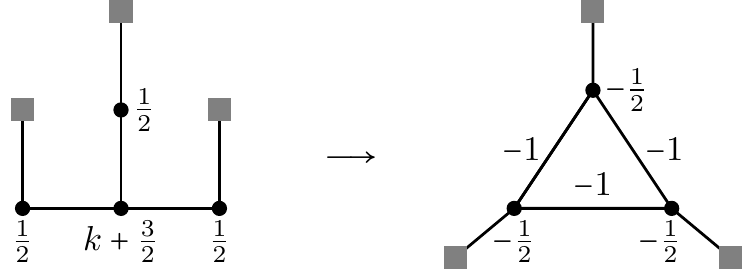}	
\esp
\end{align}
which has effective \CS levels 
\begin{align}
	k^{\eff}=
	\begin{bmatrix}
		0   & -1  &-1\\
		-1 & 0 &-1\\
		-1& -1 & 0
	\end{bmatrix}  \,.
\end{align}


\subsubsection*{Partially symmetric}

Apart from symmetric triangle graphs, we also find partially symmetric ones	
\begin{align}
	\begin{bmatrix}\begin{array}{ccc}
		\frac{2k-1}{2k+1}   & 	-\frac{2}{2k+1}  &	\frac{2}{2k+1}   \\
	-\frac{2}{2k+1}   & 	\frac{2k-1}{2k+1}    &	\frac{2}{2k+1}  \\
	\frac{2}{2k+1}   & 	\frac{2}{2k+1}  &	-\frac{2}{2k+1}   
	\end{array}
	\end{bmatrix} \,,	\quad
	\begin{bmatrix}\begin{array}{ccc}
	\frac{2k-3}{2k-1}   & 	\frac{2}{2k-1}  &	\frac{2}{2k-1}   \\
	\frac{2}{2k-1}   & 	\frac{2}{1-2k}    &	\frac{2}{1-2k}  \\
	\frac{2}{2k-1}   & 	\frac{2}{1-2k}  &	\frac{2}{1-2k}  
	\end{array} 
\end{bmatrix}  \,.
\end{align}
Integrality condition for the first matrix takes form
\begin{align}
	k=-\frac{1}{2}  + \frac{1}{n} \,,\quad n \in \mathbf{Z} \setminus  \{ 0\} \,.\label{condbi1}
\end{align}
For special values:
\[k=-\frac{3}{2}, 1, 0,\frac{1}{2}, \ldots \] 
the corresponding matrices are
\begin{align}\label{U13F32}
	\begin{bmatrix}
\begin{array}{ccc}
	2 & 1 & -1 \\
	1 & 2 & -1 \\
	-1 & -1 & 1 \\
\end{array}
	\end{bmatrix}  \,,\quad
	\begin{bmatrix}
	\begin{array}{ccc}
		3 & 2 & -2 \\
		2 & 3 & -2 \\
		-2 & -2 & 2 \\
	\end{array}
	\end{bmatrix} \,,\quad	
	\begin{bmatrix}
	\begin{array}{ccc}
		-1 & -2 & 2 \\
		-2 & -1 & 2 \\
		2 & 2 & -2 \\
	\end{array}
	\end{bmatrix} \,,\quad
	\begin{bmatrix}
	\begin{array}{ccc}
		0 & -1 & 1 \\
		-1 & 0 & 1 \\
		1 & 1 & -1 \\
	\end{array}
	\end{bmatrix}  \,, \quad \cdots 
\end{align}

For the second matrix, the integrality condition takes form
\begin{align}
	k= \frac{1}{2}  + \frac{1}{n}  \,\quad n \in \mathbf{Z} \setminus  \{ 0\}   \,.	\label{condbi2}\end{align}
For some special values:
\[k=-\frac{1}{2},0,1, \frac{3}{2},\ldots \] 
the corresponding matrices are 
\begin{align}
	\begin{bmatrix}
	\begin{array}{ccc}
		2 & -1 & -1 \\
		-1 & 1 & 1 \\
		-1 & 1 & 1 \\
	\end{array}
	\end{bmatrix}  \,,\quad
	\begin{bmatrix}
	\begin{array}{ccc}
		3 & -2 & -2 \\
		-2 & 2 & 2 \\
		-2 & 2 & 2 \\
	\end{array}
	\end{bmatrix}  \,,\quad
	\begin{bmatrix}
	\begin{array}{ccc}
		-1 & 2 & 2 \\
		2 & -2 & -2 \\
		2 & -2 & -2 \\
	\end{array}
	\end{bmatrix}  \,,\quad
	\begin{bmatrix}
	\begin{array}{ccc}
		0 & 1 & 1 \\
		1 & -1 & -1 \\
		1 & -1 & -1 \\
	\end{array}
	\end{bmatrix}  \,,\quad \cdots
\end{align}
Note that there are some overlaps $k=\pm 1/2, \pm1$, which are allowed for both symmetric and partially symmetric matrices. This implies the equivalence of theories labeled by corresponding triangle graphs.

Let us analyze partially symmetric triangle graphs for special values $k=\pm3/2, \pm 1/2$, as these also allow brane web descriptions of the original theory.  For $k= \pm 3/2$, plumbing graphs degenerate to theories with bifundamental chiral multiplets. 
In particular,  for $k=-3/2$ we obtain a mirror pair found in \cite{Dorey:1999rb}, with the following matrix of \CS levels 
\begin{align}
	k^{\eff}=
	\begin{bmatrix}
		2   & 1  &-1\\
		1 & 2 &-1\\
		-1& -1 & 1
	\end{bmatrix}  \,.\end{align}
By comparing  its triangle graph with $ST$-moved graphs for a bifundamental matter in \eqref{bifundgraph},  we see that this triangle encodes a bifundamental multiplet. Then the plumbing graph can be reduced to
\begin{align}
	\bsp
	\includegraphics[width=1.8in]{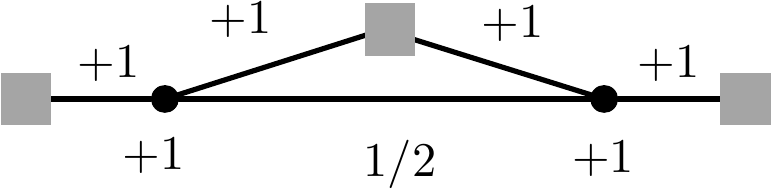}
	\esp
	\end{align}
with effective \CS levels
\begin{align}
k^{\eff} =	\begin{bmatrix}
	2   & ~1  \\
	1 &~ 2
\end{bmatrix} \,.
\end{align}
For $k=3/2$ one can also get a theory labeled by a triangle graph, with mixed \CS levels
\begin{align}	k^{\eff}=
	\begin{bmatrix}
		-1 & -1 &~1\\
		-1& -1 & ~1 \\
			1   & 1  &~0
	\end{bmatrix}   \,.  \end{align}
By comparing with \eqref{bifundgraph}, we see that it also encodes a bifundamental chiral multiplet, and the corresponding plumbing graph is	
\begin{align}
	\bsp
	\includegraphics[width=1.8in]{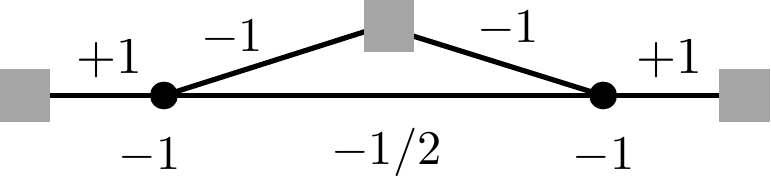}
	\esp
\end{align}
with effective \CS levels
\begin{align}
	k^{\eff} =	\begin{bmatrix}
		-1   & -1  \\
		-1 & -1
	\end{bmatrix} \,.
\end{align}


\subsubsection*{Classification}


Based on the above results, in the following table we summarize some patterns of plumbing graphs for theories $U(1)_k+3\F$, for several values of $k$:
\begin{align}
	\begin{array}{c| c| c| c| c |c| c| c |c |c| c| c }\hline
\text{level}~ k=	&~ -\frac{5}{2}~&\,{\color{red}-2~}&\,-\frac{3}{2}~& \,{\color{red}-1}~& ~~-\frac{1}{2}~&~~ ~{\color{red} 0}~~&~~ ~\frac{1}{2}~~&~ ~{\color{red}1} ~~& ~~\frac{3}{2}~~&~~{ \color{red} 2}~~& ~~\frac{5}{2}~~~   \\ \hline
		\text{symmetric}&\checkmark&\checkmark &&\checkmark&\checkmark&&\checkmark&\checkmark&&\checkmark&\checkmark \\ \hline
			\text{partially sym.}&&&\checkmark&\checkmark&\checkmark&\checkmark\checkmark&\checkmark&\checkmark&\checkmark&& \\ \hline
					\text{ bifund.}&&&\checkmark&&&&&&\checkmark&& \\ \hline
						\text{brane web}&\checkmark&&\checkmark&&\checkmark&&\checkmark&&\checkmark&&\checkmark \\ \hline
		\end{array}
	\end{align}
For $k=0$ there are two matrices, so we marks this value twice.  The values of $k$ marked in red have parity anomaly for the original theory $U(1)_k+3\F$, but we can keep them if we only consider the corresponding triangle graphs.  Note that patterns for theories with $k$ and $-k$ are the same, which confirms the existence of a duality between $U(1)_k+3\F$ and $U(1)_{-k}+3\F$, which is a manifestation of the exchange symmetry $\q \leftrightarrow \q^{-1}$, see \eqref{dualpariu1k} and \eqref{dualpairSTthry}.


\section{Brane webs}\label{secbranewebs}

In this section we analyze properties of branes webs that can be used to engineer some theories labeled by plumbing graphs, in particular the theory $U(1)_k+N_f \F+N_a\AF$. In brane web constructions we use a blue node to denote D3-brane along a perpendicular direction, which we do not draw explicitly. Hence brane webs contain two layers of 5-branes, and a D3-brane is a segment connecting these layers, see \cite{Boer:1997ts,Cheng:2021vtq,Cremonesi:2010ae,Kitao:1999aa} for more details on 3d brane webs.  
In particular,  in what follows we argue that the symmetry $\q \leftrightarrow \q^{-1}$ that exchanges the bare \CS levels $k \leftrightarrow -k$ can be interpreted as the reflection of a brane web along the horizontal line:
\begin{align}
 \q\leftrightarrow\q^{-1} ~~ =	~~ 	\text{reflection} \,,
\end{align}	
which transforms $(p,q)$-brane into $(p,-q)$-brane.  On the other hand, the rotation of a brane web by an angle $\pi/2$ transforms $(p,q)$-brane into $(q,p)$-brane. Furthermore,  $\mathcal{S}$-duality in type IIB string theory is the operation that exchanges $(p,q)$-brane and $(-q,p)$-brane, so it is basically a combination of reflection and rotation of brane webs:	
\begin{align}
	~~ 	\text{rotation} + \text{reflection} ~~ =	~~ \text{$\mathcal{S}$-duality}.
\end{align}	
In practice, one can simply view a rotation as the $\mathcal{S}$-duality operation, as it is always easy to reflect brane webs.

In some special cases, the reflection and $\mathcal{S}$-duality give rise to the same brane web.  
Some brane webs are even self-dual, i.e.  invariant under $\mathcal{S}$-duality.  For theories $U(1)_k+N_f\F+N_a\AF$,   brane webs are self-dual only for some special values of $k$. This conclusion also follows from analysis of plumbing graphs. In the manipulation of brane webs, a crucial role is played by the movement of 5-branes plays, which makes some brane webs symmetric along the diagonal direction and suggests $\mathcal{S}$-duality between theories.


\subsection{$U(1)_k + 1\F$}\label{Sduality}

To start with, we discuss special values of $k$ given in \eqref{specialk1F}. These values are related by the exchange symmetry $\q \leftrightarrow \q^{-1}$. Other values of $k$ are also meaningful, but their brane webs are not symmetric.   
For $k=\pm1/2$ one gets two dual theories of the mirror triality \eqref{triviliaty} that we discussed before:	
\begin{align}
	U(1)_{\frac{1}{2}} + 1 \F  ~~\longleftrightarrow~~ 	U(1)_{-\frac{1}{2}} + 1\F  \,.
\end{align}
The corresponding plumbing graphs are
\begin{align}\label{freenode}
	\bsp
	\btik
	\begin{scope}
\filldraw (0,0) node[below]{$1$}circle(2pt);
\draw[thick] (0,0)--(0,0.3) ;
\node at (0,0.3) {$\graybox$};
\end{scope}
\node[xshift=1cm] at (0,0) {$\leftrightarrow$};
	\begin{scope}[xshift=2cm]
	\filldraw (0,0) node[below]{$0$}circle(2pt);
	\draw[thick] (0,0)--(0,0.3) ;
	\node at (0,0.3) {$\graybox$};
\end{scope}
	\etik
	\esp
	\end{align}
where the numbers denote effective \CS levels.
Brane webs for these theories take form
\begin{align}\label{U11Fdual}
	\bsp
		\begin{tikzpicture}[scale=0.8]
		\draw[thick] (0,0)--(2,0)--(3,-1); \draw[line width=2 pt, gray] (0.5,1.5)-- (0.5,-1); \draw[thick](2,0)--(2,1.5);
		\draw[blue,fill] (0.5,0) circle(3 pt){}
		;
			\draw[dashed, orange] (-0.5,0)--(3,0);
		\node at (1.5,-2) { $ U(1)_{1/2} + 1 \F$  with $k^{\eff} =1$} ;
		\node at (5	,0.2 ) {$\xlongleftrightarrow{\text{reflection}}$};
	\end{tikzpicture}
	\begin{tikzpicture}[scale=0.8]
		\draw[thick] (0,0)--(2,0)--(2,-1.5); \draw[line width=2 pt, gray] (0.5,1)-- (0.5,-1.5); \draw[thick](2,0)--(3,1);
		\draw[blue,fill] (0.5,0) circle(3 pt){}
		;
			\draw[dashed, orange] (-0.5,0)--(3,0);
		\node at (1.5,-2.5) { $ U(1)_{-1/2} + 1 \F$  with $k^{\eff} =0$} ;
	\end{tikzpicture}
\esp
\end{align}
This mirror pair is related by reflecting the 3d brane web along the horizontal line (the dashed orange line) where the D3-brane is located. Notice that the D3-brane is invariant under such a reflection.

One can perform the $\mathcal{S}$-duality by rotating clockwise the right graph in \eqref{U11Fdual} by $\pi/2$. Here we view the D3-brane as accompanying a local conifold through a geometric transition, and from a perspective of 5d $\N=1$ brane webs, this D3-brane is also a 5-brane. Then this  $\mathcal{S}$-duality is the same as for 5d brane webs:
\begin{align}\label{U11Fdual2}
	\bsp
	\begin{tikzpicture}[scale=0.8]
	\begin{scope}
		\draw[thick] (0,0)--(2,0)--(2,-1.5); \draw[line width=2 pt, gray] (0.5,1)-- (0.5,-1.5); \draw[thick](2,0)--(3,1);
		\draw[blue,fill] (0.5,0) circle(3 pt){}
		;
		\node at (5	,0 ) {$\xlongrightarrow{\text{rotation}}$};
		\end{scope}
		\begin{scope}[rotate around={-90:(5,-3.7)}]
		\draw[thick] (0,0)--(2,0)--(2,-1.5); \draw[line width=2 pt, gray] (0.5,1)-- (0.5,-1.5); \draw[thick](2,0)--(3,1);
		\draw[dashed] (2,0) --(1,-1) ;
		\draw[blue,fill] (0.5,0) circle(3 pt){}
		;
	\end{scope}
	\end{tikzpicture}
	\esp
\end{align}
It follows that the left graph in \eqref{U11Fdual} is $\mathcal{S}$-dual to the right graph in \eqref{U11Fdual2}. We emphasize that although these two brane webs are manifestly $\mathcal{S}$-dual to each other, the theory labeled by the right graph in \eqref{U11Fdual2} is hard to identify. An interesting point is that the $\mathcal{S}$-duality only changes the position of the thick gray line and the D3-brane. 

Other special values for \CS levels are $k=\pm 3/2$, which correspond to a dual pair in \eqref{equivalence032}
\begin{align}
	\bsp
	U(1)_{\frac{3}{2}} + 1 \F  ~~\longleftrightarrow~~ 	U(1)_{-\frac{3}{2}} + 1\F \,,
	\esp
\end{align}	
whose associated plumbing graphs are
\begin{align}
	\bsp
	\btik
	\begin{scope}
		\filldraw (0,0) node[below]{$2$}circle(2pt);
		\draw[thick] (0,0)--(0,0.3) ;
		\node at (0,0.3) {$\graybox$};
	\end{scope}
	\node[xshift=1cm] at (0,0) {$\longleftrightarrow$};
	\begin{scope}[xshift=2cm]
		\filldraw (0,0) node[below]{$-1	$}circle(2pt);
		\draw[thick] (0,0)--(0,0.3) ;
		\node at (0,0.3) {$\graybox$};
	\end{scope}
	\etik
	\esp
\end{align}
Their brane webs are related by the reflection along the dashed orange line:
\begin{align}\label{U11Fdual32}
	\bsp
	\begin{tikzpicture}[scale=0.8]
				\draw[dashed,orange] (-0.8,0 )--(3, 0)  ;
		\draw[thick] (0,0)--(2,0)--(2,2); \draw[line width=2 pt, gray] (0,-0.5)-- (2.5,2); \draw[thick](2,0)--(2.5,-0.5);
		\draw[blue,fill] (0.5,0) circle(3 pt){}
		;
		\node at (1.2,-1.5) { $ U(1)_{3/2} + 1 \F$  with $k^{\eff} =2$} ;
		\node at (5,0.5) {$\xlongleftrightarrow{\text{reflect}}$};
		\end{tikzpicture}
		\begin{tikzpicture}[scale=0.8]
				\draw[dashed,orange] (-0.8,2 )--(3, 2)  ;
		\draw[thick] (0,2)--(2,2)--(2,0); \draw[line width=2 pt, gray] (0,2+0.5)-- (2.5,0); \draw[thick](2,2)--(2.5,2.5);
		\draw[blue,fill] (0.5,2) circle(3 pt){}
		;
		\node at (1.5,-1) { $ U(1)_{-3/2} + 1 \F$  with $k^{\eff} =-1$} ;
	\end{tikzpicture}
\esp
	\end{align}
Analogously, one can also rotate the brane web for $U(1)_{-3/2}+1\F$ to identify the $\mathcal{S}$-dual brane web. In comparison to the last example, the $\mathcal{S}$-dual operation only changes the position of the D3-brane from one intersection point to another one. 

The two brane webs in \eqref{U11Fdual} can be obtained by decoupling a D5-brane of \eqref{U102F} or \eqref{SQEDweb} respectively, i.e. sending its position to infinity. In what follows we show that brane webs in  \eqref{U11Fdual} and  \eqref{U11Fdual32} are special, as they are sub-webs of many dual theories. By adding more 5-branes to these sub-webs, one can produce many other dual theories.



\subsection{$U(1)_k + 2\F$}

In turn,  we consider the theory $U(1)_0 + 2 \F$, which can be engineered by two equivalent brane webs
\begin{align}\label{U102F}
	\bsp
	\btik
	\begin{scope}
		\draw[thick] (-0.2,0)--( 1,0)--(1,1)--(-0.2,1) 
		(1,0)--( 1.5, -0.5)
		(1,1)--(1.5,1.5) ;
		\draw [line width=2 pt, gray](0.2,-0.5)--(0.2,1.5) ;
		\draw[blue,fill] (0.2,0) circle (3pt);
		;
		\node at (3,0.5) {$\xlongrightarrow{\text{move} ~\F}$};
	\end{scope}
	\begin{scope}[xshift=4.5cm]
		\draw[thick] (-0.2,0)--( 1.2,0)--(1.2,1.5)
		(1.2,0)--( 1.7, -0.5)
		;
		\draw [line width=2 pt, gray](0.2,-0.5)--(0.2,1)
		(0.2,1)--(1.7,1)
		(0.2,1)--(-0.2,1.5)
		;
		\draw[blue,fill] (0.2,0) circle (3pt);
		;
	\end{scope}
	\etik
	\esp
\end{align}
for which the effective \CS level $k^{\eff} =0 +2 \times \frac{1}{2} =1$. 
The right brane web has an interesting double layer structure. It is obtained by moving a fundamental D5-brane vertically along the direction of D3-brane from the thin sub-web to the thick gray line, which we do not draw explicitly. After this move, projecting these two sub-webs along the direction of the D3-brane leads to the right brane web in \eqref{U102F}. See  \cite{Cheng:2021vtq} for more details of this move. For this theory, one can see that the $\mathcal{S}$-duality of type IIB string theory changes the vacuum of the D3-brane on brane webs in \eqref{U102F}, which is the exchange of the blue node from one intersection point to another. The plumbing graphs in the orbit of $ST$-moves are
\begin{align}\label{2Fkeql0}
	\bsp
	\includegraphics[]{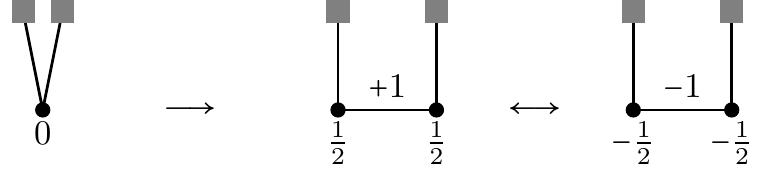}
	\esp
\end{align}
We obtain these graphs by moving the two matter fields respectively, using \eqref{ststF}.

It is also easy to draw brane webs for $k= \pm 1, \pm 2$.  Such brane webs are related by reflections of branes along the horizontal line. For these \CS values, one can get corresponding brane webs by simply adding one more D5-brane to brane webs in  \eqref{U11Fdual} and  \eqref{U11Fdual32}.


\subsection{$U(1)_k + 1\F+1\AF$}

The next example we consider is the theory $U(1)_k + 1 \F + 1 \AF$. The effective \CS level $k^{\eff} =k$ if both matter fields have positive masses. The corresponding plumbing graphs are
\begin{align}\label{1F1AFk}
	\bsp
	\includegraphics[]{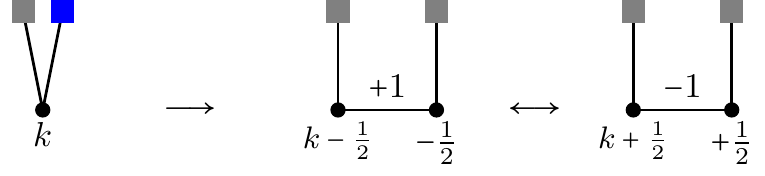}
	\esp
\end{align}
where we use the dual triality \eqref{ststAF} to move the matter. In particular,  for $k=0$  we get the theory  $U(1)_0 +1\F +1\AF$, which is the SQED in the dual pair SQED-XYZ. We show its brane webs below
\begin{align}\label{SQEDweb}
	\bsp
	\btik
	\begin{scope}
		\draw[thick,scale=0.8] (-0.2,0)--( 1,0)--(2,1)--(2.8,1) 
		(1,0)--( 1, -0.8)
		(2,1)--(2,1.8) ;
		\draw [line width=2 pt, gray,scale=0.8](0.2,-0.8)--(0.2,1.8) ;
		\draw[blue,fill] (0.16,0) circle (3pt);
		;
		\node at (3.5,0.2) {$\xlongrightarrow{\text{move}~ \AF }  $};
	\end{scope}
	\begin{scope}[xshift=5.5cm]
		\draw[thick,scale=0.8] (-0.5,0)--( 1.2,0)--(2,0.8)
		(1.2,0)--( 1.2, -0.8)
		;
		\draw [line width=2 pt, gray,scale=0.8](0.2,-0.8)--(0.2,1) --(-0.5,1)
		(0.2,1)--(1,1.8)
		;
		\draw[blue,fill] (0.16,0) circle (3pt);
		;
	\end{scope}
	\etik
	\esp
\end{align}
which corresponds to $k^{\eff}=k +1/2-1/2=\tan \theta =0 $, where $\theta$ is the relative angle between NS5-brane and NS5'-brane.  See e.g.\cite{Cheng:2021vtq} for more details on the relative angles.  These two brane webs are also related by moving a D5-brane perpendicularly; equivalently, from a perspective of 5d theories, it corresponds to moving horizontally a 7-brane attached at the infinity of this D5-brane.

The plumbing graphs for the mirror pair SQED-XYZ  are 
\begin{align}\label{SQEDXYZdualpari}
	\bsp
	\includegraphics[]{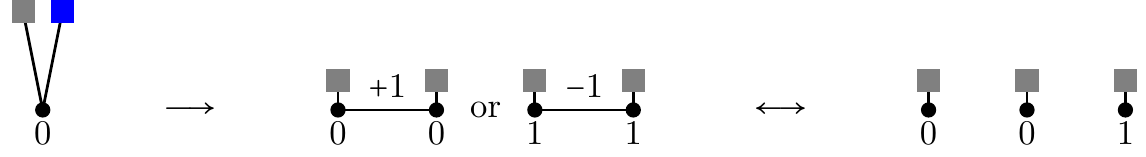}
	\esp
\end{align}
where the last graph denotes the XYZ model. The graphs in the middle are the two-node graphs, and the last graph is the three-node graph for the 2-3 move discussed in section \ref{secgauging}. One can use the superpotential triangles \eqref{trianglesgraph} to check that the SQED-XYZ dual pair only satisfies the unlinking relation. This example shows that superpotential triangles could have no mixed \CS levels between gauge nodes. 

Moreover, the right brane webs in \eqref{U102F} and \eqref{SQEDweb} are obviously $\mathcal{S}$-dual to themselves. One can decorate these $\mathcal{S}$-dual brane webs by adding 5-branes to get other self-dual brane webs. For example, the theory $U(1)_{0} +2\F+2\AF$ is also $\mathcal{S}$-dual to itself.  Its two equivalent brane webs take form
\begin{align}
	\bsp
	\btik
		\begin{scope}[xshift=-4cm]
		\draw[thick] (-0.2,0)--( 1.2,0)--(1.2,1.5)
		(1.2,0)--( 1.7, -0.5)
	(1.7,-1)--(1.7,-0.5)--(2.2,-0.5)
		;
		\draw [line width=2 pt, gray](0.2,-0.5)--(0.2,1)
		(0.2,1)--(1.7,1)
		(0.2,1)--(-0.2,1.5)
		(-0.8,1.5) -- (-0.2,1.5 )--(-0.2,2 )
		;
		\draw[blue,fill] (0.2,0) circle (3pt);
		;
	\end{scope}
	\begin{scope}
		\draw[thick,scale=0.8] (-0.5,0)--( 1.2,0)--(2,0.8)
		(1.2,0)--( 1.2, -0.8)
	(2,1.5)--(2,0.8)--(3,0.8)
		;
		\draw [line width=2 pt, gray,scale=0.8](0.2,-0.8)--(0.2,1) --(-0.5,1)
		(0.2,1)--(1,1.8)
	( 1.8,1.8)--(1,1.8)--(1,2.8)
		;
		\draw[blue,fill] (0.16,0) circle (3pt);
		;
	\end{scope}
	\etik
	\esp
\end{align}
Note that this theory can be viewed as a self-dual 3d $\N=4$ theory with gauge group $U(1)_0$ and two fundamental hypermultiplets \cite{Intriligator:1996ex}. Adding more segments would lead to non-toric intersections,  however we postulate that $\mathcal{S}$-dual theories should be $U(1)_0 + N_a \F+ N_f\AF$ with the condition that $N_a+N_f$ is an even number.


\subsection{Mirror pairs}\label{secmirrorpair}

Mirror symmetry exchanges Coulomb and Higgs branches of the moduli spaces of vacua of dual theories. In this section we show that some abelian mirror dual theories in \cite{Dorey:1999rb} have the same plumbing graphs, and some mirror dual pairs can be simply interpreted as a manifestation of $\mathcal{S}$-duality of type IIB string theory. We have discussed a mirror pair with one matter nodes before, so we start from theories with two matter nodes.

A mirror pair found in \cite{Dorey:1999rb} is	
\begin{align}\label{U12Fdual1}
	U(1)_{+1} + 1 \F +1 \AF  ~~\xlongleftrightarrow{~S~}~~ 	U(1)_{-1} + 2\F.
\end{align}
A brane web for $U(1)_{+1} + 1 \F +1 \AF$ theory takes form
\begin{align}\label{1F1AF2F}
	\bsp
	\includegraphics[width=2.5in]{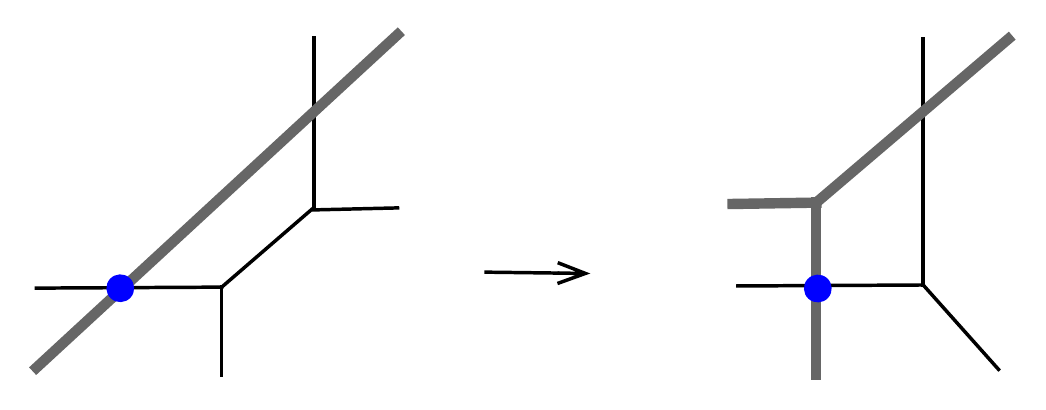}
	\esp
\end{align}
where we first attach a 7-brane at infinity of the anti-fundamental D5-brane and then Hanany-Witten move it to the left infinity to produce an equivalent brane web. 
This is equivalent to moving the 5-brane for $\AF$ perpendicularly.
A brane web for $U(1)_{-1} + 2\F$ theory is
\begin{align}\label{U12Fdual22}
	\bsp
	\includegraphics[width=2.5in]{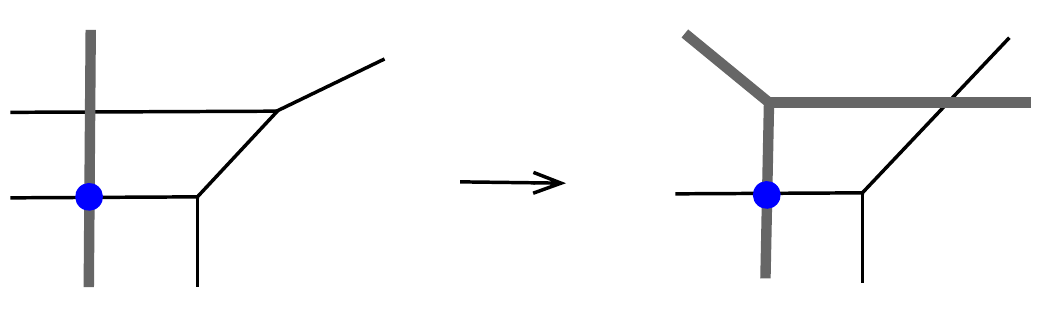}
	\esp
\end{align}
where we also move the fundamental D5-brane to the sub-web along a gray layer.  It is obvious that brane webs on the right in \eqref{U12Fdual1} and \eqref{U12Fdual22} are related by $\mathcal{S}$-duality. Since these two theories are related by mirror symmetry,  in this case it can be interpreted as the $\mathcal{S}$-duality of type IIB string theory.


The above example can be also presented in terms of plumbing graphs. Although it is not clear whether plumbing graphs may be used to distinguish and exchange different branches of moduli spaces, we find that for this mirror pair they are the same:
\begin{align}\label{U12Fkeql1}
	\bsp
\includegraphics[]{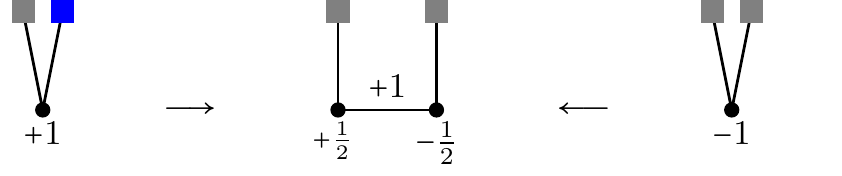}
	\esp
\end{align}
where we use \eqref{ststAF} to move a matter node. Similarly, the exchange symmetry of plumbing graphs suggests another mirror dual pair
\begin{align}\label{U12Fdual2}
	U(1)_{-1} + 1 \F + 1 \AF ~~\xlongleftrightarrow{~~}~~ 	U(1)_{+1} + 2\F \,,
\end{align}
so that corresponding plumbing graphs are also equivalent:	
\begin{align}\label{U12Fkelminus1}
	\bsp
\includegraphics[]{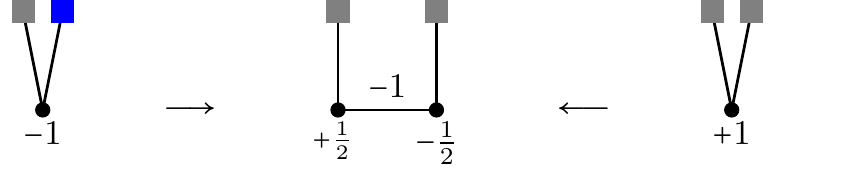}
	\esp
\end{align}
In addition,  the exchange symmetry $\q \leftrightarrow \q^{-1}$ (reflection) relates the dual pairs \eqref{U12Fdual1} and \eqref{U12Fdual2}. This extends the original mirror pair to a big dual pair containing four theories $\big{\{}U(1)_{\pm1}+2\F \,, ~U(1)_{\pm1}+1\F+1\AF\big{ \}}$, which are dual to each other through $\mathcal{S}$-duality and reflection. Brane webs for this dual pair take form: 
\begin{align}\label{mirrorwebsum}
	\bsp
	\includegraphics[width=4.5in]{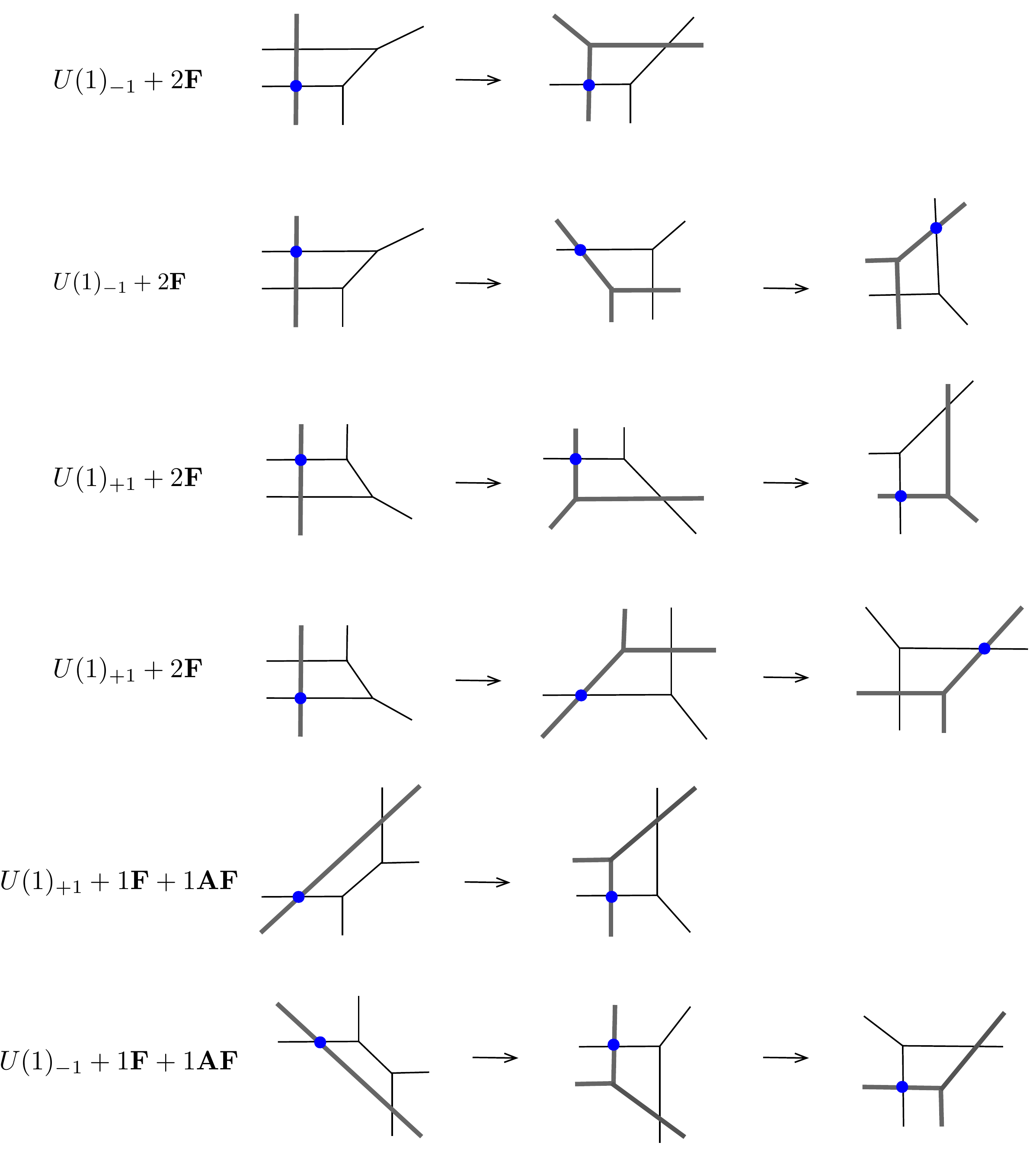}  \,.
	\esp
\end{align}
Note that for the theory $U(1)_{\pm 1} +2 \F$, the two Higgs branches of vacua correspond to two intersections on brane webs.

By adding a half-horizontal 5-brane on the brane webs in  \eqref{mirrorwebsum}, one can produce a new dual pair:
\begin{align}\label{dual2F1AF}
	U(1)_{- \frac{1}{2}} + 2 \F+1 \AF ~~\xlongleftrightarrow{~S~}~~	U(1)_{ \frac{1}{2}} + 1\F+2 \AF \,,
	\end{align}
for which the effective \CS levels for both theories also vanish $k^{\eff} =0$. One can also derive this dual pair by adding a matter node $\F$ to both sides of the dual graphs in \eqref{higgsinggraph} (or equivalently \eqref{higgsinggeneric}) and set $b=-1/2$ (or $b=-1$). The mirror triality in \eqref{ststAF} could turn this $\F$ into $\AF$. Then one gets corresponding graphs  for the dual pair in \eqref{dual2F1AF}. Note that here the operation in \eqref{higgsinggraph} should be $\mathcal{S}$-duality in the language of 3d $\N=4$ brane webs, so in total the operation we encounter for \eqref{dual2F1AF} is $\mathcal{S} \circ (ST)^2$. The dual theories in \eqref{dual2F1AF} cannot be directly related through $ST$-moves only. We note that this dual pair is given by the $\mathcal{S}$-duality of 3d brane webs:
\begin{align}\label{U12F1AFmirror}
	\bsp
	\includegraphics[width=4in]{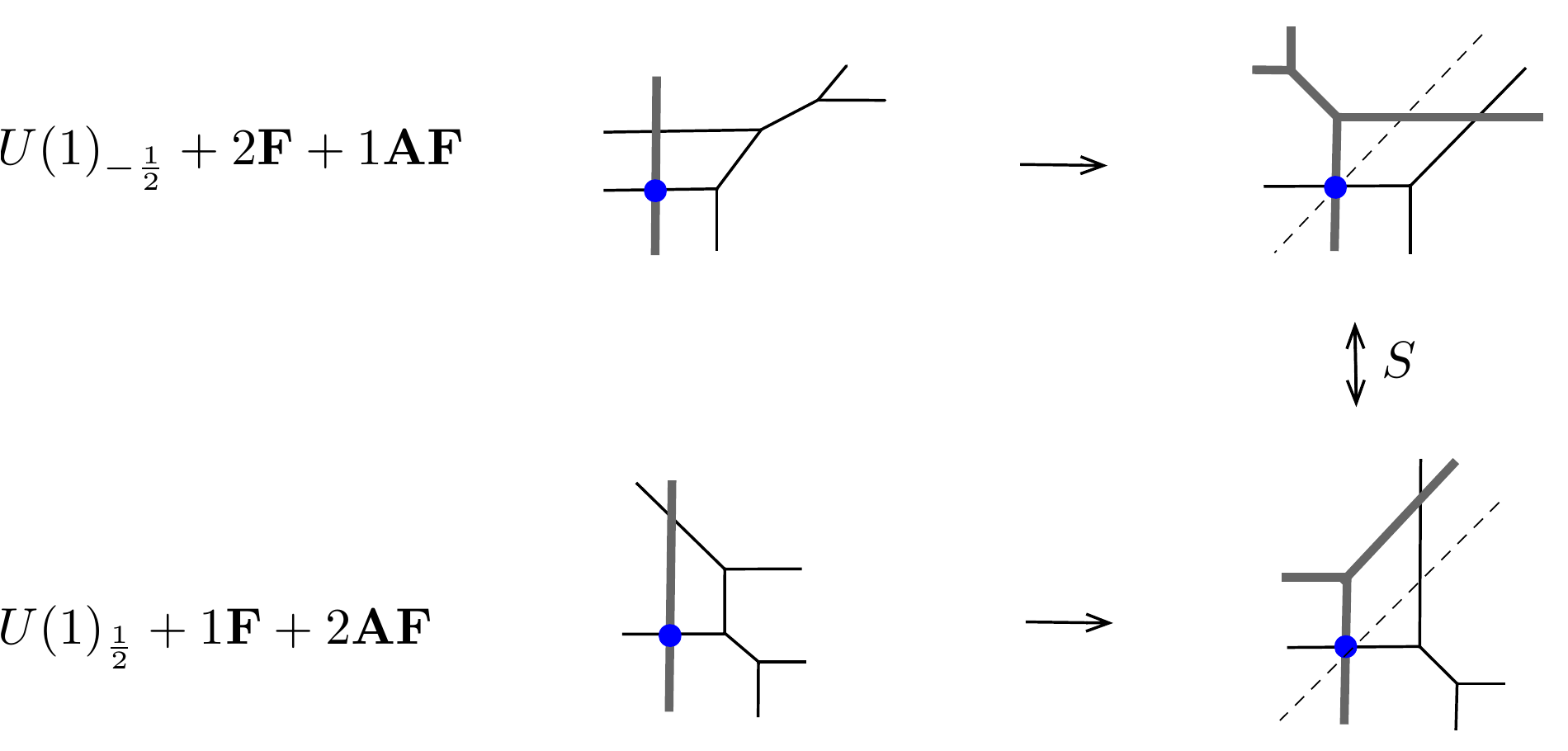} 
	\esp
\end{align}
Applying reflection leads to another dual pair
\begin{align}
	U(1)_{ \frac{1}{2}} + 2 \F+1 \AF ~~\xlongleftrightarrow{~S~}~~	U(1)_{- \frac{1}{2}} + 1\F+2 \AF ,
\end{align}
so that corresponding brane webs are the reflection of brane webs in \eqref{U12F1AFmirror}.

It appears to us that it is not possible to produce other dual pairs for theories with a single gauge group, i.e. $U(1)_k+N_f\F+N_a\AF$. In principle it is possible to apply $\mathcal{S}$-duality to any brane web,  but it may not be easy to identify the corresponding theories from dual brane webs, or dual brane webs may even not have physical interpretation. 


\subsubsection*{Abelian mirror pairs}

Generic mirror dual theories contain bifundamental matter.  It is possible to show equivalence of mirror abelian theories found in \cite{Dorey:1999rb}.  The first theory, referred to as theory A, is defined as
\begin{align}
	&\text{theory A}:~~ U(1)_{-\frac{N_f}{2}} + N_f \F .
	\end{align}
The  mirror dual theory B is a linear quiver theory with bifundamental chiral multiplets between gauge nodes and two fundamental matter fields on two tails.  Its conventional quiver diagram takes form
\begin{align}\label{quivertheoryB}
	&	\text{theory B}:~~
[1]-U(1)-U(1)-\cdots-U(1)-[1]\,,
	\end{align}
for which the $a$-th chiral multiplet has charge $(\delta_{a, i}, \delta_{a, i+1})$ under the $i$-th and $(i+1)$-th gauge nodes. This theory also has mixed \CS levels $ k_{ij}=\delta_{ij}-\frac{1}{2}\delta_{i,j-1} -\frac{1}{2}\delta_{i-1,j}$ with $i=1\,,2\,,\cdots\,, N_f-1$ and $j=1\,,\cdots\,, N_f$, which are not shown in the quiver in  \eqref{quivertheoryB}.

Let us show equivalence of these two theories. First, we perform $ST$-move for each chiral multiplet of theory A, which produces a star graph with undecorated gauge node $U(1)_0$ in the center. Then, we separate this $U(1)_0$ gauge node by introducing many new gauge nodes $U(1)_0$ using Kirby moves \eqref{deltasplit}. After integrating out some particular $U(1)_0$,  we get a plumbing graphs for theory B. We illustrate this in the example of $U(1)_{-2}+4\F$ theory:
\begin{align}
	\bsp
	\includegraphics[width=5.2in]{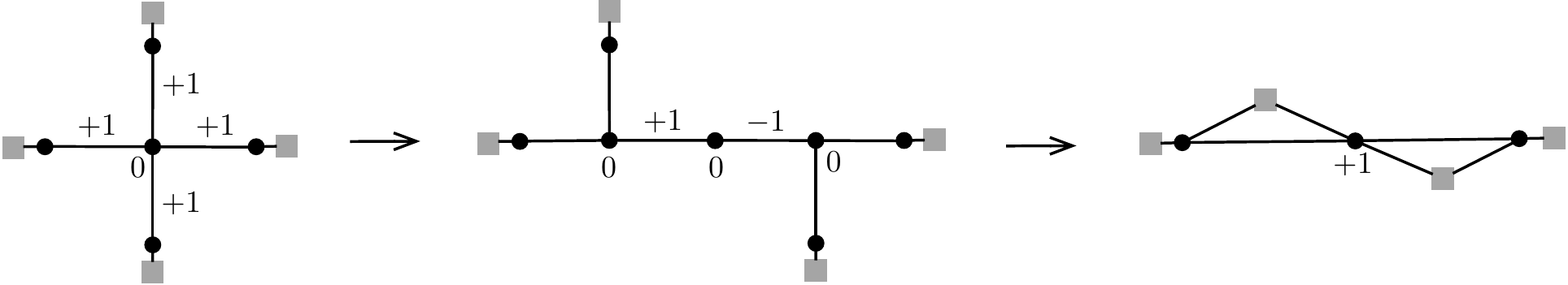}
	\esp
\end{align}
We have also made a similar check for some other mirror dual theories with generic charges, and hope that all mirror dual abelian theories have equivalent plumbing graphs.

\acknowledgments

We thank Lakshya Bhardwaj, Jin Chen,  Sung-Soo Kim, Ioannis Lavdas, Satoshi Nawata for helpful discussions.  This work has been supported by the TEAM programme of the Foundation for Polish Science co-financed by the European Union under the European Regional Development Fund (POIR.04.04.00- 00-5C55/17-00). The work of S.C. has been also supported by NSFC Grant No.11850410428.

\bigskip

\newpage
\appendix

\section{$q$-Pochhammer symbol}\label{pochappendix}

In various computations we use various identities satisfied by $q$-Pochhammer symbol $\le(x; \q  \r)_n$:
\begin{align}
&	\le(x; \q  \r)_n := \prod_{k=0}^{n-1} ( 1-x \q^k) \,, \quad 	(x;\q)_n \sim e^{ \frac{1}{\hbar} \le(   \Li_2(x) -\Li_2\le( x \q^n \r) \r) }  \,,~~ \q := e^{\hbar}.\\
& \le(x;\q^{-1}  \r)_\inf = \frac{1}{  \le( x \q;\q \r)_\inf} \,,\qquad  \le( x;\q \r)_n = \frac{ 1}{ \le( x \q^n;\q  \r)_{-n} }
 \,,\qquad  (x;\q)_n = \frac{ (x;\q)_\inf}{ (x \q^n ;\q)_\inf } \,,
	\\
&	\Li_2(x)+ \Li_2(x^{-1}) = -\frac{1}{2} \le( \log(-x) \r)^2 -\frac{\pi^2}{6} \,,\quad 	\Li_2(1) =\frac{\pi^2}{6}	\,, \\
&	\theta( x;\q )  := \frac{  \sum_{n \in \mathbf{Z}} \q^{\frac{n^2}{2}} x^n  }{(\q;\q)_\inf  }  \,,	 \quad 	\frac{\theta(\q^n z;\q) }{\theta(z;\q) } =\q^{-\frac{n^2}{2}} z^{-n} \,, \quad
	\theta(x;\q) \sim e^{-\frac{\pi^2}{6 \hbar} +\frac{\hbar}{24}}   \, e^{ \frac{1}{ \hbar} \le( -\frac{1}{2} ( \log\, x)^2 \r) }  
 \\
&		\theta ( -\q^{ \frac{1}{2}} x ) = (\q x;\q)_\inf (x^{-1};\q)_\inf \,, \quad			\theta ( -q^{ -\frac{1}{2}} x ) = (x;\q)_\inf (\q x^{-1};\q)_\inf \,, \\
& \theta(x;\q) :=\theta(x^{-1};\q)= \le(-\q^{1/2} x;\q\r)_\inf \le(-q^{1/2} x^{-1};\q\r)_\inf = \frac{ \le(-\q^{1/2} x;\q\r)_\inf}{ \le(-\q^{-1/2} x^{-1};\q^{-1}\r)_\inf } \,.
\end{align}
$q$-Pochhammer symbol is also related to MacMahon function:
\begin{align}
&  M(Q,\q) =\prod_{i,j =1}^{\inf} ( 1- Q \q^{i+j-1})  \,,\\
& (\q z;\q)_\inf = \frac{ M(z^{-1};\q) }{M( \q z^{-1};\q) } =\frac{ M(z;\q) }{M( \q z;\q) } \,,  \quad M(z;\q):= \PE\bigg[ - \frac{\q z}{ (1-\q)^2} \bigg]  \,,   \label{normal} \\
&  \frac{ M(z;\q)}{M( z^{-1};\q) } =  \frac{ M(\q z;\q)}{M( \q z^{-1};\q) } =\text{constant} \,.
	\end{align}
The doube-sine function is defined as
\begin{align}
&	s_b(x) :=   \prod_{n_1, n_2 \geqslant 0 } \frac{ n_1 b + n_2 b^{-1} + Q/2 -i\, x  }{  n_1 b + n_2 b^{-1} + Q/2 + i\, x        } \,, ~
&	s_b\left(  \frac{i Q}{2}  \pm x \right)	= \frac{ 1}{ 2\, \cosh\left( \pi Q x \right)   } \,.
	\end{align}

\section{$SL(2, \mathbf{Z})$ 	}\label{SL2Zappendex}

For non-abelian plumbing theories with gauge group $SU(2)$, the edge denotes the duality wall 	$T[SU(2)]$ theory \cite{Terashima:2011qi,Gaiotto:2008ak}, which for some manifolds such as lens spaces are interpreted as the $S$-transformation in $SL(2,\mathbf{Z})$.  $SL(2, \mathbf{Z})$ is a global symmetry of 3d gauge theories. The polarization of three manifolds $M_3$ that engineer 3d theories can be interpreted in terms of $T$ and $S$ generators in this group \cite{Dimofte:2011ju,Witten:2003ya}. 

In other notation we choose 
\begin{align}
	S=	\begin{bmatrix}
		~0~  & -1  \\
		~1~ & ~0 \\
	\end{bmatrix} \,,\quad 
	T=	\begin{bmatrix}
		~1  & ~0~  \\
		~1 & ~1 ~\\
	\end{bmatrix} \,,	
\end{align}
so that standard relations hold: $(ST)^3 =1$, $S^2=-1$, and $S^4={1}$.  Notice that there is an equivalence
\begin{align} \label{equivalence}
(ST)^2 = ST ST =(ST)^{-1} = T^{-1} S^{-1} = - T^{-1} S \,,
\end{align} where
\begin{align}
	ST=	\begin{bmatrix}
		-1  & -1  \\
		~1~ & ~0 \\
	\end{bmatrix} \,,\quad 
	TS=	\begin{bmatrix}
		~0 & ~-1~  \\
		~1 & ~-1 ~\\
	\end{bmatrix} \,,
	\quad 
	(ST)^2=	\begin{bmatrix}
		~0 & ~1~  \\
		-1 & -1 ~\\
	\end{bmatrix} \,.
\end{align}
For linear plumbing graphs, some Kirby moves can be interpreted using the equivalence \eqref{equivalence}. For example, the Kirby move
 \eqref{example1integ} can be represented as 
\begin{align}
	T^{k_1} (S T^{\mp 1}S )	T^{k_2}  ~~\rightleftarrows ~~ T^{k_1 \pm 1} S^{\pm 1} T^{ k_2 \pm 1}  \,,
\end{align}
which is obviously correct because of the equivalence	
\begin{align}
	ST^{-1} S=TST\,,\quad STS=T^{-1}S^{-1} T^{-1}  \,. \end{align}


\newpage

\bibliographystyle{JHEP}
\bibliography{ref}





\end{document}